\documentclass[reprint,aps,prl,groupedaddress,10pt]{revtex4-2}

\usepackage{packages}
\usepackage{notations}

\newcommand{\TOne}{
\begin{tikzpicture}
\draw (0,1) -- (5,1) ; 
\draw[dashed] (0,-0.1) -- (5,-0.1) ; 

\filldraw[black] (1,-0.7) circle (2pt);
\filldraw[black]  (1,-0.7) node[anchor= north east] {$y$} ;

\filldraw[black]  (5,-0.1) node[anchor= west] {$x$} ; 
\filldraw[black] (5,-0.1) circle (2pt);

\filldraw[black]  (0,1) node[anchor= east] {$0$} ;
\filldraw[black] (0,1) circle (2pt);

\draw [thick] (0 ,1)  node[anchor= east] {~} to [ curve through ={(0.5,0) ..(1,-0.7) .. (3,-0.45)} ] (5,-0.1)  ; 

\draw[thick] (0,-1.1)  node[anchor= east] {~}-- (0,1.5) node[anchor=south ] {$0$};
\draw[thick] (1,-1.1) -- (1,-0.1);
\draw[dotted] (1,-0.1) -- (1,1.5)node[anchor=south ] {$1$} ;
\draw[thick,] (5,-1.1) -- (5,1.5) node[anchor=south ] {$n$};

\filldraw[black] (-1.2,0) node[anchor= west] {$T_1=$};

\end{tikzpicture}
}

\newcommand{\TTwo}{
\begin{tikzpicture}
\draw (0,1) -- (5,1) ; 
\draw[dashed] (0,-0.9) -- (5,-0.9) ; 

\filldraw[black] (4,-0.4) circle (2pt);
\filldraw[black]  (4,-0.4) node[anchor= north east] {$y$} ;

\filldraw[black]  (5,-0.9) node[anchor= west] {$x$} ; 
\filldraw[black] (5,-0.9) circle (2pt);

\filldraw[black]  (0,1) node[anchor= east] {$0$} ;
\filldraw[black] (0,1) circle (2pt);

\draw [thick] (0 ,1)  node[anchor= east] {~} to [ curve through ={(0.5,0) ..(3,-0.05) .. (4,-0.4)} ] (5,-0.9)  ; 

\draw[thick] (0,-1.1)  node[anchor= east] {~}-- (0,1.5) node[anchor=south ] {$0$};
\draw[dotted] (4,-1.1) -- (4,-0.9);
\draw[thick] (4,-0.9) -- (4,1.5)node[anchor=south ] {$n-1$} ;
\draw[thick] (5,-1.1) -- (5,1.5) node[anchor=south ] {$n$};

\filldraw[black] (-1.2,0) node[anchor= west] {$T_2=$};

\end{tikzpicture}
}

\newcommand{\TThree}{
\begin{tikzpicture}
\draw (0,1) -- (5,1) ; 
\draw[dashed] (0,-0.5) -- (5,-0.5) ; 

\filldraw[black] (2,-0.01) circle (2pt);
\filldraw[black]  (2,-0.01) node[anchor= north east] {$y_1\in (0,x)$} ;

\filldraw[black] (3,-1) circle (2pt);
\filldraw[black]  (3,-1) node[anchor= north east] {$\substack{y_2+x,\\ y_2<0}$} ;

\filldraw[black]  (5,-0.5) node[anchor= west] {$x$} ; 
\filldraw[black] (5,-0.5) circle (2pt);

\filldraw[black]  (0,1) node[anchor= east] {$0$} ;
\filldraw[black] (0,1) circle (2pt);

\draw [thick] (0 ,1)  node[anchor= east] {~} to [ curve through ={(0.5,0) ..(1,-0.1).. (2,-0.01) .. (3,-1)} ] (5,-0.5)  ; 
\draw[thick] (0,-1.1)  node[anchor= east] {~}-- (0,1.5) node[anchor=south ] {$0$};

\draw[dotted] (2,-1.1) -- (2,-0.5);
\draw[thick] (2,-0.5) -- (2,1.5)node[anchor=south ] {$m$} ;

\draw[thick] (3,-1.1) -- (3,-0.5);
\draw[dotted] (3,-0.5) -- (3,1.5)node[anchor=south ] {$m+1$} ;

\draw[thick] (5,-1.1) -- (5,1.5) node[anchor=south ] {$n$};

\filldraw[black] (-1.2,0) node[anchor= west] {$T_3=$};

\end{tikzpicture}
}

\begin{document}

\setlength{\abovedisplayskip}{4pt}
\setlength{\belowdisplayskip}{4pt}

\title{The inverse scattering of the Zakharov-Shabat system solves\\  the weak noise theory of the Kardar-Parisi-Zhang equation\\
}

\author{Alexandre Krajenbrink}
\email{alexandre.krajenbrink@sissa.it}
\affiliation{SISSA and INFN, via Bonomea 265, 34136 Trieste, Italy}
\author{Pierre Le Doussal}
\affiliation{Laboratoire de Physique de l'\'Ecole Normale Sup\'erieure, CNRS, ENS $\&$ PSL University, Sorbonne Universit\'e, Universit\'e de Paris, 75005 Paris, France}

\date{\today}

\begin{abstract}
We solve the large deviations of the Kardar-Parisi-Zhang (KPZ) equation in one dimension at short time by introducing an approach which combines field theoretical, probabilistic and integrable techniques. We expand the program of the weak noise theory, which maps the large deviations onto
a non-linear hydrodynamic problem, and unveil its complete solvability through 
a connection to the integrability of the Zakharov-Shabat system. Exact solutions, depending on the initial
condition of the KPZ equation, are obtained using the inverse scattering method and a Fredholm determinant framework recently developed.
These results, explicit in the case of the droplet geometry, open the path to obtain the complete large deviations 
for general initial conditions.
\end{abstract}

\maketitle

Large deviation rate functions characterize rare events and play a key role in non-equilibrium statistical physics, 
as generalizations of the thermodynamic potentials \cite{TouchetteReview2018,GarrahanReviewLargeDev2018,KafriSingular2015}. They have been much studied for interacting particle models in one dimension. 
For diffusive systems, 
the macroscopic fluctuation theory (MFT) \cite{BertiniMFT2015}
provides a powerful framework to calculate the large deviation of the density, in agreement with the
available exact solutions
\cite{DerridaMFTReview2007}. For driven diffusive systems, however, such as the asymmetric exclusion process (ASEP)
\cite{DerridaReviewASEP}, there is not yet a general approach to calculate the large deviations of the current and 
density fluctuations for all geometries. Exact results from the matrix product ansatz \cite{DerridaHakimASEPMatrix1993} 
and the Bethe ansatz are available in special cases, for instance in a stationary regime, either on a finite ring,
where rate functions are found to exhibit a universal shape, for the TASEP \cite{DerridaLebowitzTASEPring,ProlhacTASEPLargeDev}, the 
ASEP \cite{LargeDevASEPstatring,ProlhacASEPFiniteTime} and the Kardar-Parisi-Zhang (KPZ) equation \cite{DerridaAppertKPZring,DerridaBrunet},
or in open geometries \cite{DerridaLebowitzASEPOpen,MallickASEPLargeDevOpen2012,OpenASEPCrampeBA}.

The KPZ equation is a prominent example of the driven diffusive class. It allows for a few exact solutions valid for all times 
\cite{CLR10,dotsenko,SS10,ACQ11,PLDSineGordon,CLDflat,QuastelFlat,SasamotoStationary,SasamotoStationary2,BCFV,gueudre2012directed,borodin2016directed,barraquand2017stochastic,KrajenbrinkReplica,BarraquandKrajenbrink2020half}, which exhibit at large times the universal typical fluctuations common to systems in the 
KPZ class \cite{CorwinKPZReview,QuastelSpohnReview,TakeuchiReview}.
Much recent attention has shifted to  its large deviation properties, at late times 
\cite{le2016large,sasorov2017large,KrajLedou2018,ProlhacKrajenbrink,corwin2020lower,JointLetter,tsai2018exact,largedev,corwin2020kpz,doussal2019large,cafasso2019riemann,ProlhacRiemannKPZ}, 
and also at short times
\cite{le2016exact,krajenbrink2017exact,krajenbrink2018large,KrajLedou2018,ProlhacKrajenbrink,krajenbrink2019beyond, NumericsHartmann,hartmannprep,Baruch,MeersonParabola,janas2016dynamical,meerson2017randomic,Meerson_Landau,Meerson_flatST,asida2019large,meerson2018large,smith2018finite,smith2019time,lin2020short,hartmann2019optimal} where two main approaches were developed. The first one uses the aforementioned exact solutions for all times,
obtained from a mapping of KPZ observables to the integrable (replica) delta Bose gas.
This allowed to obtain the short time large deviations in a few cases
\cite{le2016exact,krajenbrink2017exact,krajenbrink2018large,KrajLedou2018,ProlhacKrajenbrink,krajenbrink2019beyond}. A more versatile approach, 
closer in spirit to the MFT, is the weak noise theory (WNT) \cite{Korshunov,Baruch,MeersonParabola,janas2016dynamical,meerson2017randomic,Meerson_Landau,Meerson_flatST,asida2019large,meerson2018large,smith2018finite,smith2019time,lin2020short}.
It is a saddle point method on the dynamical field theory, which is exact at short time.
It leads to a system of two coupled non-linear differential equations, which determine
the "optimal" KPZ height field and noise producing the rare 
fluctuation. Until now however these equations have been solved only
numerically, except in some limits where useful but approximate solutions were found.
Although the existence of multi-soliton solutions was noted \cite{janas2016dynamical}, 
no exact solution allowing for the full calculation of the large deviations was obtained.

In this Letter we construct the exact solution to the weak noise theory of the KPZ equation. 
Through the integrability of the Zakharov-Shabat system, originally introduced to solve
the non-linear Schrodinger equation (NLS) \cite{ZS}, we show that the full space time dependence of the
optimal height and noise fields admit representations in terms of Fredholm determinants. 
We provide an explicit formula for the KPZ droplet initial condition, and give
the general form for a large class of initial conditions.

The KPZ equation \cite{KPZ} describes the stochastic growth in time $\tau$ of the height field $h(y,\tau)$ of an interface,
here in one space dimension $y \in \mathbb{R}$
\be
\partial_\tau h(y,\tau) = \nu \partial_y^2 h(y,\tau) + \frac{\lambda_0}{2} (\partial_y h(y,\tau))^2 + \sqrt{D} \eta(y,\tau) 
\label{eq:KPZ}
\ee
where $\eta(y,\tau)$ is a standard space time white noise, i.e. 
$\overline{\eta(y,\tau) \eta(y',\tau')}= \delta(\tau-\tau') \delta(y-y')$. 
We choose units such that $D=\lambda_0=2$, $\nu=1$ \cite{footnote100}.
We consider the probability $P(H,T)$ to observe the value 
$h(0,T)=H-H_0$ at time $\tau=T$, where $H_0$ is a constant chosen below. 
At short time, although the typical height fluctuations are
Gaussian with Edwards-Wilkinson scaling $\delta H \sim T^{1/4}$, the 
KPZ non-linearity leads to non-trivial and non-perturbative 
tails for $P(H,T)$, describing rare events.
For $T \ll 1$, it takes the large deviation 
form 
\be
P(H,T) \sim \exp( - \Phi(H)/\sqrt{T} ) 
\label{eq:HeightLargeDev}
\ee
where the exact rate function $\Phi(H)$ was obtained
for droplet, Brownian and flat initial height profiles,
from the exact solutions \cite{le2016exact,krajenbrink2017exact,Meerson_flatST,krajenbrink2019beyond}. 

We now explain how to obtain such rate function from the WNT: we first derive the
WNT equations in a way leading directly to the so-called $\{P,Q \}$ system, 
which we then analyze. To that aim, it is useful to define the rescaled time and space variables as $t=\tau/T$, $x=y/\sqrt{T}$,
where $T$, the observation time, is fixed. Through the Cole-Hopf map the KPZ field is equivalently described introducing 
$Z(x,t)=e^{h(y,\tau)+ H_0}$, which satisfies the (rescaled) stochastic heat equation (SHE) in the Ito sense
\be
\partial_t Z(x,t) = \partial_x^2 Z(x,t) + \sqrt{2} T^{1/4} \tilde \eta(x,t) Z(x,t) 
\label{eq:SHE}
\ee
where $\tilde \eta(x,t)$ is another standard space time white noise. This equation 
is now studied for $t \in [0,1]$. The noise amplitude is now of order $T^{1/4}$, hence
a short observation time $T \ll 1$ corresponds to a weak noise. Our convenient choice is $H_0=\frac{1}{2} \log T$ \cite{footnoteH0}.
It is convenient to study the following generating function
which admits a large deviation principle at short time $T \ll 1$, with $z \geq 0$ 
\be
\overline{ \exp( - \frac{z}{\sqrt{T}} e^H )  }  \sim \exp( - \frac{\Psi(z)}{\sqrt{T}} ) 
\label{eq:ZLargeDev}
\ee 
Inserting \eqref{eq:HeightLargeDev} into the expectation value over the noise in the l.h.s., we see that for $T \ll 1$, 
$\Psi(z)$ and $\Phi(H)$ are related through a Legendre transform
\be  \label{Legendre} 
\Psi(z)=\min_H ( z e^H + \Phi(H) ) 
\ee
Here we aim to calculate $\Psi(z)$ and $\Phi(H)$ using the WNT, for an initial condition 
of the
form $e^{h(y,0)} = \frac{1}{\sqrt{T}} Z_0(y/\sqrt{T})$ where $Z_0(x)$ is given, an example being the droplet initial condition $Z_0(x)=\delta(x)$.

Any average of the form \eqref{eq:ZLargeDev} can be represented using the
dynamical field theory associated to the rescaled SHE \eqref{eq:SHE} as
$\overline{ e^{ \frac{1}{\sqrt{T}} \int_\R \rmd x dt j(x,t) Z(x,t) }}=\int \mathcal{D}Z \mathcal{D}\tilde Z e^{- \frac{1}{\sqrt{T}} S[\tilde Z,Z,j]}$
with the dynamical action 
\be  \label{eq:dynaction} 
S[\tilde Z,Z,j] = \int_0^1\int_\R \rmd t \rmd x [ \tilde Z (\partial_t - \partial_x^2) Z - \tilde Z^2 Z^2 - j Z ]
\ee
where $\tilde Z/\sqrt{T}$ is the response field. In \eqref{eq:ZLargeDev} the source field
is $j(x,t)=-z \delta(x) \delta(t-1)$. For $T \ll 1$, the action is evaluated by a saddle point method.
Defining
$\tilde Z= - z P$, $Q=Z$ and $g=-z$, the saddle point equations 
of the WNT, $\frac{\delta S}{\delta \tilde Z}=0$ and $\frac{\delta S}{\delta Z}=0$,
take the form of the $\{ P,Q\}$ system
\be
\begin{split}\label{eq:PQsystem}
 \partial_t Q =& \partial_x^2 Q + 2 g P Q^2  \\
 - \partial_t P =& \partial_x^2 P + 2 g P^2 Q  
 \end{split}
\ee
a close cousin of the NLS equation
\cite{footnoteNLS}. While the $\{P,Q\}$ system is interesting in its own right, we will apply its study to
the following mixed boundary conditions, of interest for the WNT
\be \label{init}
Q(x,0) = Q_0(x) , \quad P(x,1)=\delta(x)
\ee
The source $j$ imposes this form for $P$ at $t=1$ \cite{footnoteP},
while $Q$ is
specified at $t=0$ from the initial height of the KPZ equation,
i.e. $Q_0(x)=Z_0(x)$. The function
$\Psi(z)$ in \eqref{eq:ZLargeDev} is obtained from the 
action $S$ in \eqref{eq:dynaction} at the saddle point. Using the first
equation in \eqref{eq:PQsystem} it can be written in the form
\eqref{Legendre} allowing to identify $\Phi(H) \equiv g^2 \int_0^1 \rmd t \int_\R \rmd x P^2 Q^2$, 
with $H=H^*_z:=\argmin_H ( z e^H + \Phi(H) )$, in agreement
with \cite{Baruch} (see also \cite{SM}). The "optimal shape" $h_{\rm opt}(y,\tau)$ of the 
KPZ height field from the WNT, i.e. the most probable one realizing the 
value $h_{\rm opt}(0,T)=H-H_0$ at $\tau=T$ and $y=0$, is 
obtained from the solution $Q(x,t)$ 
of \eqref{eq:PQsystem} for $t \in [0,1]$ as 
$e^{h_{\rm opt}(y,\tau)} = \frac{1}{\sqrt{T}} Q(\frac{y}{\sqrt{T}}, \frac{\tau}{T})$.

Let us first analyze the $\{P,Q \}$ system \eqref{eq:PQsystem} for general initial conditions,
and return to the WNT later. Remarkably, \eqref{eq:PQsystem} belongs to the AKNS class of integrable non-linear problems
\cite{AblowitzKaup1974},
for which there exists a Lax pair, i.e. a pair of linear
differential equations whose compatibility conditions are equivalent to \eqref{eq:PQsystem}.
Here the system reads $\partial_x \vec v= U_1 \vec v$, $\partial_t \vec v= U_2 \vec v$
where $\Vec{v}=(v_1,v_2)^\intercal$ is a two component vector (depending on $x,t,k$) 
where
\begin{equation}
U_1=
\begin{pmatrix}
-\I k/2  & - g P(x,t)\\  Q(x,t) & \I k/2 
\end{pmatrix} \quad , \quad 
U_2= 
\begin{pmatrix}
{\sf A} & {\sf B}\\
{\sf C} & -{\sf A}
\end{pmatrix}
\label{eq:LaxPairU}
\end{equation}
where ${\sf A}= k^2/2- g P Q$, ${\sf B}=g (\partial_x - \I k) P$,
${\sf C}= (\partial_x +  \I k) Q$. One can check that the compatibility condition, $\p_t U_1-\p_x U_2 +[U_1, U_2]=0$,
recovers the system \eqref{eq:PQsystem} which we solve through the following scattering problem.
Let $\Vec{v}=e^{ k^2 t/2} {\phi}$ with $ {\phi}=(\phi_1,\phi_2)^\intercal$ and $\Vec{v}=e^{- k^2 t/2} {\bar{\phi}}$
be two independent solutions of the linear problem such that 
at $x \to -\infty$, $\phi \simeq (e^{-\I k x/2},0)^\intercal$ and $\bar \phi \simeq (0,-e^{\I k x/2})^\intercal$.
Assuming from now on that $P,Q$ vanish at infinity, the $x \to +\infty$ behavior of these solutions
defines scattering amplitudes

\be \label{eq:plusinfinity} 
\phi \underset{x \to +\infty}{\simeq}
\begin{pmatrix}
a(k,t)e^{-\frac{\I  kx}{2}}\\b(k,t)e^{\frac{\I  kx}{2}}
\end{pmatrix}  ~,~
\bar{\phi} \underset{x \to +\infty}{\simeq}
\begin{pmatrix}
\tilde{b}(k,t)e^{-\frac{\I  kx}{2}}\\ -\tilde{a}(k,t)e^{\frac{\I  kx}{2}}
\end{pmatrix}
\ee 
Plugging this form into the $\partial_t$ equation of the Lax pair at $x \to +\infty$,
one finds 
a very simple time dependence, 
$a(k,t)=a(k)$ and $b(k,t)=b(k)e^{-k^2 t}$, 
$\tilde a(k,t)=\tilde a(k)$ and $\tilde b(k,t)=\tilde b(k)e^{k^2 t}$.
Another relation is obtained from the Wronskian of the two solutions 
, $a(k) \tilde{a}(k)+b(k)\tilde{b}(k)=1$ \cite{footnoteWronskien}.

Before providing explicit formula for these scattering amplitudes
let us show how to obtain from them the solution for the $\{ P,Q\} $ system, i.e. how to construct 
the inverse scattering transform. The spatial part of the Lax pair is a 1D Dirac equation 
called the ZS system, originally introduced to solve the NLS equation \cite{ZS,zakharov1979integration}, 
and extended by AKNS \cite{AblowitzKaup1974}. 
It was shown very recently \cite{krajenbrink2020painleve,bothner2021atlas} that 
the inverse scattering problem can be solved by the means of Fredholm determinants (FD).
Introducing the two reflection coefficients $r(k)=b(k)/a(k)$ and $\tilde r(k)=\tilde b(k)/(g \tilde a(k))$
one defines two functions \cite{footnotereal} 

\be \label{2fonctions} 
A_t(x)\!=\!\! \int_\R \!\frac{\rmd k}{2 \pi} r(k) e^{\I k x-k^2 t}  ,  B_t(x)\!=\!\! \int_\R \!\frac{\rmd k}{2 \pi} \tilde r(k) e^{k^2 t-\I k x}
\ee
and two linear operators from $\mathbb{L}^2(\mathbb{R}^+)$ to $\mathbb{L}^2(\mathbb{R}^+)$ with respective kernels
\be \label{kernels} 
{\cal A}_{xt}(v,v')= A_t(x+v+v') ~,~ {\cal B}_{xt}(v,v')= B_t(x+v+v')
\ee
Note that these functions and kernels obey the simple heat equation in space time, and we assume that $A_t(x)$, $B_t(x)$ vanish
fast enough for $x \to +\infty$.
The solutions $P,Q$ \cite{krajenbrink2020painleve,bothner2021atlas} are reconstructed as
\bea \label{soluQP} 
&& Q(x,t)= \bra{ \delta } {\cal A}_{xt} (I + g {\cal B}_{xt} {\cal A}_{xt})^{-1} \ket{\delta} \\
&& P(x,t)= \bra{ \delta } {\cal B}_{xt} (I + g {\cal A}_{xt} {\cal B}_{xt})^{-1} \ket{\delta} \nonumber
\eea
where $\ket{\delta}$ is the vector with component $\delta(v)$ so that $\bra{\delta} {\cal O}\ket{\delta}=\mathcal{O}(0,0)$ for
any operator $\mathcal{O}$. The product $P Q$, which is a conserved charge, i.e. $\partial_t \int_\R\rmd x P Q =0$ as
easily verified from \eqref{eq:PQsystem}, can be expressed from a FD as
$g P Q = \partial_x^2 \log \Det( I + g {\cal B}_{xt} {\cal A}_{xt} )$. 
The formula \eqref{soluQP} thus provides the general solution of the $\{ P,Q\} $ system, parameterized
by the two functions $A_t$ and $B_t$, equivalently, by the scattering amplitudes. Although these are
in one-to-one correspondence with the $\{P,Q\}$ boundary data, making it explicit is non-trivial,
and is our aim below. Particular cases are such that ${\cal A}_{xt}$ and ${\cal B}_{xt}$ are operators of finite ranks,
leading to solitonic type solutions \cite{SM}. The simplest one leads to $g P Q = \frac{(\kappa+\mu)^2}{4 \cosh^2 \frac{1}{2} (\kappa+\mu) (x-x_0(t)) }$.
In the context of WNT this soliton has been used as an approximate solution for the right tail for $H \to +\infty$, and another rank one 
family was noticed in \cite{janas2016dynamical}. However this is insufficient to obtain the full rate function $\Phi(H)$ which requires the  (infinite-rank) general solution obtained in this work.

Let us now apply this to the WNT, i.e. for the boundary data in \eqref{init}, and characterize the scattering amplitudes. Integrating the
$\partial_x$ equation of the Lax pair at $t=1$ for $\bar{\phi}$ 
using \eqref{init} allows to obtain \cite{SM} $\tilde{b}(k)=g e^{-k^2}$, 
and the relation $\tilde{a}(k)=1-g \int_{0}^{+\infty} \rmd x' Q(x',1)e^{-\I k x'}$.
From the same equation at $t=1$ for $\phi$, one obtains
$a(k)=1- g \int_{-\infty}^{0} \rmd x' Q(x',1)e^{-\I kx'}$,
where at this stage $Q(x,1)$ is unknown. If the initial 
condition $Q(x,0)$ is even in $x$ (which we assume from now on)
then $\tilde{a}(k)=a(-k)=a^*(k^*)$
and $b(k)$ is real and even. This leads to the form
\begin{equation}
a(k)=e^{- \I \varphi(k)} \sqrt{1-g b(k) e^{-k^2}}
\end{equation}
where we still have two unknown functions, a phase $\varphi(k)$, which is odd $\varphi(k)=-\varphi(-k)$, and $b(k)$.

To determine them we derive integral equations for the functions $A_t,B_t$.
Let us set $t=1$ and ignore the time index. Differentiating the
vector $\ket{\beta_x} := (I+g {\cal A}_x {\cal B}_x)^{-1} \ket{\delta}$,
with respect to $x$ and using the identity $\partial_x ({\cal A}_x {\cal B}_x) = - {\cal A}_x \ket{\delta} \bra{\delta} {\cal B}_x$
we obtain
\begin{equation}
\p_x \ket{\beta_x}=gP(x,t=1) (I+g{\cal A}_x{\cal B}_x)^{-1} {\cal A}_x\ket{\delta}
\end{equation}
Since $P(x,t=1)=\delta(x)$, integrating one finds
\begin{equation} 
\ket{\beta_x}=\ket{\delta}-g \Theta(-x) (I+g{\cal A}_0{\cal B}_0)^{-1} {\cal A}_0\ket{\delta}
\label{eq:solutionBetaX}
\end{equation}
Hence, for $x>0$, $\ket{\beta_x}=\ket{\delta}$ which, inserted in the 
definition of $\ket{\beta_x}$, leads to $g {\cal A}_x {\cal B}_x \ket{\delta}=0$ for $x>0$.  Taking another derivative w.r.t $x$ one obtains 
$A_1(x') B_1(x)=0$ for $x'>x>0$, which implies that $B_1(x>0)=0$, hence
that the operator ${\cal B}_{x1}$ is zero for $x>0$. This in turns yields
$Q(x>0,1)=A_1(x)$, which for $Q$ even implies $Q(x,1)=A_1(|x|)$.
Next, inserting \eqref{eq:solutionBetaX} for general $x$ into the definition of $\ket{\beta_x}$, taking a derivative w.r.t $x$, and finally using the fact 
that ${\cal A}_{x1} {\cal B}_{x1} {\cal A}_{x1} \ket{\delta}=0$ as a consequence of $B_1(x>0)=0$, we 
obtain $B_1(x)=\delta(x)+g\Theta(-x) \bra{\delta} {\cal B}_{x,1} {\cal A}_{0,1}\ket{\delta}$.
This is the sought for integral equation, which reads more explicitly
\be \label{AQ} 
B_1(x) = \delta(x) + g \Theta(-x) \int_0^{+\infty}\! \! \!  \rmd v
B_1(x+v) A_1(v) 
\ee 
where $A_1(v)=(p * A_0)(v):=\int_\R  \rmd y p(v-y) A_0(y)$ 
denoting the heat kernel at unit time $p(z):=\frac{e^{- z^2/4}}{\sqrt{4 \pi}}$.

{\it Droplet initial condition}.
Let us now specialize to $Q_0(x)=\delta(x)$. The solution of 
\eqref{eq:PQsystem} then clearly satisfies the symmetry 
$Q(x,t)=P(x,1-t)$. This in turns implies that $A_t(x)=B_{1-t}(x)$
and $r(k)=e^{k^2} \tilde r(-k)$, also implying $b(k)=1$,
which we use below. Hence in \eqref{AQ} we can replace $A_1(v)$ by $(p * B_1)(v)$
and we obtain a closed {\it non-linear} integral equation for the function $B_1$
(which equals $A_0$). This equation still looks formidable,
however, for readers familiar with random walks, it has a flavor of another famous integral equation, the Hopf-Ivanov (HI) equation
\cite{Hopf1934,Ivanov1994}, which however is {\it linear}, and reads
\be \label{Blin} 
B_1(x) = \delta(x) + g\Theta(-x)  \int_{-\infty}^0 \rmd y p(x-y) B_1(y)
\ee 
Amazingly, we found that these two equations, \eqref{Blin} and \eqref{AQ} are
{\it equivalent}. This can be tested in perturbation in $g$, and is shown to all orders 
in \cite{SM}. The HI equation arises in survival probabilities of random walks
\cite{Pollaczek1952,Spitzer1957,MajumdarIvanovReview,Mounaix2018}. Indeed writing $B_1(x)$ as a series, 
$B_1(x)=\delta(x)+\sum_{n=1}^{+\infty} g^n B_{1,n}(x)$,
and inserting in \eqref{Blin}, leads to the recursion $B_{1,n}(x)=\int_{y<0} p(x-y) B_{1,n-1}(y)$.
The interpretation is then straightforward. Consider $X(j) \in \mathbb{R}$ a discrete time random walk, 
$X(j+1)=X(j)+z_j$, with $z_j$ i.i.d. with jump probability $p(z)$. Then $B_{1,n}(x)$ is the probability that 
the walk starting at $X(0)=0$ arrives at $X(n)=x$ in $n \geq 1$ steps, while remaining negative,
$\{ X(j) \leq 0\}_{j=0,\dots,n}$ (and $\int_{-\infty}^0 \rmd x B_{1,n}(x)=\binom{2n}{n}2^{-2n}$ is given by the universal Sparre-Andersen theorem \cite{SparreAndersen,footnoteProlhac}).
To show that the solution of \eqref{Blin} also solves \eqref{AQ} then 
amounts to split the walk into two independent parts, upon crossing the level $x$ for the last time, \cite{SM}. Introducing the Laplace transform $\hat B_1(s)=\int_{-\infty}^0 e^{s x} B_1(x)$, the 
solution of the HI equation is known to be \cite{Ivanov1994}
\be  \label{IvanovSolu} 
\hat B_1(s) = \exp\big( - \int_\R \frac{\rmd q}{2 \pi} \frac{s}{s^2+ q^2} \log(1 - g \tilde p(q)) \big)
\ee 
where $\tilde p(k)$ is the Fourier transform of $p(z)$, here $\tilde p(k)=e^{-k^2}$. Going 
from Laplace to Fourier, from \eqref{2fonctions} one finds 
$r(k)=e^{k^2} \tilde r(-k)= \hat B_1(s)|_{s=- \I k + 0^+}$. Using $\frac{1}{s + \I q} \to PV (\frac{\I}{k-q}) +  \pi \delta(k-q))$ 
we obtain from \eqref{IvanovSolu},
the reflection coefficient $r(k)$ and its phase $\varphi(k)$ 
\be \label{soluphase} 
r(k)= \frac{e^{ \I \varphi(k)}}{\sqrt{1- g e^{-k^2}}}  ~, ~
\varphi(k)= \dashint_\R \frac{\rmd q}{2\pi} \, \frac{k \log(1-g e^{-q^2} )}{q^2-k^2}
\ee 
which completes the solution of \eqref{eq:PQsystem} for droplet IC.

To extract $\Phi(H) \equiv g^2 \int_\R  \int_0^1 \rmd x\rmd t \, P^2 Q^2$ from our solution
appears to require the computation of a difficult space time integral. 
This is overcome by relating it, as well as $\Psi(z)$, to conserved 
quantities. Indeed, we can use the construction of ZS \cite{ZS} to generate 
all conserved quantities $C_n$ for the $\{ P,Q\}$ system. Recall the Lax pair
of equations \eqref{eq:LaxPairU} for $\vec v$ and define $\Gamma(x,t,k)=\frac{v_2}{v_1}$. The  
$\partial_t$ equation leads to a conservation equation 
$\partial_t  (- g P \Gamma) = \partial_x J$, where $J={\sf A} - k^2/2 + {\sf B} \Gamma$.
Let us expand $\Gamma(x,t,k) = \sum_{n \geq 1} \frac{\Gamma_n(x,t)}{(\I k)^n}$.
Each $C_n = - g \int_\R \rmd x  P \Gamma_n$ is thus a total conserved charge, i.e. $\frac{d}{dt} C_n=0$. 
The $\Gamma_n$ are obtained recursively from the $\partial_x$ Ricatti equation
$\partial_x \Gamma = \I k \Gamma + Q + g P \Gamma^2$, with $\Gamma_1=-Q$.
This procedure generates
$C_1= g \int_\R\rmd x P Q$, $C_2=g \int_\R\rmd x P \p_x Q$, $C_3=g (\int_\R\rmd x P \p_x^2Q + g P^2 Q^2)$ . 
The {\it values} $C_n(g)$ taken by these conserved charge are encoded in $a(k)$.
Indeed integrating 
the equation for $\partial_x \log \phi_1$ over $x$ in $\R$ gives
$\log a(k) = - g \int_\R \rmd x P\Gamma$. Hence the 
Laurent expansion of $\log a(k)$ yields the values of the conserved charges,
$\log a(k) = \sum_{n \geq 1} \frac{C_n(g)}{(\I k)^n}$. Until now this is
general for any initial condition of the $\{ P,Q\} $ system. Now recall that for the droplet IC we obtained $\log a(k) = -  \I \varphi(k) + \frac{1}{2} \log(1 - g e^{- k^2})$.
Since the second term has vanishing Laurent expansion, we find that $- \I \varphi(k) = \sum_{n \geq 1} \frac{C_n(g)}{(\I k)^n}$.
Expanding in powers of $1/k$ in \eqref{soluphase} 
we obtain \cite{footnoteeven} 
\begin{equation} \label{C1C3} 
 C_1(g)=\frac{1}{\sqrt{4\pi}}{\rm Li}_{\frac{3}{2}}(g) \; ,\;  
C_3(g)= \frac{-1}{\sqrt{16\pi}}\mathrm{Li}_{\frac{5}{2}}(g)
\end{equation}
Since $C_1 = g \int_\R\rmd x P Q$ is time independent, evaluated at $t=1$ it
leads to $C_1(g)=g Q(0,1) = g e^H$. On the other hand, differentiating the Legendre transform in \eqref{Legendre}
w.r.t. $z$ gives $\Psi'(z)=e^H$. This implies that $C_1(-z) = - z \Psi'(z)$, and by integration
\be  \label{Psi} 
\Psi(z)= \Psi_0(z) := - \frac{1}{\sqrt{4\pi}}\mathrm{Li}_{5/2}(-z) 
\ee 
which allows to determine $\Phi(H)$ parametrically as 
$\Phi(H)= \Psi(z) - z \Psi'(z)$, $e^H=\Psi'(z)$ (it can
also be obtained from $C_3(g)$, see \cite{SM}).
Our WNT result \eqref{Psi} agrees with \cite{le2016exact}, without relying on an exact solution of the KPZ equation. 

Until now we 
assumed $z = - g \geq 0$ corresponding to $H \leq \hat H_0=-\frac{1}{2} \log(4 \pi)$, 
the most probable value of $H$
such that $\Phi'(\hat H_0)=0$ \cite{footnotePsi}. However 
\eqref{eq:PQsystem} also holds for any $H>\hat H_0$
\cite{SM,janas2016dynamical}, corresponding to the attractive regime $g>0$ of the $\{ P,Q\}$ system.
Indeed, $\Psi(z)$ can be analytically continued to $z<0$, allowing to 
determine $\Phi(H)$ for any $H$ \cite{le2016exact}. For 
$H \in (-\infty,H_c]$, \eqref{Psi} holds, with 
$z=-g$ varying from $+\infty$ down to $z=-1$. For $H>H_c=\log \frac{\zeta(3/2)}{\sqrt{4\pi}}$,
a second continuation is needed, $\Psi(z)=\Psi_0(z)+ \Delta(z)$, with $\Delta(z)= \frac{4}{3} (\log(-\frac{1}{z}))^{3/2}$
with $z \in [-1,0)$ as $H \in [H_c,+\infty)$. These continuations correspond to two branches of solutions of the $\{ P,Q\}$ system for $0<g \leq 1$.
One finds \cite{SM} that the second branch corresponds to the spontaneous generation of a 
solitonic part in the solution, of rapidity $\kappa_0$
with $g=e^{-\kappa_0^2}$, 
which dominates
the large deviations for $H \to +\infty$. It is described by
$A_t(x) = A_t(x)|_{\phi(k) \to \phi(k)+\Delta \phi(k)} 
+2 \kappa_0 e^{-\kappa_0 x+\kappa_0^2 t+\I\varphi(\I\kappa_0)}$,
where $\Delta \varphi(k)= 2 \,  \arctan(\frac{\kappa_0}{k})$
and $B_t(x)=A_{1-t}(x)$. The values of the odd
conserved charges are increased by $\Delta C_n(g)=\frac{2}{n}\kappa_0^n$, 
which for $n=1$ induces the additional part $\Delta(z)$. 
\begin{figure}[h!]
    \centering
    \includegraphics[scale=0.5]{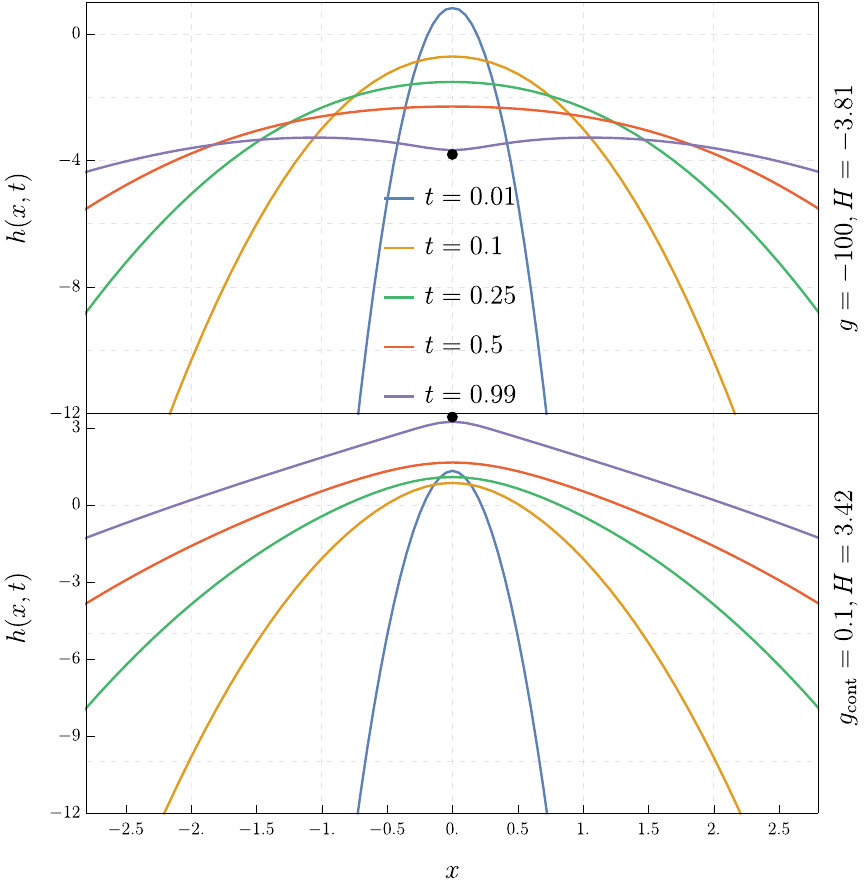}
    \caption{The optimal height $h(x,t)$ for droplet initial condition
    plotted for various times $t$ for final values (black dot) $H=-3.81$ (\textbf{top}) and $H=3.42$ (\textbf{bottom}).}
    \label{fig:hxt}
\end{figure}

{\it General initial condition}. For general even $Q_0(x)$,
the only difference is that $b(k)$ is non-trivial,
with $r(k)= b(k) e^{k^2} \tilde r(-k)$, and now 
$A_t(x)=(\hat b * B_{1-t})(x)$ where $\hat b(x)$ denote the Fourier
transform of $b(k)$ and $*$ the convolution. Equation~\eqref{AQ}, replacing
$A_1(v)=(p*\hat b*B_1)(v)$, is again equivalent to a linear HI equation for $B_1(x)$,
obtained by simply replacing $p$ by $p * \hat b$ in \eqref{Blin},
with the same random walk interpretation for a new
jump probability $p(z) \to (p * \hat b)(z)$. Thus this leads to $r(k)= \frac{b(k) e^{\I \varphi(k)}}{\sqrt{1- g b(k) e^{-k^2}}}$
and $\varphi(k)= \dashint_\R \frac{\rmd q}{2\pi} \, \frac{k \log(1-g b(q) e^{-q^2} )}{q^2-k^2}$.
One now finds $C_1(g) = \int_\R \frac{\rmd q}{2 \pi} {\rm Li}_1(g b(q) e^{-q^2})$  leading to \cite{footnoteb}
\be
z \Psi'(z)=  \int_\R \frac{\rmd q}{2 \pi} \log(1+z b(q) e^{-q^2})
\ee 
Interestingly, such a form was observed to describe all
known exact solutions \cite{footnotegeneral}, e.g. flat IC with
$b(q) \to 1/q^2$ \cite{footnoteprove}.
Thus, for a 
general initial condition we reduced the problem to 
computing a single unknown function $b(k)$, and relating it to
$Q_0(x)$, a question left for the future.
In \cite{SM} we give a formula relating
$b(k)$, $Q_0(x)$ and $P(x,1)$ allowing for
expansions around the droplet solution. 

In conclusion our solution allows to calculate the optimal height and noise for arbitrary values of $H$, previously inaccessible. The Fredholm approach provides a novel analytical and numerical scheme for the solution of the integrable $\{ P,Q\}$ system as shown in Fig.~\ref{fig:hxt}, see \cite{SM}. The present work demonstrates that inverse scattering methods
can successfully address optimal fluctuation theory of stochastic systems, 
leading to analytic results and interesting phenomena such as spontaneous soliton generation.

\begin{acknowledgments}
\paragraph{Acknowledgments.}  We thank T. Bothner, S. N. Majumdar, B. Meerson, G. Schehr and N. Smith for discussions on
related topics. AK acknowledges support from ERC under Consolidator grant number 771536 (NEMO).  PLD acknowledges support from the ANR grant ANR-17-CE30-0027-01 RaMaTraF.  
\end{acknowledgments}

\newpage{\pagestyle{empty}\cleardoublepage}

\newpage

.

\newpage

\begin{widetext} 

\makeatletter
\renewcommand{\theequation}{S\arabic{equation}}
\renewcommand{\thefigure}{S\arabic{figure}}

\setcounter{section}{0}
\renewcommand{\thesubsection}{S-\Alph{subsection}}

\setcounter{secnumdepth}{2}

\begin{large}
\begin{center}

Supplementary Material for\\  {\it The inverse scattering of the Zakharov-Shabat system solves  the weak noise theory of the Kardar-Parisi-Zhang equation }

\end{center}
\end{large}

We give the principal details of the calculations described in the main text of the Letter. 
We also give additional information about the results displayed in the text.

\subsection{Large deviation rate functions, analytic continuations, and $\{P, Q \}$ system}
\label{sec:ratefunctions} 

We review here the general properties of the large deviation rate functions $\Phi(H)$ and $\Psi(z)$, and the explicit results in the case of the droplet initial condition. These functions are defined in the text from the 
probability $P(H,t)$ of the (shifted) KPZ height $H=h(0,T) + H_0$, which takes the following
leading asymptotics for $H=\mathcal{O}(1)$, $z=\mathcal{O}(1)$ and $T \ll 1$
\be \label{def1} 
P(H,T) \sim e^{- \frac{\Phi(H)}{\sqrt{T}} }  \quad , \quad \int_\R \rmd H\,  P(H,t) e^{ - \frac{z}{\sqrt{T}} e^H } \sim e^{ - \frac{\Psi(z)}{\sqrt{T}} } 
\ee
Inserting the first estimate into the integral in the second term, we see that the rate functions $\Psi(z)$ and $\Phi(H)$ are related through the Legendre transform
\be  \label{Legendre2} 
\Psi(z)=\min_{H \in \mathbb{R}} ( \Phi(H) + z e^H) = (\Phi(H)+ z e^H )|_{H=H_z}   \quad , \quad H = H_z \quad \Leftrightarrow \quad \Phi'(H) = - z e^H
\ee
Since it is expected that $\Phi(H)$ is convex, with $\Phi(\pm \infty)=\infty$, there is a unique solution to the minimization condition $\Phi'(H) = - z e^H$, denoted $H^*_z$
in the text, and simply $H_z$ here for convenience \cite{footnoteHz}.
From $\Phi(H)$ one thus obtains $\Psi(z)$ as in \eqref{Legendre2}. However in the known solvable cases, it is $\Psi(z)$ which can be computed, and $\Phi(H)$
is obtained from it. Formally this is done as follows. Taking a derivative of \eqref{Legendre2} one obtains $\Psi'(z) = e^{H_z}$, 
and one obtains $\Phi(H)$ in a parametric form
\be \label{inversion} 
e^H = \Psi'(z)  \quad , \quad \Phi(H) =  \Psi(z) - z \Psi'(z)
\ee
However, the definition \eqref{def1} of $\Psi(z)$ is obviously valid only for $\Re (z) \geq 0$, since otherwise the integral over $H$ diverges.
For $z \geq 0$, $\Phi'(H_z) = - z e^{H_z} \leq 0$, hence $H_z$ varies in $(-\infty,\hat H_0]$, 
where $\hat H_0$ is the most probable value of $H$ defined by $\Phi'(\hat H_0)=0$ \cite{footnoteHz}. Thus $z \geq 0$ corresponds to the left side of $P(H,t)$. 
However, the function $\Psi(z)$ is the generating function of the cumulants of $e^H$, and can be analytically continued for $z<0$. This allows to
obtain $\Phi(H)$ for all $H$, including $H > \hat H_0$, using \eqref{inversion}. Let us now 
explain this point further on a specific example.\\

These functions are known for a very limited number of initial conditions, see \cite[Table 7.1]{krajenbrink2019beyond}.
For the droplet initial condition, as found in \cite{le2016exact} and obtained here by a completely different method, one has
\be 
\Psi(z)= \Psi_0(z) := - \frac{1}{\sqrt{4\pi}}\mathrm{Li}_{5/2}(-z) 
\ee 
where we recall that $\mathrm{Li}_{a}(x) = \sum_{n \geq 1} \frac{1}{n^a} x^n$ is the polylogarithm function. The function
$\Psi_0(z)$ is analytic in $z$ except for a branch cut on the real axis for $z<-1$. The most probable value corresponds to $z=0$ and from \eqref{inversion}
reads $e^{\hat H_0}= \Psi'(0)=1/\sqrt{4 \pi}$, hence $z \leq 0$ corresponds to $H = H_z \leq \hat H_0=-\frac{1}{2} \log(4 \pi)$. 
The function $\Psi(z)$ can be analytically continued to $z \in [-1,0)$. Similarly to the logarithm function, 
on top of the principal value of the polylogarithm, analytic continuations can also be defined on higher Riemann sheets by 
adding a jump function. Here in order to obtain $\Phi(H)$ for $H \in (-\infty,+\infty)$ we need the two branches
\be \label{2branches} 
\Psi(z) = \Psi_0(z) \quad , \quad \Psi(z)= \Psi_0(z) + \Delta(z) \quad , \quad \Delta(z)= \frac{4}{3} (\log(-\frac{1}{z}))^{3/2}
\ee
It is then possible to invert the Legendre transform \eqref{Legendre2} in the form
\be  \label{Legendre3} 
\Phi(H)= \begin{cases} \max_{z \in [-1,+\infty[} ( \Psi_0(z) - z e^H ) \quad &, \quad H < H_c \\
\min_{z \in [-1,0]} ( \Psi_0(z) + \Delta(z) - z e^H ) \quad &, \quad H > H_c 
\end{cases} 
\ee
which leads to the same parametric determination \eqref{inversion} upon inserting either $\Psi(z)= \Psi_0(z)$ (first line in \eqref{Legendre3})
or $\Psi(z)= \Psi_0(z)+\Delta(z)$ (second line in \eqref{Legendre3}). Here $H_c=\log \frac{\zeta(3/2)}{\sqrt{4\pi}}$ 
corresponds to the value $H=H_z$ where $z=-1$, i.e. $e^{H_c}= \Psi'(-1) = \Psi_0'(-1)$. Indeed $\Delta'(-1)=0$ which ensures continuity at the turning point,
and in fact all derivatives of $\Phi(H)$ at $H_c$ are found to be continuous. 
We will refer to $\Psi_0(z)$ as the first branch, or main branch, and $\Psi_0(z) + \Delta(z)$ as the second branch.

\begin{figure}[t!]
    \centering
    \includegraphics[scale=0.28]{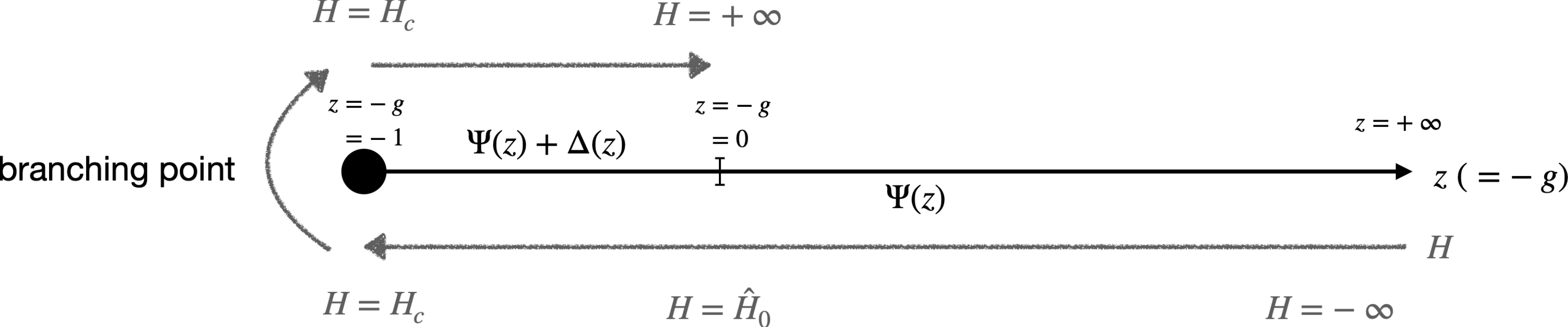}
    \caption{Schematic representation of the parametric solution of the optimization problem. For
$H < H_c$ one uses the function $\Psi$ to invert the Legendre transform taking the parameter $z$ to decrease
from $+\infty$ to $-1$. At $H = H_c$ or $z =-1$ , one needs to turn around the branching point and replace
$\Psi$ by its continuation $\Psi+\Delta$ to determine all $H > H_c$ by increasing the parameter $z$ from $-1$ to 0.}
    \label{fig:my_labelbranch}
\end{figure}

This leads to the following asymptotic behaviors for $\Phi(H)$
\be \label{behaviors} 
\Phi(H) \simeq 
\begin{cases} 
c_-|H|^{5/2} \quad &, \quad H \to -\infty \\
 c_2 (H- \hat H_0)^2 \quad &, \quad |H- \hat H_0| \ll 1 \\
 c_+H^{3/2} \quad &, \quad H \to +\infty 
\end{cases} 
\ee 
where for the droplet initial condition we have \cite{le2016exact,MeersonParabola}
\begin{equation}
    c_-=\frac{4}{15\pi}, \quad c_+=\frac{4}{3},\quad  c_2=\frac{1}{\sqrt{2\pi}}
\end{equation}
The asymptotics \eqref{behaviors} are in fact quite general, i.e. valid for other initial conditions with different values of the amplitudes $c_\pm,c_2$. 
For $H$ around the most probable value $\hat H_0$, inserting the estimate \eqref{behaviors} into \eqref{def1} gives the typical fluctuations, which are Gaussian and
obey the Edwards-Wilkinson scaling $H - \hat H_0 \sim T^{1/4}$. However in the tails the non trivial exponents $5/2$ (left) and $3/2$ emerge (right).
To derive \eqref{behaviors} from \eqref{Legendre3} one uses the asymptotics of the polylogs $\mathrm{Li}_n(-z)\underset{z\to +\infty}{\sim}-\frac{1}{\Gamma(1+n)}(\log z)^n+\mathcal{O}((\log z)^{n-2})$,
see details in Refs.~\cite{le2016exact,krajenbrink2019beyond}. 

The existence of $H_c$ and of two branches in $\Psi(z)$ is quite general (not special to the droplet initial condition), as was found in Refs.~\cite{krajenbrink2017exact,krajenbrink2018large,krajenbrink2019beyond}.
It is useful for the following to indicate a general recipe to obtain the jump function $\Delta(z)$. Assume one can write the function $\Psi(z)$ to be continued as
\be   \label{delta0}
\Psi(z) = \int_I \rmd y F(y) \frac{z}{y(y+z)}\, ,
\ee  
where $I$ is some interval. In that case one has \cite{krajenbrink2019beyond}
\be  \label{delta1} 
\Delta(z) = 2 \I \pi F(-z) \, .
\ee 
This reproduces the above result for the droplet initial condition. We will return to this procedure below in Section \ref{app:continuation}. 
Let us now discuss this structure in terms of the $\{P, Q \}$ system \eqref{eq:PQsystem}. The $\{P, Q \}$ system was derived in
the text, as the saddle point equations in the field variables $Z, \tilde Z$ of the dynamical action $S[\tilde Z,Z,j]$ in Eq.~\eqref{eq:dynaction}. More precisely it was shown that the large deviation function $\Psi(z)$ defined in \eqref{def1} can be written from the saddle point values of the fields, as
\be 
\Psi(z) = S[\tilde Z,Z,j]|_{Z_{\rm sp}=Q, \tilde Z_{\rm sp} =  g P , j=g \delta(x) \delta(t-1), g=-z}
= \int_0^1\int_\R \rmd t \rmd x [ g P (\partial_t - \partial_x^2) Q - g^2 P^2 Q^2 ] - g Q(0,1)|_{g=-z}
\ee 
and where $P,Q$ are the solutions of \eqref{eq:PQsystem} with $g=-z$ and the boundary data \eqref{init}. Using $(\partial_t - \partial_x^2) Q=2 g P^2 Q^2$ 
this leads to
\be  \label{psisp} 
\Psi(z) =  \Phi(H) + z e^{H}|_{H=H_z} = g^2 \int_0^1\int_\R \rmd t \rmd x  P^2 Q^2  - g Q(0,1)|_{g=-z} 
\ee 
The middle expression is from \eqref{Legendre2}. The last term can be identified with $z e^{H_z}$ with $Q(0,1) = Z_{\rm sp}(0,1) = e^{H}|_{H=H_z}$, i.e.
the saddle point (i.e. "optimal") value of the solution of the SHE at $x=0$ and $t=1$, $Z(0,1)=e^H$, where there $H$ denotes the fluctuating KPZ variable. 
From \eqref{psisp} we see that we can thus identify the rate function $\Phi(H)$ as
\be \label{eq:PhiIntegral}
 \Phi(H)|_{H=H_z} = g^2 \int_0^1\int_\R \rmd t \rmd x  P^2 Q^2 |_{g=-z} 
\ee 

Hence, assuming that we know the solution $P,Q$ of the $\{P, Q \}$ system \eqref{eq:PQsystem} with the boundary data 
$P(x,1)=\delta(x)$ and $Q(x,0)=Q_0(x)$, upon setting $g=-z$, the above method allows to obtain $\Psi(z)$ and $\Phi(H)$.
$\Psi(z)$ can be obtained e.g. from the value of $Q(0,1)$ as a function of $z$, since $\Psi'(z)=e^{H_z}=Q(0,1)$. Alternatively 
$\Phi(H_z)$ can be obtained computing the integral \eqref{eq:PhiIntegral}. For the droplet initial condition, the first is done in the text,
the second in Section~\ref{app:C3}, in both cases using conserved charges. 
To obtain $\Phi(H)$ for $H>H_c$ one uses an analytical continuation in $z$, with two branches, as explained above in the case of the droplet initial condition.
It is illustrated in Fig.~\ref{fig:gh} (left) where the relation $g$ versus $H$ is plotted, obtained
by inverting the relation $H=\log(\Psi'(-g))$, with $\Psi(z)$ given in Eq.~\eqref{2branches}. As we see on the figure, 
for $0<g \leq 1$ there are two values of $H$ corresponding to the same $g$. This corresponds to two distinct solutions of the $\{P, Q \}$ system 
in the attractive regime: they are different since the values of $Q(0,1)=e^H$ are different. As discussed in the text and below in
Section~\ref{app:continuation} the second branch contains an additional solitonic part, and for $H \to +\infty$, although $g \to 0$ the solution remains non trivial. \\

\begin{figure}[t!]
    \centering
    \includegraphics[scale=0.52]{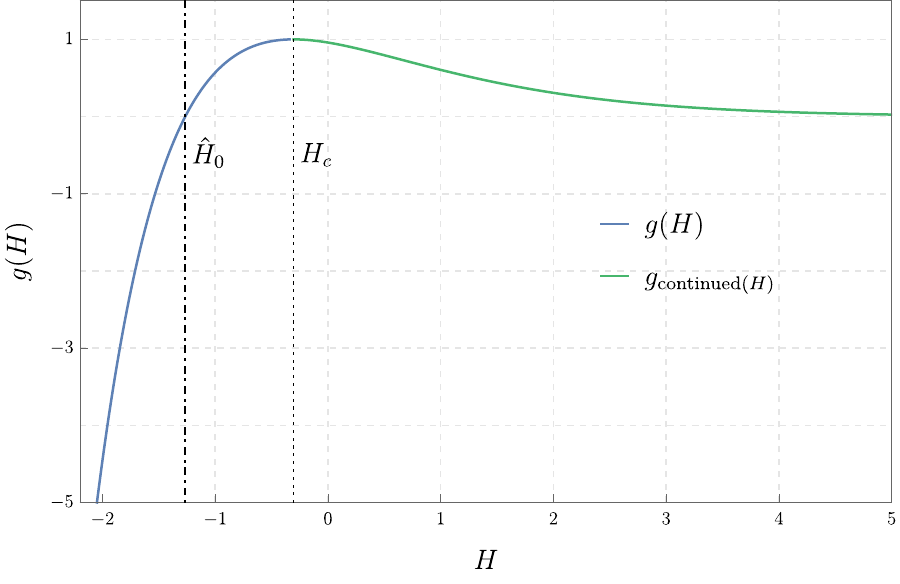}~~
\includegraphics[scale=0.52]{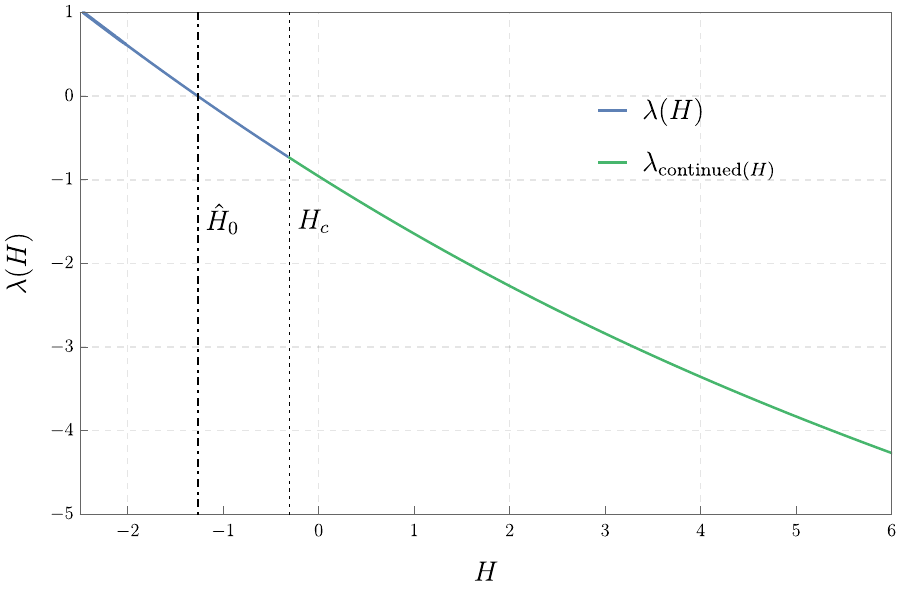}
    \caption{\textbf{Left.} Plot of the coupling $g$ versus $H$, where $g$ is the coupling of the $\{P, Q \}$ system \eqref{eq:PQsystem} with boundary data 
    \eqref{init} which should be used to obtain the large deviation rate function $\Phi(H)$ for a given $H$. For $0 < g \leq 1$ two values of $H$ correspond to
    the same $g$, leading to the two branches of solutions discussed in the text. 
    \textbf{Right.} Plot of the parameter $\lambda$ versus $H$, where the $\{P, Q \}$ system \eqref{eq:PQsystem} should now be solved with boundary data 
    \eqref{boundary2}. The function $\lambda(H)$ is monotonous. In this parametrization, the two branches of solutions in the left plot can be seen as a single branch. The notation \textit{continued} is used to describe the second branch of the functions.
     }
    \label{fig:gh}
\end{figure}

To see this phenomenon in a different light, instead of using $z$ as a parameter, it is possible to use $H$. 
More precisely, we can consider the ensemble where the parameter conjugated to $H$, called $\lambda$, is fixed
(while $z$ is the parameter conjugated to $e^H$). This amount to consider the generating function $G(\lambda)$
and its associated rate function $\psi(\lambda)$ defined as
\be \label{Glambda}
G(\lambda)= \int_\R \rmd H P(H,T) e^{- \frac{\lambda}{\sqrt{T}} H  } \simeq e^{- \frac{\psi(\lambda)}{\sqrt{T}}} \quad , \quad 
\psi(\lambda) = \min_{H \in \mathbb{R}} (\Phi(H) + \lambda H) 
\ee 
The relation between $\lambda$ and $H$ is now $\Phi'(H)=- \lambda$. Since $\Phi(H)$ is convex, $\lambda(H)$ is now monotonous, and there is 
a unique $H \in (-\infty,\infty)$ for each $\lambda \in (-\infty,+\infty)$. In Fig.~\ref{fig:gh} (right) we have plotted $\lambda(H)$ for the droplet initial condition, obtained parametrically
as $\lambda = z e^H = z \Psi'(z)$, $H=\log(\Psi'(z))$. The two branches for $\Psi(z)$ can now be seen as a single branch for $\lambda(H)$.
As is detailed in the next section, working at fixed $\lambda$ corresponds to solving the $\{P, Q \}$ system \eqref{eq:PQsystem}
with boundary data
\be \label{boundary2} 
Q(x,0)= Q_0(x) \quad , \quad g P(x,1) Q(x,1) = - \lambda \delta(x) 
\ee 
Note that $g$ can be absorbed in a redefinition $g P \to P$, hence it can be set to e.g. $g=-1$, so now the only parameter is $\lambda$, which varies 
on the whole real axis. In this parametrization ($\lambda$ instead of $g=-z$), the solution of the $\{P, Q \}$ system with boundary data 
\eqref{boundary2} can be seen as a single branch. Although this parameterization looks conceptually simpler, it hides the fact that something 
non trivial happens at $H=H_c$, i.e. the emergence of a solitonic part in the solution, which can be seen as the  emergence of 
a bound state in the inverse scattering method.

\subsection{KPZ dynamical action and and $\{P, Q \}$ system in $\lambda$ parametrization}

\label{app:dynH} 

In this section we derive the $\{P, Q \}$ system as the WNT on the dynamical action associated to the KPZ field $h$. It complements the short 
derivation given in the text using the dynamical action associated to the SHE field $Z$ and, being very similar, it provides the necessary details for that
derivation as well.
In addition it shows that the $\{P, Q \}$ system allows to solve the large deviation problem for any value of $H$, as claimed in the text. The derivation
is a variation on the one of \cite{Baruch,janas2016dynamical}, but here we work in the fixed $\lambda$ ensemble (see previous section),
which allows to give a clear interpretation of the parameter $\Lambda$ used there (analogous to our $\lambda$, see below).\\ 

Let us consider the KPZ equation \eqref{eq:KPZ} of the text, and rewrite it in the rescaled coordinates $x=y/\sqrt{T}$
and $t=\tau/T$, where we recall that $T$ is a fixed observation time. It is convenient to define the shifted field 
$h(y,\tau)+H_0$, where $H_0$ is a constant (defined as $H_0= \frac{1}{2} \log T$). We will slightly abuse notations
and use the same name for the new function, i.e. $h(x,t)=h(y,\tau)+ H_0$. It satisfies
\be  \label{KPZ2}
\partial_t h = \partial_x^2 h + (\partial_x h)^2 + \sqrt{2} T^{1/4} \tilde \eta(x,t) 
\ee 
where $\tilde \eta(x,t)$ is a standard space-time white noise. Let us denote $H=h(0,1)$. The 
generating function $\overline{ \exp( - \frac{\lambda}{\sqrt{T}} H ) }$, where $\overline{\dots}$ denotes
the average over the noise, of measure $\mathcal{D}\tilde{\eta} \exp(-1/2 \iint_{xt} \tilde{\eta}(x,t)^2)$, can be written using the standard MSR method \cite{MSR,kamenev2011field,TauberBook}. One expresses
the equation of motion \eqref{KPZ2} using delta functions, and their integral representations
by multiplying \eqref{KPZ2} by the response field $\frac{1}{\sqrt{T}} \tilde h(x,t)$, leading to 
\be
\begin{split}
 G(\lambda)&= \overline{  \exp( - \frac{\lambda}{\sqrt{T}} H ) } \\
 &= 
 \int \mathcal{D} \tilde h \mathcal{D} h\mathcal{D}\tilde{\eta} \exp\left( - \frac{1}{\sqrt{T}} [ S_0[\tilde h, h] + \lambda H - \int_0^{+\infty} \int_\R \rmd t   \rmd x 
 (\sqrt{2} T^{1/4} \tilde h(x,t) \tilde \eta(x,t) -\frac{\sqrt{T}}{2} \tilde{\eta}(x,t)^2)] \right) \\
& = \int \mathcal{D}\tilde h \mathcal{D} h  \exp\left( - \frac{1}{\sqrt{T}} ( S[\tilde h,h] + \lambda H) \right) \label{actionMSRh} 
\end{split}
\ee
where
\be 
S_0[\tilde h,h] = \int_0^{+\infty} \int_\R \rmd t    \rmd x \tilde h ( \partial_t h - \partial_x^2  h - (\partial_x h)^2 ) \quad , \quad 
S[\tilde h, h]= S_0[\tilde h,h] - \int_0^{+\infty} \int_\R \rmd t   \rmd x \tilde h^2 
\ee 
One has $G(0)=1$ automatically since the MSR path integral has unit normalisation as is checked using the proper time discretization
(here we use a fixed initial condition $h(x,0)$). \\

For $T \ll 1$ this can be evaluated by the saddle point (SP) method, leading to the SP equations obtained
by functional derivatives w.r.t. $\tilde h(x,t)$ and $h(x,t)$ of \eqref{actionMSRh} respectively
\bea
 \partial_t h &&= \partial_x^2 h + (\partial_x h)^2 + 2 \tilde h \label{wnth1} \\
 - \partial_t \tilde h && = \partial^2_x \tilde h - 2 \partial_x( \tilde h \partial_x h) - \lambda \delta(x) \delta(t-1) 
\label{wnth2} 
\eea
and the saddle point solution obeys $\tilde h(x,t)=0$ for $t>1$.
The functional derivative w.r.t. $h(x,t)$ of the term in \eqref{actionMSRh} containing $\lambda$ produces a source 
term in the second equation. By integrating over time $t \in [1^-,1^+]$ in \eqref{wnth2} and using that
$\tilde h(x,1^+)=0$, we see that it can be removed if we declare that the equations are to be studied on $t \in [0,1]$
and supplemented by the boundary condition at $t=1$
\be \label{boundaryhtilde} 
\tilde h(x,t=1) = -  \lambda \delta(x) 
\ee
the boundary condition for $\tilde h$ being free at $t=0$. The boundary condition is given for $h$ at $t=0$ and is free 
at $t=1$. The solution of the SP equation depends on a single parameter $\lambda$. Let us denote this solution $h_\lambda(x,t)$ and $\tilde h_\lambda(x,t)$, and
insert them back into Eq.~\eqref{actionMSRh}. One obtains 
\be
G(\lambda)= \overline{ \exp( - \frac{\lambda}{\sqrt{T}} H ) } \sim \exp\left( - \frac{1}{\sqrt{T}} ( \int_0^1 \int_\R \rmd t  \rmd x  \tilde h_\lambda^2  + \lambda h_\lambda(0,1) ) \right)
\ee
Hence, from \eqref{Glambda}, this implies
\be
\psi(\lambda) = \min_{H \in \mathbb{R}} ( \Phi(H) + \lambda H) = \int_0^1 \int_\R \rmd t  \rmd x \tilde h_\lambda^2  + \lambda h_\lambda(0,1)
\ee
Let us denote $H_\lambda = h_\lambda(0,1)$. Taking a derivative on each side and using that it is a saddle-point, we obtain that
$H_\lambda$ is also equal to $H_\lambda = {\rm arg min} ( \Phi(H) + \lambda H)$, that is the solution of
$\Phi'(H_\lambda)= - \lambda$, Hence, we also have the identity
\be \label{aga} 
\Phi(H_\lambda) = \int_0^1 \int_\R \rmd t  \rmd x \tilde h_\lambda^2
\ee

\paragraph{Equivalence with the $\{ P,Q \}$ system.} The saddle point equations \eqref{wnth1}, \eqref{wnth2} can be equivalently written as the $\{ P,Q \}$ system. Let us
define
\be \label{PQ2} 
Q(x,t)= e^{h(x,t)} \quad , \quad g P(x,t) Q(x,t) = \tilde h(x,t) 
\ee
Substituting into \eqref{wnth1}, \eqref{wnth2} we obtain the $P,Q$ system equation of the text \eqref{eq:PQsystem}.
However, from \eqref{boundaryhtilde} and \eqref{PQ2}, the boundary data are now given by \eqref{boundary2}, and the solution
depends on the single parameter $\lambda$ (since $g$ can be absorbed in $P$ simultaneously in the equations \eqref{eq:PQsystem} and in the boundary data \eqref{boundary2}). 
The $\{ P,Q \}$ system is thus naturally obtained here in the $\lambda$ parametrization. As discussed in the previous section,
it is equivalent to the parametrization using $g=-z$. Indeed, one can identify $H_z=H_\lambda$, $\lambda = z e^{H_z}$, and $\Phi(H_\lambda)$ from \eqref{aga}
with $\Phi(H_z)$ from \eqref{eq:PhiIntegral}. The relation between $\lambda$ and $g=-z$ however is not bijective and, for the droplet initial condition, can be
read parametrically from Fig.~\ref{fig:gh}, by eliminating $H$.

\paragraph{Remark.} In the text we used Ito time discretization to define the SHE \eqref{eq:SHE} as is customary as a proper
definition of the KPZ equation via the Cole-Hopf mapping \cite{CorwinKPZReview}. Going from $Z(x,t)$
to $h(x,t)=\log Z(x,t)$ adds a constant term in the KPZ equation \eqref{KPZ2}. For a smoothed noise correlator
$\overline{\tilde \eta(x,t) \tilde \eta(x',t')}= R(x-x') \delta(t-t')$ this constant term is $R(0)$
(which diverges for white noise). We have ignored this term in the above calculation. 
Alternatively one could instead work using Stratonovich.

\subsection{Comparison with the notations used in Meerson et al. works}

It is useful to give a dictionary with the common notations used in the WNT works of Refs.
\cite{Baruch,MeersonParabola,janas2016dynamical,meerson2017randomic,Meerson_Landau,Meerson_flatST,asida2019large,meerson2018large,smith2018finite,smith2019time}.
In these works the WNT equations involve the KPZ height field ${\sf h}(x,t)$ and the noise field $\rho(x,t)$. As here, $x,t$ are the rescaled space time variables, with 
$t \in [0,1]$ and $T$ denotes the observation time.
We denote ${\sf h}$ the KPZ field $h$ used in these works, ${\sf H}$ its value at $x=0$, $t=1$ ($t=T$ in the unrescaled original time),
and ${\sf S} = \frac{1}{2} \iint  \rmd x  \rmd t \rho^2$ the
saddle point action there. It is a function ${\sf S}(\sf H)$ of the observed KPZ height. The correspondence with our notations goes as follows
\be \label{dictionary} 
h(x,t) = - {\sf h}(x,t)/2 \quad , \quad  \tilde h(x,t)=-\rho(x,t)/4 \quad , \quad H = - {\sf H}/2 \quad , \quad \Phi(H)= \frac{1}{8} {\sf S}({\sf H}) 
\quad , \quad \lambda = \frac{1}{4} \Lambda 
\ee  
The saddle point equation $\Phi'(H)= - \lambda$ is thus equivalent to ${\sf S}'({\sf  H})=\Lambda$. 
Note that the definition of $\sf H$ in these works includes a shift (analogous to our $H_0$) which, however we have not
attempted to pin down. The centering of $\Phi(H)$ around the most probable value allows to make the correspondence
in each case. 

\subsection{General solutions of the $\{P,Q\}$ system and the solitonic (finite-rank) solutions} \label{app:finiterank} 

As mentioned in the text the general solution of the $\{P,Q\}$ system \eqref{eq:PQsystem} are of the form \cite{krajenbrink2020painleve,bothner2021atlas}
\be \label{soluQP2} 
Q(x,t)= \bra{ \delta } {\cal A}_{xt} (I + g {\cal B}_{xt} {\cal A}_{xt})^{-1} \ket{\delta} \quad , \quad 
P(x,t)= \bra{ \delta } {\cal B}_{xt} (I + g {\cal A}_{xt} {\cal B}_{xt})^{-1} \ket{\delta} 
\ee
in terms of two space-time dependent linear operators ${\cal A}_{xt}$, ${\cal B}_{xt}$ acting from $\mathbb{L}^2(\mathbb{R}^+)$ to $\mathbb{L}^2(\mathbb{R}^+)$,
where ${\cal O}^{-1}$ denotes the inverse of the operator ${\cal O}$. The kernel of these operators have the form, with $v,v' \geq 0$
\be \label{formAB} 
{\cal A}_{xt}(v,v')= A_t(x+v+v') ~,~ {\cal B}_{xt}(v,v')= B_t(x+v+v')
\ee
where the functions $A_t(x)$ and $B_t(x)$ are two solutions of the standard heat equation, i.e. forward $\partial_t A_t(x)=\partial_x^2 A_t(x)$, 
and backward $- \partial_t B_t(x)=\partial_x^2 B_t(x)$. These operators thus act on functions $f(v)$ as $({\cal A}_{xt} f)(v)= \int_0^{+\infty}  \rmd v' A_t(x+v+v') f(v')
= \langle \delta_v | {\cal A}_{xt}|f\rangle$.
We often use the bra-ket notation $f(v) = \bra{\delta_v} f \rangle$, where 
$\ket{\delta_{w}}$ is the vector with component $\delta(v-w)$ so that $\bra{\delta_v} {\cal O}\ket{\delta_{v'}} =\mathcal{O}(v,v')$,
and in \eqref{soluQP2} we denote $\ket{\delta}= \ket{\delta_{0}}$.\\

One can check, e.g. by expansion in $g$ that the $\{P,Q\}$ system \eqref{eq:PQsystem} is obeyed order by order: at lowest order $g^0$
the functions $P(x,t),Q(x,t)$ are simply identical to $B_t(x),A_t(x)$. 

This operator construction requires that the functions $A_t(x),B_t(x)$ vanish sufficiently fast at 
$x \to +\infty$ so that all integrals over the variables $v,v'$ converge. In the limit $x \to +\infty$ the operators ${\cal A}_{xt}$, ${\cal B}_{xt}$
thus become "small" and $P(x,t),Q(x,t)$ are thus asymptotically equivalent to $B_t(x),A_t(x)$. Hence Eq.~\eqref{soluQP2} can only describe solutions $P(x,t),Q(x,t)$ which decay
at $x \to +\infty$ (there are no restrictions however on their behavior for $x \to -\infty$). Note that a mirror-image construction can also be performed exchanging the roles of $x>0$ and $x<0$.\\

The product operator ${\cal B}_{xt} {\cal A}_{xt}$ which occurs in \eqref{soluQP2} thus reads in components $({\cal B}_{xt} {\cal A}_{xt})(v,v')=\int_0^{+\infty}  \rmd v'' B_t(x+v+v'') A_t(x+v''+v')$.
Taking a derivative w.r.t. $x$ we see that the integrand becomes a total derivative w.r.t. $v''$ which upon integration gives simply $- B_t(x+v) A_t(x+v')$, which is a projector.
This leads to the important identities used in the text
\be \label{identity} 
\partial_x ( {\cal B}_{xt} {\cal A}_{xt} ) = - {\cal B}_{xt} \ket{\delta} \bra{\delta} {\cal A}_{xt} \quad , \quad 
\partial_x ( {\cal A}_{xt} {\cal B}_{xt} ) = - {\cal A}_{xt} \ket{\delta} \bra{\delta} {\cal B}_{xt}\, .
\ee

It allows to express the product $g P Q$ in terms
of a Fredholm determinant. Indeed one has, using \eqref{identity}

\be \label{derivation} 
\begin{split}
 \partial_x \log \Det( I + g {\cal B}_{xt} {\cal A}_{xt} )_{\mathbb{L}^2(\R_+)} &= \partial_x {\rm Tr} \log( I + g {\cal B}_{xt} {\cal A}_{xt} )\\
&= - g {\rm Tr} [ ( I + g {\cal B}_{xt} {\cal A}_{xt} )^{-1} {\cal B}_{xt} \ket{\delta} \bra{\delta} {\cal A}_{xt} ]  \\
&=  - g \bra{\delta} {\cal A}_{xt}  ( I + g {\cal B}_{xt} {\cal A}_{xt} )^{-1} {\cal B}_{xt} \ket{\delta} \\
&= - g \bra{\delta} ( I + g {\cal A}_{xt} {\cal B}_{xt} )^{-1} {\cal A}_{xt} {\cal B}_{xt} \ket{\delta} \\
&=  \bra{\delta} ( ( I + g {\cal A}_{xt} {\cal B}_{xt} )^{-1} - I )   \ket{\delta} 
\end{split}
\ee
Taking another derivative, using that
\begin{equation}
    \p_x ( I + g {\cal A}_{xt} {\cal B}_{xt} )^{-1}=-g( I + g {\cal A}_{xt} {\cal B}_{xt} )^{-1} \p_x({\cal A}_{xt} {\cal B}_{xt} ) ( I + g {\cal A}_{xt} {\cal B}_{xt} )^{-1}
\end{equation}
and using again \eqref{identity} we obtain
\be \label{gPQDet} 
\begin{split}
 \partial_x^2 \log \Det( I + g {\cal B}_{xt} {\cal A}_{xt} )_{\mathbb{L}^2(\R_+)} &= g \bra{\delta}  ( I + g {\cal A}_{xt} {\cal B}_{xt} )^{-1} 
{\cal A}_{xt} \ket{\delta} \bra{\delta} {\cal B}_{xt} ( I + g {\cal A}_{xt} {\cal B}_{xt} )^{-1}    \ket{\delta} \\
&= g Q(x,t) P(x,t)
\end{split}
\ee 
where we rearranged the operators to obtain the last equality. Note that, since the product $C_1 = \int_\R  \rmd x g P Q$ is a conserved quantity, i.e. $\partial_t C_1=0$ as
easily verified from \eqref{eq:PQsystem}, $C_1$ equals to minus the value reached by the last term in \eqref{derivation} 
at $x=-\infty$ (since it vanishes for $x=+\infty$). As a consequence, this value is finite and time-independent.\\

In general the inversion of the operators in \eqref{soluQP2} is a non-trivial task. However there exists solutions for which this inversion is 
possible explicitly, these are the (multi) solitonic solutions. They correspond to the case where at least one of the operators ${\cal A}_{xt}$ and ${\cal B}_{xt}$ is a finite-rank operator. 


{\bf Rank-one}. The simplest solution is when ${\cal A}_{xt}$ and ${\cal B}_{xt}$ are both rank-one operators. Since they are of the form \eqref{formAB} and
must obey the heat equation, they can be written as
\be \label{rank1soliton} 
{\cal A}_{xt}(v,v') = q_\kappa(x,t) e^{- \kappa (v+v')} \quad , \quad {\cal B}_{xt}(v,v') = p_\mu(x,t) e^{- \mu (v+v')}
\ee
where we denote $q_\kappa=q_\kappa(x,t) = \tilde q e^{- \kappa x + \kappa^2 t}$
and $p_\mu = p_\mu(x,t) = \tilde p e^{- \mu x - \mu^2 t}$ two (evanescent) plane wave solutions of the heat equation,
where $\tilde q,\tilde p$ are amplitudes.
One can equivalently write ${\cal A}_{xt}= q_\kappa(x,t) \ket{\kappa} \bra{\kappa}$ and 
${\cal B}_{xt}= p_\mu(x,t) \ket{\mu} \bra{\mu}$ were we denote $\bra{\delta_v} \kappa \rangle = e^{- \kappa v}$, 
$\bra{\delta_v} \mu \rangle = e^{- \mu v}$. In order for the operator product ${\cal A}_{xt} {\cal B}_{xt}$ to be 
convergent we must impose that ${\Re}(\kappa+\mu) >0$. Here we will focus on the case where $\kappa,\mu$ are real and the product $g \tilde p \tilde q >0$, where the
solutions for $P,Q$ can be defined for all times. It is possible to choose these parameter complex and obtain
solutions on finite time intervals but we will not consider it here. The scalar product is then 
$\langle \kappa | \mu \rangle = \frac{1}{\kappa+\mu}$.
Since ${\cal A}_{xt} {\cal B}_{xt}  = \frac{q_\kappa p_\mu}{\kappa+\mu} |\kappa \rangle  \langle \mu|$ is also rank-one, and similarly for
${\cal B}_{xt} {\cal A}_{xt}$ the operator inversion in \eqref{soluQP2} is easy to perform using the Sherman Morrison formula
$( I + |a \rangle \langle b| )^{-1} = I - \frac{|a \rangle \langle b|}{1 + \langle b| a \rangle}$
and one obtains the simplest solution of the $\{ P,Q\} $ system 
\be 
\label{Q1} 
Q(x,t) =  \frac{\tilde q e^{- \kappa x + \kappa^2 t} }{1 + g \frac{\tilde p \tilde q}{(\kappa+\mu)^2} 
e^{- (\kappa+\mu) x + (\kappa^2-\mu^2) t } } \quad , \quad
 P(x,t) = \frac{\tilde p e^{- \mu x - \mu^2 t} }{1 + g \frac{\tilde p \tilde q}{(\kappa+\mu)^2} 
e^{- (\kappa+\mu) x + (\kappa^2-\mu^2) t } } 
\ee
For $\kappa,\mu$ real this corresponds to the standard one-soliton solution with 
\be \label{gPQ} 
g P Q  = 
\frac{(\kappa+\mu)^2}{ 4 \cosh^2 \left( \frac{1}{2} (\kappa+\mu) (x-x_0(t)) \right)}
\quad , \quad 
x_0(t) = (\kappa-\mu) t + y_0
\ee
with $y_0=\frac{2}{\kappa+\mu} \log \frac{\sqrt{g \tilde p \tilde q}}{\kappa+\mu}$.
Note that the matrix determinant lemma for a rank-one perturbation also
gives $\det( I + g {\cal B}_{xt} {\cal A}_{xt}) = \det( I + g \frac{q_\kappa p_\mu}{\kappa+\mu} |\mu \rangle  \langle \kappa|)=
1 + g \frac{q_\kappa(x,t) p_\mu(x,t)}{(\kappa+\mu)^2}$.
We see that \eqref{gPQ} then agrees with the general formula \eqref{gPQDet}. In particular, integrating
over space, one finds that the conserved charge for the soliton is $C_1 = \int_\R  \rmd x g P Q = \kappa+\mu$. Hence $C_1$
does not depend on $g$, which shows the non-perturbative nature of such solutions. The higher-order  conserved quantities read
 \begin{equation}
     C_2=\frac{1}{2}(\mu^2-\kappa^2), \quad C_3=\frac{1}{3}(\mu^3+\kappa^3), \quad C_n=\frac{1}{n}(\mu^n-(-\kappa)^n).
 \end{equation}
Note that soliton solutions of the $\{P,Q\}$ system with $\kappa+\mu<0$ also exist, and can be deduced from the present ones by $x \to -x$,
However in terms of operators, they require the mirror-image construction. 
\\

{\bf General finite-rank}. Consider now the case where ${\cal A}_{xt}$ and ${\cal B}_{xt}$ are rank $n_1$ and $n_2$ operators, respectively, i.e.
${\cal A}_{xt} = \sum_{j=1}^{n_1} q_{\kappa_j} |\kappa_j \rangle \langle \kappa_j|$ and ${\cal B}_{xt} = \sum_{i=1}^{n_2} p_{\mu_i} |\mu_i \rangle  \langle \mu_i|$
and $q_{\kappa_j}=q_{\kappa_j}(x,t) = \tilde q_j e^{- \kappa_j x + \kappa_j^2 t}$ and $p_{\mu_i}=p_{\mu_i}(x,t) = \tilde p_i e^{- \mu_i x - \mu_i^2 t}$
are plane waves. An elementary calculation, not reproduced here, using either the Woodbury matrix identity, or resummation of the
expansion in $g$, gives 
\bea
&& Q(x,t) = \sum_{i,j=1}^{n_1} q_{\kappa_i}  (I + g \sigma \gamma)^{-1}_{ij}  \quad , \quad P(x,t) = \sum_{i,j=1}^{n_2} p_{\mu_i}  (I + g \sigma^T \gamma^T)^{-1}_{ij}  \\
&& \gamma_{ij}= \frac{p_{\mu_i} q_{\kappa_j}}{\mu_i + \kappa_j} \quad , \quad 
\sigma_{ij}= \frac{1}{\kappa_i+\mu_j} 
\quad , \quad q_{\kappa_j} = \tilde q_j e^{- \kappa_j x + \kappa_j^2 t} 
\quad , \quad
p_{\mu_j} = \tilde p_j e^{- \mu_j x - \mu_j^2 t} \\
&& \nonumber 
\eea
For small ranks it is easy to check with Mathematica that these obey the $\{ P, Q \}$ system.
Using the matrix determinant lemma one also has
\be
\Det( I + g {\cal B}_{xt} {\cal A}_{xt} ) = \det_{n_1 \times n_1} (I + g \sigma \gamma) = \det_{n_2 \times n_2} (I + g \sigma^T \gamma^T) =  \Det( I + g {\cal A}_{xt} {\cal B}_{xt} )
\ee
and one can check explicitly again the relation \eqref{gPQDet}.
Hence the determinant which appears in these finite-rank soliton solutions is, more explicitly
\be
\det_{1 \leq i,j \leq n_1} (\delta_{i,j} + g \sum_{k=1}^{n_2} 
\frac{1}{\kappa_i+\mu_k}  \frac{p_{\mu_k}(x,t) q_{\kappa_j}(x,t)}{\mu_k + \kappa_j} )  
= \det_{1 \leq i,j \leq n_2} (\delta_{i,j} + g \sum_{k=1}^{n_1} 
\frac{1}{\mu_i+\kappa_k}  \frac{ q_{\kappa_k}(x,t) p_{\mu_j}(x,t)}{\kappa_k+ \mu_j } ) 
\ee 

One can check that the family of solutions pointed out in \cite[Appendix B]{janas2016dynamical}
using the Hirota method 
is equivalent to our particular family $n_1=N-1$, $n_2=1$ 
and considering $q_{\kappa_1},\dots,q_{\kappa_{N-1}},\frac{1}{p_\mu}$ that is
$\kappa_j=c_j$, $j=1,N-1$ and $c_N=-\mu$. So it is a rank-one family.

\subsection{Derivation of the general solution to the $\{P,Q\}$ system}
In this appendix, we show that the functions defined in the text
\bea \label{soluQP-appendix} 
&& Q(x,t)= \bra{ \delta } {\cal A}_{xt} (I + g {\cal B}_{xt} {\cal A}_{xt})^{-1} \ket{\delta} \\
&& P(x,t)= \bra{ \delta } {\cal B}_{xt} (I + g {\cal A}_{xt} {\cal B}_{xt})^{-1} \ket{\delta} \nonumber
\eea
in terms of the operators 
\be \label{kernels-appendix} 
{\cal A}_{xt}(v,v')= A_t(x+v+v') ~,~ {\cal B}_{xt}(v,v')= B_t(x+v+v')
\ee
with $v,v' \in \mathbb{R}^+$, and which satisfy
\be 
\label{time-derivative-appendix} 
 \p_t \mathcal{A}_{xt}=\p_x^2\mathcal{A}_{xt} \quad , \quad \p_t \mathcal{B}_{xt}=-\p_x^2\mathcal{B}_{xt}
\ee
are solutions to the $\{P,Q\}$ system
\be
\begin{split}\label{eq:PQsystem-appendix}
 \partial_t Q =& \partial_x^2 Q + 2 g P Q^2  \\
 - \partial_t P =& \partial_x^2 P + 2 g P^2 Q  
 \end{split}
\ee
{\bf Remark.} If $\{P,Q\}$ is a solution of \eqref{eq:PQsystem-appendix} for a coupling constant $g$, 
then $\{(\tilde P = (g/g') P,Q\}$ is a solution of \eqref{eq:PQsystem-appendix}  for a coupling constant $g \to g'$. 
That solution can also be obtained from \eqref{soluQP-appendix} upon rescaling the operator $\B \to (g'/g) \B$.
Hence it is sufficient to show the above statement for a single value of $g$, and we will solely focus on $g=-1$ in the following.
\subsubsection{Preliminary tools and definitions}
We introduce the appropriate tools and results to derive the solution of \eqref{eq:PQsystem-appendix}. Define a hierarchy of four functions indexed by an integer $p \geq 0$ \cite{krajenbrink2020painleve,bothner2021atlas}
\bea
&& q_p(x,t) = \bra{\delta } \A^{(p)} (I - \B \A)^{-1} \ket{\delta}  \quad , \quad 
u_p(x,t) = \bra{\delta } \A^{(p)} (I - \B \A)^{-1} \B\ket{\delta} \\
&& \tilde q_p(x,t) = \bra{\delta } \B^{(p)} (I - \A\B)^{-1} \ket{\delta} \quad , \quad 
\tilde u_p(x,t) = \bra{\delta } \B^{(p)} (I - \A \B)^{-1} \A \ket{\delta}
\eea
where $\A^{(p)}(v,v')=A^{(p)}_t(x+v+v')$ has the same kernel differentiated $p$-times, and the same for $\B^{(p)}$.
Note that $u_0 = \tilde u_0$ since the operators are symmetric.  Note that
\begin{equation}
\label{eq:AKNS-Marcenko-solution}
Q(x,t)=q_0(x,t), \quad P(x,t)=\tilde{q}_0(x,t) \,.
\end{equation}

As a consequence of the additive structure of the kernels \eqref{kernels-appendix} one has the following algebraic identities
\bea 
            \partial_x ( {\cal A}_{xt} {\cal B}_{xt} ) &=&  - {\cal A}_{xt} \ket{\delta} \bra{\delta} {\cal B}_{xt}     \label{algebraic1}  \\
                    \p_x ( I - {\cal A}_{xt} {\cal B}_{xt} )^{-1} & = & ( I - {\cal A}_{xt} {\cal B}_{xt} )^{-1} \p_x({\cal A}_{xt} {\cal B}_{xt} ) ( I - {\cal A}_{xt} {\cal B}_{xt} )^{-1} \nonumber \\
                    &=& -( I - {\cal A}_{xt} {\cal B}_{xt} )^{-1} {\cal A}_{xt} \ket{\delta} \bra{\delta} {\cal B}_{xt}( I - {\cal A}_{xt} {\cal B}_{xt} )^{-1}
                    \label{algebraic2} 
 \eea
so the space derivative of the operator  $( I - {\cal A}_{xt} {\cal B}_{xt} )^{-1}$ is a rank-one operator. The same identities hold with
$\A$ and $\B$ exchanged. Note that in \eqref{algebraic1} one can replace $\A$ by any of its derivative, and the same for $\B$. \\

From this, we deduce a hierarchy of differential equations for the functions $q_p, u_p, \tilde{q}_p, \tilde{u}_p$ which reads for all $p\geq 0$
\bea \label{systemqu} 
&& \partial_x q_p=q_{p+1}- q_0 u_p \quad , \quad \partial_x u_p=- \tilde q_0 q_p \\
&& \partial_x \tilde q_p=\tilde q_{p+1}- \tilde q_0 \tilde u_p
\quad , \quad \partial_x \tilde u_p=- q_0 \tilde q_p
\eea 
The equation for $\partial_x q_p$ and $\partial_x \tilde q_p$ are straightforward consequences of 
the definitions and \eqref{algebraic2}. The equations for $\partial_x u_p$ and $\partial_x \tilde u_p$
are derived as follows. 

\begin{equation}
\begin{split}
\p_x\bra{\delta } \A^{(p)} (I - \B \A)^{-1} \B\ket{\delta}&=\p_x\bra{\delta } D^p \A (I - \B \A)^{-1} \B\ket{\delta}\\
&=\p_x\bra{\delta } D^p  (I - \A \B)^{-1} \A\B\ket{\delta}\\
&=-\p_x\bra{\delta } D^p  \ket{\delta}+\p_x\bra{\delta } D^p  (I - \A \B)^{-1} \ket{\delta}\\
&=-\bra{\delta } D^p  ( I - \A \B )^{-1} \A \ket{\delta} \bra{\delta} \B( I - \A \B )^{-1} \ket{\delta}\\
&=-\bra{\delta } D^p \A ( I - \B \A )^{-1}  \ket{\delta} \bra{\delta} \B( I - \A \B )^{-1} \ket{\delta}\\
&=-\bra{\delta }  \A^{(p)} ( I - \B \A )^{-1}  \ket{\delta} \bra{\delta} \B( I - \A \B )^{-1} \ket{\delta}\\
\end{split}
\end{equation}
where $D$ is the derivative operator which acts as $(Df)(u)=f'(u)$and where we have used that $\p_x D^p =0$. For a recent rigorous derivation of this system of equation, see \cite{bothner2021atlas}.\\ 

The above system \eqref{systemqu} admits an infinite set of conserved quantities, and as a consequence 
for all $k \geq 1$ one has
\bea
(-1)^k u_{k-1} + \tilde u_{k-1} =   \sum_{\ell=1}^{k-1} (-1)^{k-\ell-1} (\tilde u_{\ell-1} u_{k-\ell-1} 
- \tilde q_{\ell-1} q_{k-\ell-1} )
\eea 
This is shown by checking that taking the derivative $\partial_x$ one finds an identity and
using that all these functions vanish at $x \to +\infty$.  In particular, we will use below only the case $k=2$, which reads
 \be
 u_1 + \tilde u_1 = \tilde u_0 u_0 - \tilde q_0 q_0 = u_0^2 - \tilde q_0 q_0 
 \ee
 recalling that $\tilde u_0=u_0$. We now require time-derivative relations to complete these preliminaries results. From the time-derivative structure \eqref{time-derivative-appendix} we have that  
 
\be 
   \p_t ( I - {\cal A}_{xt} {\cal B}_{xt} )^{-1}=( I - {\cal A}_{xt} {\cal B}_{xt} )^{-1} \p_t({\cal A}_{xt} {\cal B}_{xt} ) ( I - {\cal A}_{xt} {\cal B}_{xt} )^{-1}
\ee    
and using \eqref{algebraic1} for the derivatives
    \begin{equation}
    \begin{split}
        \p_t({\cal A}_{xt} {\cal B}_{xt})&=\p_x \left[(\p_x{\cal A}_{xt}) {\cal B}_{xt}-{\cal A}_{xt} \p_x{\cal B}_{xt})\right]\\
        &=-\p_x{\cal A}_{xt} \ket{\delta} \bra{\delta} {\cal B}_{xt}+{\cal A}_{xt} \ket{\delta} \bra{\delta} \p_x{\cal B}_{xt}
        \end{split}
    \end{equation}
    as well as
      \begin{equation}
    \begin{split}
        \p_t({\cal B}_{xt} {\cal A}_{xt})&=\p_x \left[ - (\p_x{\cal B}_{xt}) {\cal A}_{xt} + {\cal B}_{xt} \p_x{\cal A}_{xt}\right]\\
        &=\p_x{\cal B}_{xt} \ket{\delta} \bra{\delta} {\cal A}_{xt}-{\cal B}_{xt} \ket{\delta} \bra{\delta} \p_x{\cal A}_{xt}
        \end{split}
    \end{equation}

which is the same with $\mathcal{A}_{xt}$, $\mathcal{B}_{xt}$ exchanged.\\
    
    \subsubsection{Derivation of the $\{ P, Q \}$ system}
    
We can now obtain the $\{ P, Q \}$ system \eqref{eq:PQsystem-appendix} with $g=-1$. 
Let us start with the equation for $q_0=Q$. One has, using the above algebraic relations 
    \begin{equation}
\begin{split}
    \p_t q_0 &= \p_t  \bra{\delta } \A (I - \B \A)^{-1} \ket{\delta}  \\
    & =  \bra{\delta } \p_t \A (I - \B \A)^{-1} \ket{\delta}+ \bra{\delta } \A \p_t (I - \B \A)^{-1} \ket{\delta}  \\
        & =  q_2- \bra{\delta } \A  (I - \B \A)^{-1} \left[{\mathcal B}_{xt}\ket{\delta} \bra{\delta}\p_x{\mathcal A}_{xt}  - \p_x{\mathcal B}_{xt}\ket{\delta} \bra{\delta}{\mathcal A}_{xt} \right] (I - \B \A)^{-1}\ket{\delta}  \\
        &= q_2- u_0q_1+ \tilde{u}_1q_0\\
        &= \partial_x q_1+q_0 u_1- u_0q_1+ \tilde{u}_1q_0\\
        &= \partial_x^2 q_0+ \partial_x (q_0u_0)+q_0 u_1- u_0q_1+ \tilde{u}_1q_0\\
        &= \partial_x^2 q_0+(q_1u_0-q_0u_0^2-q_0^2\tilde{q}_0)+q_0 u_1- u_0q_1+ \tilde{u}_1q_0\\
        &= \partial_x^2 q_0-q_0u_0^2-q_0^2\tilde{q}_0+q_0 u_1+ \tilde{u}_1q_0\\
        &= \partial_x^2 q_0-q_0u_0^2-q_0^2\tilde{q}_0+(u_0^2-q_0\tilde{q}_0)q_0\\
        &= \partial_x^2 q_0-2q_0^2\tilde{q}_0\\
    \end{split}
\end{equation}
which gives the first equation of the $\{ P, Q \}$ system. The equation for $\tilde q_0=P$ is derived in the same way
    \begin{equation}
\begin{split}
  -   \p_t \tilde q_0 &= - \p_t  \bra{\delta } \B (I - \A \B)^{-1} \ket{\delta}  \\
    & = -  \bra{\delta } \p_t \B (I - \A \B)^{-1} \ket{\delta} - \bra{\delta } \B \p_t (I - \A \B)^{-1} \ket{\delta}  \\
        & =  \tilde q_2- \bra{\delta } \B  (I - \A \B)^{-1} \left[{\mathcal A}_{xt}\ket{\delta} \bra{\delta}\p_x{\mathcal B}_{xt}  - \p_x{\mathcal A}_{xt}\ket{\delta} \bra{\delta}{\mathcal B}_{xt} \right] (I - \A \B)^{-1}\ket{\delta}  \\
        &= \tilde q_2- \tilde u_0 \tilde q_1+ u_1 \tilde q_0\\
        &= \partial_x \tilde q_1+ \tilde q_0 \tilde u_1- \tilde u_0 \tilde q_1+ u_1 \tilde q_0\\
        &= \partial_x^2 \tilde q_0+ \partial_x (\tilde q_0 \tilde u_0)+\tilde q_0 \tilde u_1- \tilde u_0 \tilde q_1+ u_1\tilde q_0\\
        &= \partial_x^2 \tilde q_0+(\tilde q_1 \tilde u_0- \tilde q_0 \tilde u_0^2- \tilde q_0^2 q_0)+\tilde q_0 \tilde u_1- \tilde u_0 \tilde q_1+ u_1 \tilde q_0\\
        &= \partial_x^2 \tilde q_0- \tilde q_0 \tilde u_0^2- \tilde q_0^2 q_0+\tilde q_0 \tilde u_1+ u_1 \tilde q_0\\
        &= \partial_x^2 \tilde q_0- \tilde q_0 \tilde u_0^2-\tilde q_0^2 q_0+(u_0^2-q_0\tilde{q}_0) \tilde q_0\\
        &= \partial_x^2 \tilde q_0-2\tilde q_0^2 q_0\\
    \end{split}
\end{equation}
recalling that $\tilde u_0=u_0$. This gives the second equation of the $\{ P, Q \}$ system. \\

{\bf Remark.} The operator structure \eqref{soluQP2} as a solution of the $\{P,Q\}$ system \eqref{eq:PQsystem} can also be verified
from a mapping of the Lax pair \eqref{eq:LaxPairU} of the $\{P,Q\}$ system to a Riemann-Hilbert problem uniquely solvable by the means of a Fredholm determinant, see \cite{krajenbrink2020painleve,bothner2021atlas} for recent works on this approach.\\   
    
{\bf Remark.} There is another approach to derive the solutions \eqref{soluQP-appendix} for the $\{ P,Q\}$ system and more generally for a system in the AKNS class. One can introduce the Gelfand-Levitan-Marcenko equations, see \cite{AblowitzKaup1974} for review, and proceed to a formal operator inversion to obtain \eqref{soluQP-appendix}.

\subsection{Scattering problem at $t=1$}
\label{sec:scatt1} 

Here we show the relations obtained in the text for the scattering amplitudes 
for the WNT with a general initial condition. Although $Q(x,0)$ is as yet unspecified,
we can use the data at $t=1$, $P(x,1)=\delta(x)$ from Eq.~\eqref{init}.
\\

{\bf Equation for $\bar{\phi}$ at $t=1$ }. This equation allows to determine $\tilde b(k)$ and
gives some relation involving $\tilde a(k)$. We call $\bar{\phi}_{1,2}(x,t)$ the two components of $\bar{\phi}$. 
Let us recall that at $x \to -\infty$, $\bar \phi \simeq (0,-e^{\I k x/2})^\intercal$. The first equation of
the Lax pair given in the text, $\partial_x \vec v= U_1 \vec v$ with $\Vec{v}=e^{- k^2 t/2} {\bar{\phi}}$ and $U_1$ given in \eqref{eq:LaxPairU} , reads in components at $t=1$,
using that $P(x,1)=\delta(x)$ from \eqref{init},
\begin{equation} \label{eq:barphi12} 
\p_x \bar{\phi}_1=-\I \frac{k}{2} \bar{\phi}_1- g \delta(x)\bar{\phi}_2 \quad , \quad \p_x \bar{\phi}_2=\I \frac{k}{2} \bar{\phi}_2+ Q(x,1)\bar{\phi}_1\\
\end{equation}
Let us integrate the first equation. Since $\bar{\phi}_1$ vanishes at $-\infty$ it gives
\be \label{eq:phi11} 
\bar{\phi}_1(x,1)=- g e^{-\I \frac{k}{2} x}\Theta(x) \bar{\phi}_2(0,1)
\ee 
Taking the limit $x \to +\infty$, we thus obtain, from the asymptotics \eqref{eq:plusinfinity} that
\be \label{eq:btilde11} 
\tilde{b}(k,t=1)=- g \bar{\phi}_2(0,1) 
\ee
To determine $\bar{\phi}_2(0,1)$ we can integrate the second equation in \eqref{eq:barphi12}, which gives, using 
\eqref{eq:phi11} and \eqref{eq:btilde11}
\begin{equation}
\begin{cases} \label{eq:phibsolu}
e^{-\I \frac{k}{2} x}\bar{\phi}_2(x,1)=\bar{\phi}_2(0,1) + \tilde{b}(k,1)\int_{0}^x \rmd x' Q(x',1)e^{-\I k x'}, \quad x>0\\
\bar{\phi}_2(x,1)=-e^{\I \frac{k}{2} x}, \quad x<0
\end{cases}
\end{equation}
where in the second equation we have used that $\bar{\phi}_2(x,1)\simeq -e^{\I \frac{k}{2} x}$ for $x \to -\infty$.
Assuming continuity of $\bar{\phi}_2(x,1)$ at $x=0$, this leads to $\bar{\phi}_2(0,1)=-1$ and to
\be  \label{eq:tildeb} 
\tilde{b}(k,t=1)= g \quad   \Rightarrow \quad \tilde b(k)= g e^{-k^2} 
\ee 
since we recall that $\tilde b(k,t)=\tilde b(k) e^{k^2 t}$, see text. Taking the $x\to +\infty$ limit
of \eqref{eq:phibsolu} and using the asymptotics \eqref{eq:plusinfinity} we also obtain the relation given in the text
\begin{equation} \label{eq:tildea} 
\tilde{a}(k,1) = \tilde a(k) = 1- g \int_{0}^{+\infty} \rmd x' Q(x',1)e^{- \I k x'}
\end{equation}

{\bf Equation for $\phi$ at $t=1$ }. This equation allows to obtain some relation involving $a(k)$ and $b(k)$. 
The first equation of the Lax pair, $\partial_x \vec v= U_1 \vec v$ with $\Vec{v}=e^{k^2 t/2} {\phi}$ as given
in the text now reads, in components and at $t=1$, using that $P(x,1)=\delta(x)$ from \eqref{init},
\begin{equation} \label{eq:barphi12n} 
\p_x {\phi}_1=-\I \frac{k}{2} {\phi}_1- g \delta(x) {\phi}_2 \quad , \quad \p_x {\phi}_2=\I \frac{k}{2} {\phi}_2+ Q(x,1) {\phi}_1\\
\end{equation}
i.e. the same equations as \eqref{eq:barphi12} but the boundary conditions are different. Let us recall that at $x \to -\infty$, 
$\phi \simeq (e^{-\I k x/2},0)^\intercal$. The equations \eqref{eq:barphi12n} can be rewritten as 
\begin{equation} \label{2eq} 
\begin{split}
[e^{\I \frac{k}{2} x}\phi_1(x,1)]'&=- g \delta(x) \phi_2(x,1)e^{\I \frac{k}{2} x}, \qquad [e^{-\I \frac{k}{2} x}\phi_2(x,1)]'=Q(x,1)\phi_1(x,1)e^{-\I \frac{k}{2} x}
\end{split}
\end{equation}
Integrating these two equations, and using the asymptotics \eqref{eq:plusinfinity} at $x \to +\infty$ we obtain
\begin{equation}
\begin{split} \label{solusolu} 
\phi_1(x,1)&=e^{-\I \frac{k}{2} x}(\Theta(-x)+a(k)\Theta(x)), \quad a(k)-1=- g \phi_2(0,1)\\
\phi_2(x,1)&=e^{\I \frac{k}{2} x }\int_{-\infty}^x \rmd x' Q(x',1)e^{- \I k x'}(\Theta(-x')+a(k)\Theta(x'))
\end{split}
\end{equation}
where we used that $a(k,t)=a(k)$, see the main text. Setting $x=0$ in the second equation we obtain the relation displayed in the text
\begin{equation} \label{eq:a} 
\phi_2(0,1)=\int_{-\infty}^0 \rmd x' Q(x',1)e^{-\I k x'}, \quad 
a(k)=1- g \int_{-\infty}^0 \rmd x' Q(x',1)e^{- \I k x'}
\end{equation}

Note that integrating the second equation in \eqref{2eq} for $\phi_2(x,1)e^{-\I \frac{k}{2} x}$ between 0 and $+\infty$ and using the asymptotics \eqref{eq:plusinfinity} 
leads to an expression for $b(k)$, however this expression is equivalent to the one obtained from the relation
$a(k) \tilde a(k) + b(k) \tilde b(k)=1$ obtained from the Wronskian (see text) together with the above results for
$\tilde b(k),\tilde a(k), a(k)$. 


As noted in the text, at this stage $Q(x,1)$ is unknown. If the initial 
condition $Q(x,0)$ is even in $x$, then so is $Q(x,1)$ and \eqref{eq:tildea} and \eqref{eq:a} imply that $\tilde{a}(k)=a(-k)=a^*(k^*)$.
From the Wronskian relation and \eqref{eq:tildeb} one thus gets $b(k) g e^{-k^2} = 1 - a(k) a(-k) = 1 - |a(k)|^2$, hence
$b(k)$ is real and even in $k$. Alternatively one sees that $|a(k)|$ is fixed by $b(k)$ so one can write
\begin{equation}
a(k)=e^{- \I \varphi(k)} \sqrt{1-g b(k) e^{-k^2}}
\end{equation}
where $\varphi(k)$ is a real and odd function $\varphi(k)=-\varphi(-k)$. 

\subsection{Scattering problem: general $t$}
\label{sec:scattgeneral} 

Here we obtain some relations from the scattering problem at any $t$ always valid for the $\{ P,Q \} $ system 
(for arbitrary boundary conditions, i.e. beyond the WNT). In the case of the WNT with the boundary 
conditions \eqref{init}, when specified to $t=1$ they lead to the same results as in the previous
section. When specified to $t=0$ they give some (useful but not too explicit) formula for $b(k)$ 
for general initial condition. For the droplet initial condition they give $b(k)=1$. \\

Let us return to the $\partial_x$ equation of the Lax pair for $\phi(x,t)$, which we write at a fixed $t$, in the form
(here we also indicate the dependence in $k$) 
\begin{equation}
\begin{split}
\partial_x [e^{\I \frac{k}{2} x}\phi_1(x,t,k)]&=- g P(x,t) \phi_2(x,t,k)e^{\I \frac{k}{2} x}, \qquad \partial_x [e^{-\I \frac{k}{2} x}\phi_2(x,t,k)] = Q(x,t)\phi_1(x,t,k)e^{-\I \frac{k}{2} x}
\end{split}
\label{eq:SuppMatLaxPairEqGeneral}
\end{equation}
Integrating the first equation of \eqref{eq:SuppMatLaxPairEqGeneral} from $x=-\infty$ to $x=+\infty$, and using that $a(k,t)=a(k)$ we have
\begin{equation}
a(k)-1=-g \int_{-\infty}^{+\infty} \rmd x \, P(x,t)\phi_2(x,t,k)e^{\I \frac{k}{2} x}
\end{equation}
and from $-\infty$ and a value $x$
\begin{equation}
e^{\I \frac{k}{2}  x}\phi_1(x,t,k)=1- g \int_{-\infty}^{x} \rmd x' \, P(x',t)\phi_2(x',t,k)e^{\I \frac{k}{2} x'}
\end{equation}

Integrating the second equation of \eqref{eq:SuppMatLaxPairEqGeneral} between $-\infty$ and $+\infty$, we have 
\begin{equation}
b(k) e^{-k^2 t} =  \int_{-\infty}^{+\infty} \rmd x Q(x,t)\phi_1(x,t,k)e^{-\I \frac{k}{2} x}
\end{equation}
and from $-\infty$ and a value $x$
\begin{equation}
e^{-\I \frac{k}{2} x}\phi_2(x,t,k)= \int_{-\infty}^{x} \rmd x' \, Q(x',t)\phi_1(x',t,k)e^{-\I \frac{k}{2} x'}
\end{equation}
Inserting, this gives two integral equations for $\phi_{1,2}$
\begin{equation}
\begin{split}
e^{\I \frac{k}{2} x}\phi_1(x,t,k)&=1-g  \int_{-\infty}^{x} \rmd x' \int_{-\infty}^{x'} \rmd x'' \, P(x',t) Q(x'',t)\phi_1(x'',t,k)e^{\I k x'-\I \frac{k}{2} x''}\\
e^{-\I \frac{k}{2} x}\phi_2(x,t,k)&=\int_{-\infty}^{x} \rmd x' \, e^{-\I k x'} Q(x',t)\left(1- g \int_{-\infty}^{x'} \rmd x'' \, P(x'',t)\phi_2(x'',t,k)e^{\I \frac{k}{2} x''}\right)
\end{split}
\end{equation}
Iteration of these equations gives a series representation for $b(k)$ as a sum of alternating products of terms $Q (Q P)^{n-1}$, $n \geq 1$, integrated over ordered sectors as
\bea \label{agaa} 
&& b(k) e^{-k^2 t} = \\
&& \sum_{n=0}^\infty (-g)^n \int_{x_{2n+1}<x_{2n}<\dots < x_1} \prod_{j=1}^{2n+1}  \rmd x_j e^{ \I k (\sum_{j=1}^n x_{2j}- \sum_{j=0}^n x_{2j+1})}Q(x_1,t)P(x_2,t)\dots P(x_{2n},t)Q(x_{2n+1},t) \nonumber
\eea
as well as a a series representation for $a(k)$ as a sum of alternating products of terms $(Q P)^{n}$, $n \geq 0$, integrated over ordered sectors as
\begin{equation} \label{agaa2} 
a(k)= \sum_{n=0}^\infty (-g)^n \int_{x_{2n}<x_{2n}<\dots < x_1} \prod_{j=1}^{2n} \rmd x_j e^{\I k \sum_{j=1}^n (x_{2j+1}-x_{2j})}P(x_1,t) Q(x_2,t) \dots P(x_{2n},t)Q(x_{2n},t)
\end{equation}
These relations are valid for any $t$, and any boundary condition for the $\{ P,Q \} $ system (beyond its application to WNT).\\

Let us apply them to the WNT with the boundary conditions \eqref{init}. Setting $t=1$ in \eqref{agaa2} we see that $P(x,1)=\delta(x)$
implies that only the first two terms, $n=0$ and $n=1$, survive in the series. Indeed for $n \geq 2$ the $\delta$ function
implies that in the integral all odd $x_{2n+1}=0$ and the integration over the even $x_{2n}$
will be restricted to a vanishing small interval, leading to a vanishing result since $Q(x,1)$ is a smooth function. Hence,
from $P(x,1)=\delta(x)$ we obtain
\be 
a(k) = 1 - g \int_{x_2<x_1}  \rmd x_1  \rmd x_2 P(x_1,1) Q(x_2,1) e^{\I k (x_1-x_2) } = 1 - g \int_{x_2<0}  \rmd x_2 Q(x_2,1) e^{- \I k x_2 }
\ee 
which is precisely \eqref{eq:a}. For the droplet initial condition $Q(x,0)=\delta(x)$, setting $t=0$ in \eqref{agaa}, by the
same argument we see that only the first term $n=0$ remains, leading to 
\be 
b(k) = \int_\R \rmd x_1 e^{- \I k x_1} Q(x_1,0) = 1 
\ee 
Of course this result can be also obtained as in the previous section considering $\phi$ at $t=0$.  Although we will not present it here, by considering the equation for $\bar \phi$ one arrives at similar equations,
and similar series expansions for $\tilde a(k)$ (of the form $(Q P)^{n-1}$) and $\tilde b(k)$ (of the form $P (Q P)^{n-1}$). 
Specified at $t=1$ they reproduce the results of the previous
section, i.e. $\tilde b(k)= g e^{-k^2}$ and \eqref{eq:tildea} for $\tilde a(k)$. 
\\

{\it Expansion near the droplet IC}. Although complicated looking, formula \eqref{agaa} allows for an expansion around the droplet initial condition.
Let us define, as in the text, the Fourier transform $\hat b(x) = \int_\R \frac{\rmd k}{2 \pi} b(k) e^{i k x}$. Consider \eqref{agaa} at $t=0$. It implies
\be 
\hat b(x) = Q_0(x) + \int_{x_3<x_2<x_1} \delta(x+x_2-x_1-x_3)  Q_0(x_1) P(x_2,0) Q_0(x_3) + \dots 
\ee 
Suppose now that $Q_0(x)=\delta(x) + \epsilon f(x)$ where $\epsilon$ is small. To first order in $\epsilon$ one finds that
all terms with more integrations than in the second term vanish, hence we have formally
\be 
\hat b(x) = Q_0(x) + \epsilon \int \rmd x_2  \left( \Theta(-x) \Theta(-x_2) + \Theta(x) \Theta(x_2) \right) f(x+x_2) P^{\rm drop}(x_2,0) + \mathcal{O}(\epsilon^2)
\ee 
where the superscript "drop" denotes the value for the droplet initial condition. Note that
$P^{\rm drop}(x_2,0)=Q^{\rm drop}(x_2,1)= A_1(|x_2|)$, hence to this order we obtain an
explicit formula in terms of known quantities. If we choose $f(x)$ even we further obtain the (even) result
\be 
\begin{split}
\hat b(x) &= \delta(x) + \epsilon \left( f(x) +  \int_0^{+\infty} \rmd y f(x+y) A^{\rm drop}_1(y) \right) + \mathcal{O}(\epsilon^2)\\
&=
\delta(x) + \epsilon \left( f(x) + \int_{-\infty}^0 \rmd z  A^{\rm drop}_0(z) \tilde f(z;x) \right) + \mathcal{O}(\epsilon^2)
\end{split}
\ee 
where $\tilde f(z;x)=\int_0^{+\infty} \rmd y f(x+y) p(z-y)$, $p(z)=\frac{1}{\sqrt{4 \pi}} e^{-z^2}$ and $A^{\rm drop}_t(x)$ is given in \eqref{ABdroplet}.

\subsection{From a non-linear integral equation to the Hopf-Ivanov linear equation}
\label{sec:Ivanov} 

Consider the non-linear integral equation \eqref{AQ} in the text. For the droplet initial condition one 
has $A_0(x)=B_1(x)$. This equation then becomes a closed equation for the function $B_1(x)$
\be \label{AQnew} 
B_1(x) = \delta(x) + g \Theta(-x) \int_0^{+\infty}  \rmd v
B_1(x+v) \int_{-\infty}^{+\infty}\rmd yp(v-y) B_1(y)
\ee 
Here we have defined $p(z):=\frac{e^{- z^2/4}}{\sqrt{4 \pi}}$, although the present considerations 
extend to any continuous symmetric function $p(z)=p(-z)$ normalized to unity. Now consider the linear Hopf-Ivanov (HI) equation \eqref{Blin}, which reads
\be \label{Blin2} 
B_1(x) = \delta(x) + g\Theta(-x)  \int_{-\infty}^0\rmd yp(x-y) B_1(y)
\ee 

Let us separate, in both cases, the delta function part and write 
\be
B_1(x) = \delta(x) + \Theta(-x) \tilde B_1(x) 
\ee 
where $\tilde B_1(x)$ is smooth. It is defined only for $x \leq 0$, hence all equations below
are understood for $x \in \mathbb R^-$. We then obtain from \eqref{AQnew} the non-linear integral equation to be solved for $\tilde B_1(x)$ 
\be
\! \!  \tilde B_1(x) = g  p(x) + 
g \int_{-\infty}^0\rmd y p(x+y) \tilde B_1(y) 
+ g \int_0^{-x}  \rmd v  p(v) \tilde B_1(x+v)  +
g  \int_0^{-x}  \rmd v \tilde B_1(x+v)  
\int_{-\infty}^{0} dy  p(v-y) \tilde B_1(y) \label{AQ2}
\ee 
while Hopf-Ivanov's equation \eqref{Blin2} leads instead to
\be \label{Alin2} 
\tilde B_1(x) = g  p(x)
+ g \int_{-\infty}^0\rmd yp(x-y) \tilde B_1(y)
\ee 
Note that we will use the symmetry $p(-z)=p(z)$ in several places. To show that the solutions of \eqref{Alin2} are also solutions of \eqref{AQ2} we rewrite these two equations as
recursion relations by expanding $\tilde B_1(x)$ in series of $g$
\be 
\tilde B_1(x) = \sum_{n=1}^{+\infty} g^n B_{1,n}(x) 
\ee 
We already note that for both equations 
\be \label{eq:11} 
B_{1,1}(x)=p(x) 
\ee

The recursion relation for the $B_{1,n}$ from \eqref{AQ2} is, 
for $n \geq 2$ 
\bea \label{a1}
&& B_{1,n}(x) = \int_{-\infty}^0\rmd yp(x+y) B_{1,n-1}(y) + 
\int_{x}^0\rmd yp(x-y) B_{1,n-1}(y) \\
&& +
\sum_{m=1}^{n-2} \int_{x}^0 dy_1 \int_{-\infty}^0 dy_2 p(y_1-y_2-x) B_{1,m}(y_1) B_{1,n-1-m}(y_2)  
\eea
where we recall $x \leq 0$, while Hopf-Ivanov's recursion is
\be \label{a2}
B_{1,n}(x) = \int_{-\infty}^0\rmd yp(x-y) B_{1,n-1}(y) 
\ee 
We can already check the agreement of the two recursions to order $2$, indeed, from \eqref{a1} we have
\be
B_{1,2}(x) = \int_{-\infty}^0\rmd yp(x+y) p(y) + 
\int_{x}^{0}\rmd yp(x-y) p(y) = \int_{-\infty}^{-x}\rmd yp(x+y) p(y) = \int_{-\infty}^{0}\rmd yp(y) p(y-x)
\ee
which is exactly the result for $B_{1,2}(x)$ from \eqref{a2}, using \eqref{eq:11}. Note that this agreement is independent on precise shape of $p(z)$,
provided $p(-z)=p(z)$.\\

To show the coincidence to all orders, it is convenient to recall the random walk interpretation of the Hopf-Ivanov equation,
as mentionned in the text. Consider $X(j) \in \mathbb{R}$ a discrete time random walk, 
$X(j+1)=X(j)+z_j$, with $z_j$ i.i.d. with a symmetric jump probability $p(z)$. The recursion relation 
\eqref{a2} simply expresses that $B_{1,n}(x)$ is the probability that 
the walk starting at $X(0)=0$ arrives at $X(n)=x$ in $n \ge 1$ steps, while remaining negative,
$\{ X(j) \leq 0\}_{j=0,\dots,n}$.  Let us now rewrite the recurrence \eqref{a1} as a sum of three terms, $B_{1,n}(x) = T_1 + T_2 + T_3$, with
\bea \label{a1n}
&& T_1= \int_{-\infty}^0\rmd yp(x+y) B_{1,n-1}(y) = \int_{-\infty}^x\rmd yp(y) B_{1,n-1}(y-x)\\
&& T_2= \int_{x}^0 \rmd y p(x-y) B_{1,n-1}(y) \\
&& T_3= \sum_{m=1}^{n-2} \int_{x}^0 \rmd y_1 \int_{-\infty}^0 \rmd y_2 p(y_1-y_2-x) B_{1,m}(y_1) B_{1,n-1-m}(y_2)
\eea
This recursion and the Hopf-Ivanov recursion \eqref{a2} become identical if one notices that $m$ in $T_3$ is actually 
the step just before the last time that the level $x$ is crossed by the walk, before reaching $x$ at its end. 
Indeed then one can decompose the walk in three parts, (i) before and up to $m$, (ii) the single step
between $m$ and $m+1$ which must then necessarily cross level $x$, hence it must be
above $x$ at $m$ and below $x$ at $m+1$. And (iii) the third part is then by reflection exactly the same
as going from zero (the endpoints being shifted by $x$) to $y-x$ with $y<x$). This is further explained in
Fig.~\ref{Fig_Ivanov}. 


\begin{figure}[t!]
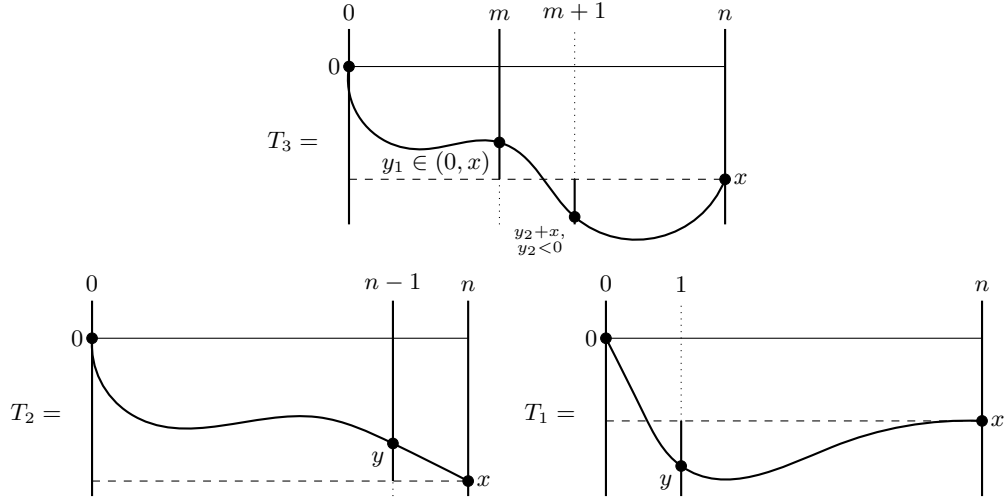

    \centering
    \TThree \\
    \TTwo
    \TOne
    \caption{Illustration of the three terms $T_1, T_2, T_3$ which enter into the recurrence \eqref{a1n} for $B_{1,n}(x) = T_1 + T_2 + T_3$.
    We recall that $B_{1,n}(x)$ is the probability that a random walk of $n$ steps starting at $0$ arrives at $x<0$ and
has all points negative. One defines $m=0,\dots,n-1$ the step such that the level $x$ is crossed
{\it for the last time} between $m$ and $m+1$. Then the term $T_3$ is the generic term, the
term $T_2$ corresponds to $m=n-1$ and the term $T_1$ corresponds to $m=0$.} 
    \label{Fig_Ivanov}
\end{figure}

\subsection{Conserved charges}
\label{sec:conserved} 

The generating function of the conserved charges is $\Gamma(x,t,k)$. It is defined from 
a particular solution of the Lax pair $\partial_x \vec v= U_1 \vec v$, $\partial_t \vec v= U_2 \vec v$ 
with $U_1,U_2$ given in \eqref{eq:LaxPairU}. More precisely it is defined as 
$\Gamma= \frac{v_2}{v_1}= \frac{\phi_2}{\phi_1}$, 
where $\Vec{v}=(v_1,v_2)^\intercal$ is the solution of the Lax pair of the form 
$\Vec{v}=e^{ k^2 t/2}\phi$ where ${\phi}=(\phi_1,\phi_2)^\intercal$ is as defined in the text, i.e. such that 
at $x \to -\infty$, $\phi \simeq (e^{-\I k x/2},0)^\intercal$ and that it behaves as \eqref{eq:plusinfinity} 
for $x \to +\infty$. The $\partial_x$ equation gives
\be \label{eq:Lax2} 
 \partial_x v_1 = - \I \frac{k}{2} v_1 - g  P v_2 \quad , \quad  \partial_x v_2 = \I \frac{k}{2} v_2 + Q v_1 
\ee 
which implies that $\Gamma$ obeys a Ricatti equation
\be \label{eq:RicattiGamma} 
\partial_x \Gamma = \I k \Gamma +  Q + g  P \Gamma^2 
\ee

From \eqref{eq:LaxPairU} using $\Vec{v}=e^{ k^2 t/2}{\phi}$ we see that $\log \phi_1$ satisfies
\be  \label{eq:logphi1} 
\partial_x \log \phi_1 = - \I \frac{k}{2} - g  P \Gamma \quad , \quad \partial_t \log \phi_1   = {\sf A} - \frac{k^2}{2} + {\sf B} \Gamma
\ee 
where the second equation comes from the first component of the $\partial_t$ equation of the Lax pair, i.e. $\partial_t (e^{ k^2 t/2} {\phi}) = U_2 e^{ k^2 t/2} {\phi}$
where $U_2$ is given in \eqref{eq:LaxPairU} and ${\sf A}= k^2/2- g P Q$, ${\sf B}=g (\partial_x - \I k) P$. Taking the cross derivatives in 
\eqref{eq:logphi1} we obtain the conservation equation 
\bea  \label{gasp} 
&& \partial_t  (- g  P \Gamma) = \partial_x ({\sf A} - \frac{k^2}{2} + {\sf B} \Gamma)
\eea 
To obtain the conserved charges and the associated currents one expands $\Gamma$ in a Laurent series $\Gamma = \sum_{n \geq 1} \frac{\Gamma_n(x,t)}{(\I k)^n}$
as explained in the text. The conserved quantities $C_n$ are then 
\be \label{eq:cons} 
C_n = \int_\R  \rmd x \,  \tilde C_n  \quad , \quad \tilde C_n = - g P \Gamma_n   \quad , \quad \partial_t \tilde  C_n  = \partial_x J_n
\ee 
where the associated currents are obtained from ${\sf A} - \frac{k^2}{2} + {\sf B} \Gamma = \sum_n \frac{J_n(x,t)}{(\I k)^n} $, leading to
\bea
J_n = g (\partial_x P) \Gamma_n - g  P \Gamma_{n+1}
\eea 

Now the $\Gamma_n$ can be obtained recursively as follows. From the Riccatti equation \eqref{eq:RicattiGamma} one obtains the following recurrence for $n \geq 1$
\be
\Gamma_{n+1}  = \partial_x \Gamma_n - g P \sum_{p=1}^{n-1} \Gamma_p \Gamma_{n-p}  \quad , \quad 
\Gamma_1 = - Q
\ee

Explicit calculation then gives the conserved quantities 
\bea \label{eq:conscharges} 
&& \Gamma_1 = -  Q \quad , \quad C_1= g \int_\R  \rmd x P Q \quad , \quad  \Gamma_2 = -  \p_x Q  \quad ,\quad C_2 = g \int_\R  \rmd x P \p_x Q \\
&& 
C_3= \int_\R  \rmd x (g P \p^2_x Q + g^2 P^2 Q^2   )
\quad , \quad 
C_4 = \int_\R  \rmd x g P \left(g Q^2 \p_x P+4 g P Q \p_x Q+ \partial_x^3 Q \right) \nonumber 
\eea 
and the associated currents 
\be \label{eq:currents} 
\begin{split}
& J_1= g (P \p_x Q- \p_x P Q) \quad , \quad J_2 = g (P\p^2_x Q - \p_x P \p_x Q) + g^2 P^2 Q^2 \, \\
& J_3 = g (P \p^3_x Q - \p_x P \p^2_x Q) + 4 g^2 P^2 Q \p_x Q 
\end{split}
\ee
Note that the conservation equations $\frac{d}{ \rmd t} C_n=0$ holds if the current $J_n$ vanishes at infinity, which is the case here,
since we assume that the functions $P,Q$ both vanish at infinity. \\

Finally, there is a relation between the scattering amplitude $a(k)$ and the {\it values} taken by the conserved quantities.
Indeed, integrating the equation \eqref{eq:logphi1} for $\partial_x \log \phi_1$ from $x=-\infty$ to $x=+\infty$ gives
\be
\log a(k) = [ \log (\phi_1 e^{\I \frac{k}{2} x} ) ]_{-\infty}^{+\infty} =
- g \int_\R  \rmd x P(x,t) \Gamma(x,t,k) = \sum_{n \geq 1} \frac{C_n}{(\I k)^n}
\ee 
where we have used that $\phi_1 \simeq e^{-\I k x/2}$ as $x \to -\infty$,
and \eqref{eq:plusinfinity} as $x \to -\infty$. Thus if one knows the solution for $a(k)$ one can 
expand in a Laurent series in $\I k$ to obtain the values of the $C_n$, as done in the text where they
are denoted $C_n(g)$. Alternatively the Laurent series of $a(k)$ can be reconstructed from the knowledge of 
the values taken by the conserved quantities.

\subsection{Calculation of $\Phi(H)$ from $C_3$}
\label{app:C3} 

Here we calculate the space time integral of the solution of the $P,Q$ system
\be
g^2 \int_0^1 \rmd t  \int_\R \rmd x \, P^2 Q^2 \equiv \Phi(H)
\ee
where the $\equiv$ means that it also defines $\Phi(H)$ if the system is parameterized by $H$,
which is equal to $\Phi(H_z)$ if the system is parameterized by $z=-g$, as in the text. We recall
that $H_z$ denotes the value of $H$ where the minimum in $\Psi(z)= \min_H (\Phi(H) + z e^H)$ is attained.
Here we show that for the droplet initial condition the resulting $\Phi(H)$ agrees with the Legendre transform of
$\Psi(z)$ obtained in \eqref{Psi}. 

We first note that the third conserved quantity can also be rewritten as
\be \label{eq:C3new} 
C_3 = \int_\R  \rmd x (g^2 P^2 Q^2 + g P \p^2_x Q) = \frac{g^2}{2} \int_\R  \rmd x  P^2 Q^2 + \frac{1}{2} \int_\R \rmd x  J_2 
\ee
where $J_2$ is the current given in \eqref{eq:currents}, where we used integration by part to replace
the term $- \int_\R \rmd x \p_x P \p_xQ$ by $\int_\R \rmd x P \p^2_xQ$. Now since $\frac{d}{ \rmd t} C_3=0$ we can integrate this equation
over $t \in [0,1]$ 
\be
C_3 = \frac{g^2}{2} \int_0^1  \rmd t \int_\R  \rmd x  P^2 Q^2 + \frac{1}{2} \int_0^1  \rmd t \int_\R  \rmd x  J_2 
\ee
The last integral can be transformed using an integration by part and the conservation law \eqref{eq:cons} 
\be
\int_0^1  \rmd t \int_\R  \rmd x J_2 = - \int_0^1  \rmd t \int_\R  \rmd x x \partial_x J_2 = - \int_0^1  \rmd t \int_\R  \rmd x x \partial_t \tilde C_2 
= g \int_\R \rmd x x  P Q' |_{t=0} 
\ee 
where we have used that $\tilde C_2= g P Q'$ from 
\eqref{eq:conscharges}, and that $\int_\R \rmd x x P Q' |_{t=1} = 0$ since $P(x,t=1)=\delta(x)$ 
(and $Q(x,t=1)$ is a smooth function). Performing an additional integration by part we obtain
\be 
\int_0^1  \rmd t \int_\R  \rmd x J_2 = - g \int_\R  \rmd x Q(x,0) ( x  P'(x,0) + P(x,0)) 
\ee 
Until now this is valid for any initial condition since we used only that $P(x,t=1)=\delta(x)$. 

Let us now specify to the droplet initial condition, i.e. $Q(x,0)=\delta(x)$. Since $P(x,0)$ is smooth we obtain
\be  \label{eq:J2C1} 
\int_0^1  \rmd t \int_\R  \rmd x J_2 = - g \int_\R  \rmd x Q(x,0) P(x,0) = - C_1 
\ee 

Putting together \eqref{eq:J2C1} and \eqref{eq:C3new} we obtain that for the droplet initial condition
\be 
g^2 \int_0^1 \rmd t  \int_\R \rmd x \, P^2 Q^2 = 2 C_3(g) + C_1(g)  
\ee 
In the text we have obtained
\be \label{C1C3new} 
C_3(g=-z)= \frac{{\rm Li}_{5/2}(-z)}{\sqrt{16 \pi}} = \frac{1}{2} \Psi(z) \quad , \quad 
C_1(g=-z)=\frac{1}{\sqrt{4\pi}}{\rm Li}_{3/2}(-z) = - z \Psi'(z) 
\ee
and we note that $- 2 g \partial_g C_3(g)=C_1(g)$.
Hence we now obtain, with $g=-z$
\be 
g^2 \int_0^1 \rmd t  \int_\R \rmd x \, P^2 Q^2 = \Psi(z) - z \Psi'(z) = \Phi(H=H_z) 
\ee 
The last equality is the parametric solution of the Legendre transform
$\Psi(z)= \min_H (\Phi(H) + z e^H)$, which is inverted by 
applying $d/dz$ leading to $\Psi'(z)=e^{H_z}$ where $H_z$
is the value of $H$ where the minimum is attained.

\subsection{Solutions of the WNT for general $H$, and solitonic solutions of the $\{P,Q \}$ system in the attractive case} 
\label{app:continuation} 

In the text, we described the solutions of the $\{P,Q \}$ system applied to the WNT, i.e. with boundary conditions 
\eqref{init}. It has the form \eqref{soluQP}, \eqref{kernels}, in terms of the two functions $A_t(x)$, $B_t(x)$.
In \eqref{2fonctions} we assumed that the integral over $k$ is on the real axis. This is correct for $H<H_c$,
and we recall (see Section \ref{sec:ratefunctions}) that as $H$ varies from $-\infty$ to $H_c$, the coupling $g$
varies from $-\infty$ to $g=1$. We recall below the expressions and properties of these functions $A_t(x)$, $B_t(x)$ in that regime. \\

For $H>H_c$ however the solution of the $\{P,Q \}$ system admits solitonic contributions. 
The existence of solitonic contributions is a known scenario in the inverse scattering method \cite{ZS}, which occurs when
$r(k)$, has a pole for ${\Im k}>0$, or $\tilde r(k)$, has a pole for ${\Im k}<0$. This corresponds in the
$(v_1,v_2)$ system to a discrete part in the spectrum which is interpreted as a bound state of the Dirac operator. In this case we have 
to return to the formula for the $A_t(x),B_t(x)$ functions \eqref{2fonctions} in the text.
This formula, where $k$ is on the real axis, can be violated in the attractive case $g>0$, as we find here for $H>H_c$. In this case, 
the functions $A_t(x),B_t(x)$ are the sum of two parts, one which is analogous to \eqref{2fonctions} 
with an integral over $k$ on the real axis, usually called the radiative part, and the other 
being a solitonic part (i.e. of finite-rank). We derive below these two parts. 

\subsubsection{Droplet initial condition}

Let us start with the droplet initial condition, and first recall the solution for $H<H_c$ obtained in the text.
In this case $Q(x,0)=Q_0(x)=\delta(x)$ and the symmetry $Q(x,t)=P(x,1-t)$ holds for $t \in [0,1]$. 
In addition, $P(x,t)$ and $Q(x,t)$ are both even functions. 
\\

{\bf Solution for $H<H_c$}. Let us first recall the expressions for the functions $A_t(x),B_t(x)$ for $H<H_c$
\be \label{ABdroplet} 
 A_t(x)=\int_\R \frac{\rmd k}{2\pi}e^{\I k x-k^2 t+\I \varphi(k)} \frac{1}{\sqrt{1-g e^{-k^2}}}   \quad , \quad  B_t(x)=\int_\R \frac{\rmd k}{2\pi}e^{-\I k x-k^2(1- t)-\I \varphi(k)} \frac{1}{\sqrt{1-g e^{-k^2}}}
\ee
where $g=g(H)$ is obtained by inverting $H=\log(\Psi_0'(-g))$, with $\Psi_0(z)$ given in \eqref{2branches}, see Fig.~\ref{fig:gh}, and
\be \label{soluphase2} 
\varphi(k)= \dashint_\R \frac{\rmd q}{2\pi} \, \frac{k \log(1-g e^{-q^2} )}{q^2-k^2}
\ee 
The function $\varphi(k)$ is odd and is plotted in Fig.~\ref{fig:varphik}. It decays as $1/k$ at large $k$, and expanding \eqref{soluphase2} in $k$ and $g$ it is easy to see that it 
admits the following series representation in $1/k$ at large $|k|$
\be \label{phiseries} 
\varphi(k)= \frac{1}{2 \pi} \sum_{m \geq 0} \frac{1}{k^{1+2 m}} \Gamma(\frac{1}{2} + m) {\rm Li}_{\frac{3}{2} + m}(g) 
\ee 
Using that $- \I \varphi(k) = \sum_{n \geq 1} \frac{C_n(g)}{(\I k)^n}$ we obtain the values of all the conserved charge $C_n$ defined in the text and in Section
\ref{sec:conserved}. One finds $C_n(g)=0$ for even $n$, and for odd $n=2 m + 1$ with $m \geq 0$
\be \label{conservedall} 
C_{2 m+1}(g) = (-1)^m \frac{1}{2 \pi} \Gamma(\frac{1}{2} + m) {\rm Li}_{\frac{3}{2} + m}(g) 
\ee 
which for $m=\{ 0,1\}$ is the result given in the text in \eqref{C1C3}. One can also obtains its small $k$ behavior which is
\be 
\varphi(k)= k p_1(g) + k |k| p_2(g) + \mathcal{O}(k^3) \quad , \quad p_1(g)=\int_2^{+\infty} \frac{du}{2 \pi u^{3/2}} \log \left(\frac{1-g e^{-u}}{1-g} \right)
\quad p_2(g)= \frac{g ({\rm arcsinh}(1)-\sqrt{2})}{\pi  (g-1)} 
\ee

\begin{figure}[t!]
    \centering
    \includegraphics[scale=0.53]{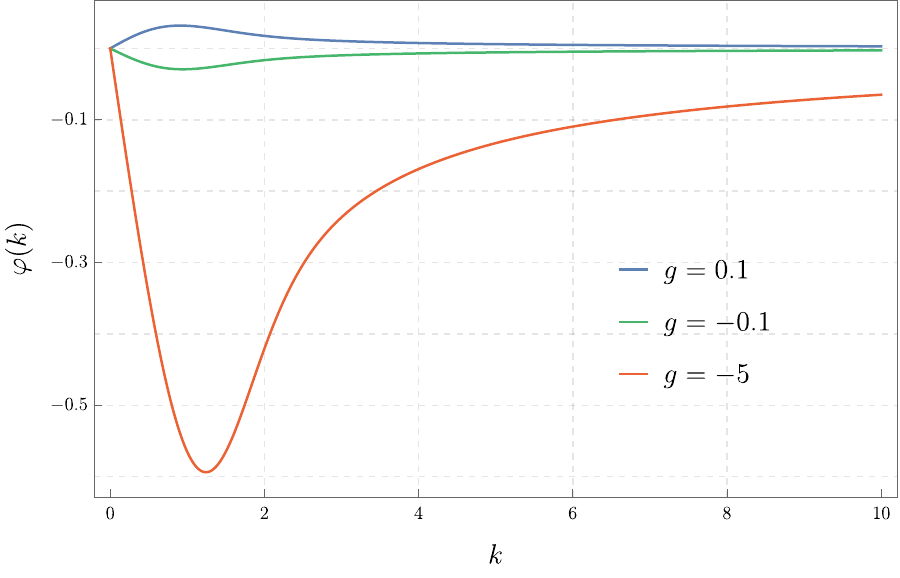}
    \caption{The phase $\varphi(k)$ defined in \eqref{soluphase2} plotted versus $k$ for various values of $g$}
    \label{fig:varphik}
\end{figure}

\begin{figure}[t!]
    \centering
    \includegraphics[scale=0.5]{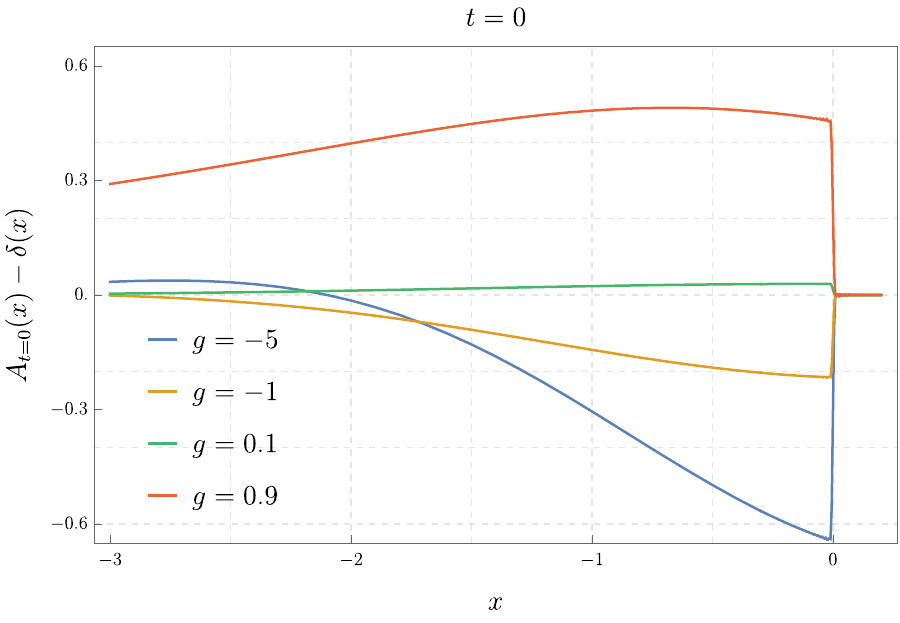}
    \includegraphics[scale=0.485]{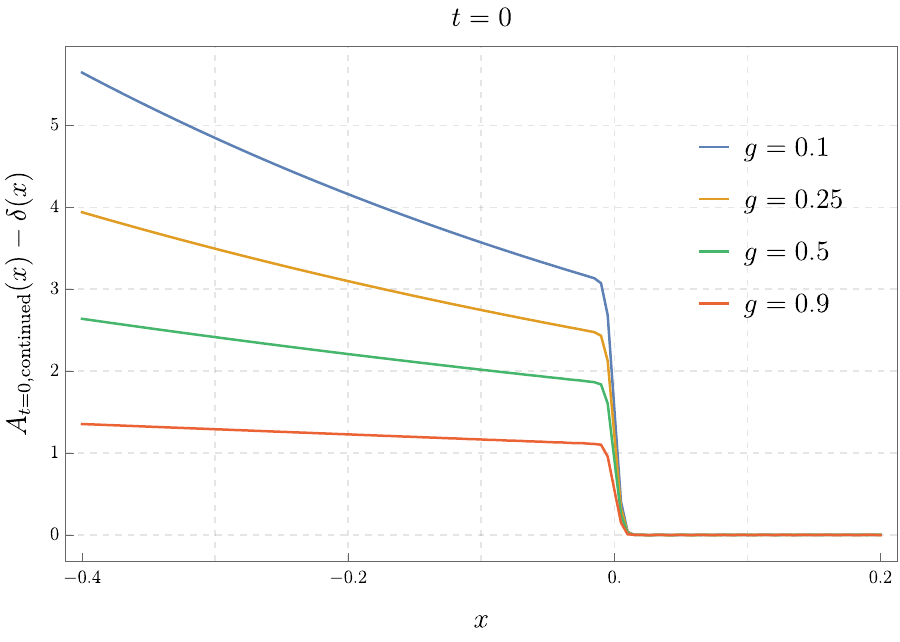}
    \caption{\textbf{Left:} Plot of the smooth part $A_0(x)-\delta(x)$ of $A_0(x)$ from \eqref{ABdroplet} plotted versus $x$ for various values of $g$. One sees that it vanishes for $x>0$.
    \textbf{Right:} the same plot for the second branch \eqref{eq:Acont} containing a solitonic contribution discussed below. The vanishing property at $x>0$ still holds for this branch.}
    \label{fig:A0}
\end{figure}

\begin{figure}[t!]
    \centering
    \includegraphics[scale=0.5]{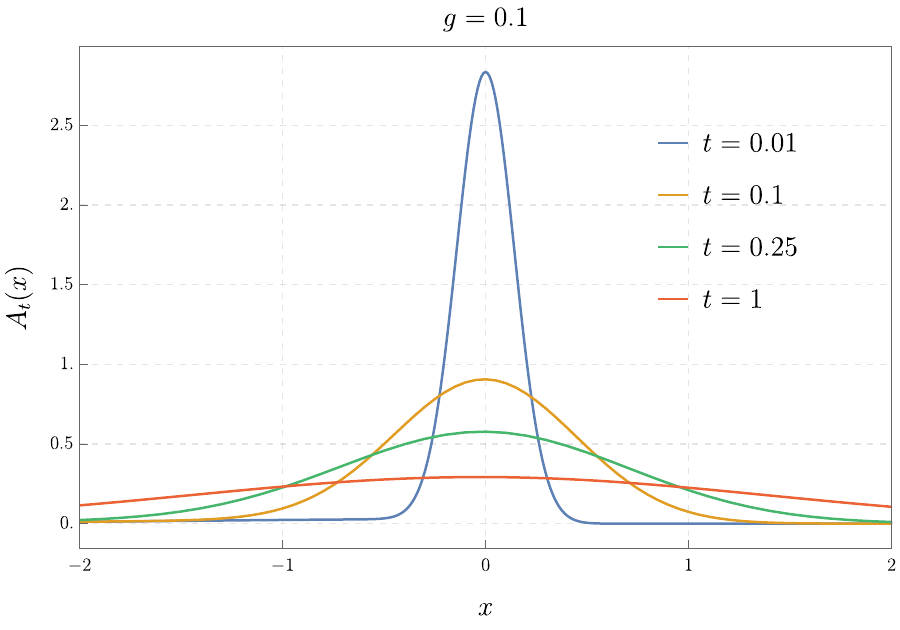}~~
\includegraphics[scale=0.5]{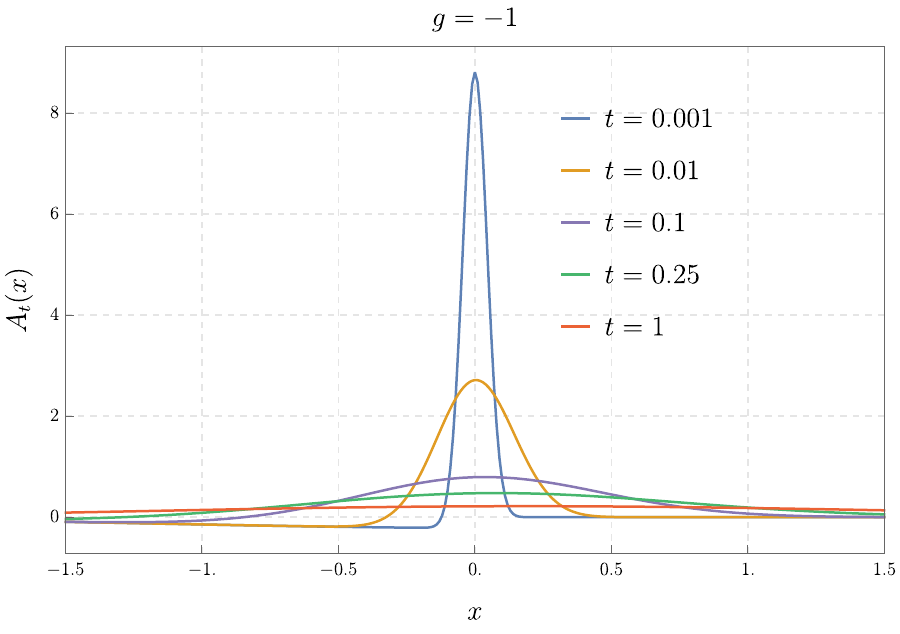}
    \caption{Plots of $A_t(x)$ from \eqref{ABdroplet} for various $t$ and for some values of $g$. We see the broadening at $t>0$ of the delta peak present at $t=0$.}
    \label{fig:At}
\end{figure}

The functions $A_t(x)$ and $B_t(x)$ in \eqref{ABdroplet} are forward and backward solutions of the standard heat equation. They are related through $A_t(x)=B_{1-t}(x)$, a 
property specific to the droplet initial condition. For $g=0$ they are simply equal to the heat kernel, i.e. $A_t(x)=\frac{1}{\sqrt{4 \pi t}} e^{- \frac{x^2}{4 t}}$,
but at $g \neq 0$ they are non trivial, and very asymmetric in $x$. 
As found in the text (see also Section \ref{sec:Ivanov}), $A_{t=0}(x)$ is the sum of a delta function and of a smooth part, i.e. $A_0(x)=\delta(x) + \Theta(-x)\tilde A_0(x)$.
Remarkably, the smooth part, which is $\mathcal{O}(g)$ at small $g$, vanishes for $x>0$ for all $g$. 
The same property holds for $B_{t=1}(x)$ at $t=1$. It is easy to see on the integral representation \eqref{ABdroplet} that there should be a delta function part: 
since $\varphi(k)$ vanishes at large $k$, the integrand at $t=0$ becomes simply $e^{\I k x}$ at large $|k|$ (i.e. the
reflection amplitude $r(k) \to 1$ at large $|k|$ in \eqref{2fonctions}). However it is non-trivial to see on the Fourier representations that these 
functions vanish for all $x>0$.  These functions are plotted
for various $t$ and $g$ in the Figures \ref{fig:A0} (left) and \ref{fig:At}. In Fig.~\ref{fig:A0} (left) the smooth part at $t=0$ has been plotted (i.e. the $\delta(x)$ piece
was substracted) and we see that indeed it takes the form $\Theta(-x)\tilde A_0(x)$, i.e. it exhibits a jumps at $x=0$ and it vanishes for $x>0$, a quite non trivial check. 
At $t>0$ as we can see in Fig.~\ref{fig:At}, the delta peak in $A_t(x)$ broadens diffusively.
\\

{\bf Solution for $H>H_c$}. We now construct the second branch of solutions for $0 < g \leq 1$, which corresponds to $H>H_c$.
Let us recall that for that branch the relation $g=g(H)$ is obtained by inverting $H=\log(\Psi'(-g))$, with $\Psi(z)=\Psi_0(z)+\Delta(z)$ given in \eqref{2branches}, 
see also \eqref{Legendre3} and Fig.~\ref{fig:gh}. We start again with the formula \eqref{IvanovSolu} for the 
Laplace transform $\hat B_1(s)= \int_{-\infty}^0  \rmd x e^{s x} B_1(x)$, which we now use to probe the structure of $r(k)$ in the
complex plane for $k$. It reads
\be  \label{IvanovSolu2} 
\hat B_1(s) = \exp( - f(s) ) \quad , \quad f(s) := \int_\R \frac{\rmd q}{2 \pi} \frac{s}{s^2+ q^2} \log(1 - g e^{-q^2}) 
\ee 
It is convenient to consider the following derivative of $f(s)$, obtained as
\begin{equation}
    g\p_g  f(s)=2\int_{0}^{+\infty} \frac{\rmd q}{2 \pi} \frac{s}{s^2+ q^2} \frac{-g e^{-q^2}}{1-g e^{-q^2}}
\end{equation}
Upon the change of variable $y=e^{q^2}$, with $\rmd y/y=2q \rmd q$, we can rewrite it as
\begin{equation}
    g\p_g  f(s)=-\int_{1}^\infty \frac{\rmd y}{2 \pi} \frac{s}{s^2+ q(y)^2} \frac{g}{y}\frac{1}{q(y)} \frac{1}{y-g}
\end{equation}
This integral has the form \eqref{delta0} with $z=-g$ which, according to \eqref{delta1}, leads to a jump $\Delta f(s)$ which obeys 
\begin{equation}
    g\p_g \Delta f(s)=\I \frac{s}{s^2 + q(g)^2}\frac{1}{q(g)}
\end{equation}
where $q(g)$ is the solution of $g= e^{q^2}$. Using that $g\p_g =1/(2 q) \p_q$, this is integrated into
\begin{equation}
    \Delta f(s)=2\I \arctan(\frac{q(g)}{s})
\end{equation}
where, to ensure continuity at the branching point $g=1$, we required that $\Delta f(s)\vert_{g=1}=0$, using that $q(g=1)=0$.
The new branch thus corresponds to the following continuation of the Laplace transform of $B_1$, which contains an additional rational factor
as compared the first branch.
\be \label{eq:Bcont} 
\hat B_1(s)   \to \hat B_1(s) e^{-2\I \arctan(\frac{q(g)}{s}) } = \hat B_1(s) \frac{s - \I q(g)}{s + \I q(g)} = \hat B_1(s) \frac{s + \kappa_0}{s - \kappa_0} 
\ee
where  $q(g)=  \I \kappa_0$, for $0<g \leq 1$, in terms of the real positive parameter $\kappa_0$, which satisfies 
$g= e^{-\kappa_0^2}$ and has the interpretation of a rapidity (see below). One goes from Laplace to Fourier by 
comparing the definition $\hat B_1(s)= \int_{-\infty}^0  \rmd x e^{s x} B_1(x) = \int_{\mathbb{R}}  \rmd x e^{s x} B_1(x)$
with the inverse Fourier transform of \eqref{2fonctions}, i.e. $\tilde r(-k) e^{-k^2}= \int_{\mathbb{R}}  \rmd x e^{- \I k  x} B_1(x)$, leading to
$r(k)=e^{k^2} \tilde r(-k)= \hat B_1(s)|_{s=- \I k + 0^+}$. To obtain the continuation of $\tilde r(k)$ in the complex plane we must thus change $k \to -k$, i.e. 
substitute $s= \I k + 0^+$ in \eqref{eq:Bcont}, leading to
\be \label{rk2} 
\tilde r(k) \to \tilde r(k) \frac{k-\I \kappa_0}{k+\I \kappa_0}
\ee
We see that this continuation has a pole at $k=- \I \kappa_0$ in the lower half plane. According to the general theory \cite{ZS} it
implies that the function $B_t(x)$ has two parts,
\begin{equation} \label{eq:Bcont2} 
B_t(x)=\int_\R \frac{\rmd k}{2\pi}e^{-\I k x-k^2(1- t)-\I \varphi(k)}\frac{k-\I \kappa_0}{k+\I \kappa_0}\frac{1}{\sqrt{1-g e^{-k^2}}}+2 \kappa_0 e^{-\kappa_0 x+\kappa_0^2 (1-t)-\I\varphi(-\I\kappa_0)}
, \quad g=e^{-\kappa_0^2} 
\end{equation}
To obtain the first term (radiative part) we have used \eqref{rk2} and the formula for the first branch of $\tilde r(k)$ given in the text and in \eqref{ABdroplet}.
The second term originates from the pole of $\tilde r(k)$ from \eqref{rk2} in the complex plane at $k=- \I \kappa_0$, and has the (rank-one) solitonic
form $e^{-\kappa_0 x +\kappa_0^2 (1-t)}$ described in Section \ref{app:finiterank}. To obtain its amplitude we can focus on $t=1$ in \eqref{ABdroplet} and
compute the residue 
\bea
&& - {\rm Res}_{k= - \I \kappa_0} \tilde r(k) e^{k^2} \frac{k-\I \kappa_0}{k+\I \kappa_0} 
= - {\rm Res}_{k= - \I \kappa_0}  \hat B_1(s = \I k + 0^+)  = 2  \kappa_0 
e^{- \I\varphi(-\I\kappa_0) } \\
&& \I\varphi(-\I\kappa_0) = f(s=\kappa_0) =  \int_\R \frac{\rmd q}{2 \pi} \frac{\kappa_0}{\kappa_0^2+ q^2} \log(1 - g e^{-q^2}) \label{resid} 
\eea

Note the extra minus sign due to the fact that the contour around $- \I \kappa_0$ must be clockwise. Note also 
that the factor $1/\sqrt{1- g e^{-k^2}}$ is only present for $k$ on the real axis, hence it is not present in
the residue of the pole at $k=- \I \kappa_0$. Similarly we obtain $r(k) \to r(k) \frac{k+\I \kappa_0}{k-\I \kappa_0}$ and
\begin{equation} \label{eq:Acont} 
A_t(x)=\int_\R \frac{\rmd k}{2\pi}e^{\I k x-k^2 t+\I \varphi(k)}\frac{k+\I \kappa_0}{k-\I \kappa_0}\frac{1}{\sqrt{1-g e^{-k^2}}}+2 \kappa_0 e^{-\kappa_0 x+\kappa_0^2 t+\I\varphi(\I\kappa_0)}
\end{equation}
with the relation $A_t(x)=B_{1-t}(x)$, which is consistent with the symmetry $Q(x,t)=P(x,1-t)$ of the solution for the droplet initial condition. 

The function $A_t(x)$ for the second branch of solutions, given by \eqref{eq:Acont},
is plotted in Fig.~\ref{fig:Acontinued}. On the left it is plotted for $g=0.95$ for various $t$. For finite $t$ one distinguishes the soliton shape,
but as $t\to 0$ and for $x>0$ it becomes again peaked around $x=0$. Eventually, as we have checked in Fig.~\ref{fig:A0} (right),
at $t=0$ it vanishes for $x>0$, which again is non trivial to see on the integral representation \eqref{eq:Acont}.
Again $A_0(x)$ is the sum of a delta function (with unit coefficient) and a continuous part, and we see that for
$x>0$, the positive part of the soliton exactly cancels with the radiative part.

In Fig.~\ref{fig:Acontinued} (right) the two branches are plotted for $t=0.25$ around the turning point at $g=1$. One sees
that the transition occurs smoothly. As $g \to 1$ along the first branch (i.e. $H \to H_c^-$) the function $A_t(x)$ decays more
slowly for $x \to -\infty$, and for $g=1$ is goes to a constant. This coincides with the generation of the soliton along the
second branch, the function $A_t(x)$ is now growing as $x \to -\infty$.\\

\begin{figure}[t!]
    \centering
    \includegraphics[scale=0.5]{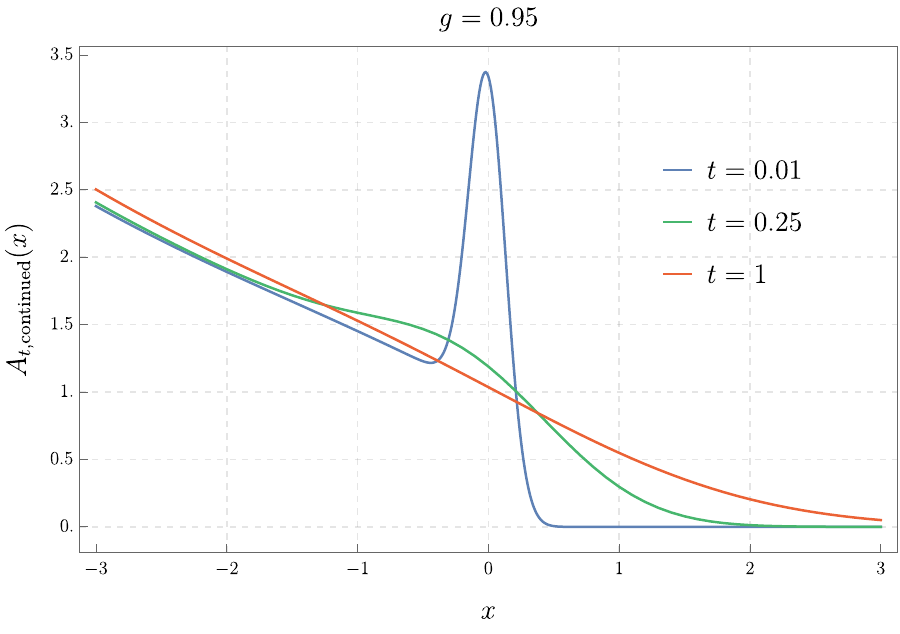}
    \includegraphics[scale=0.5]{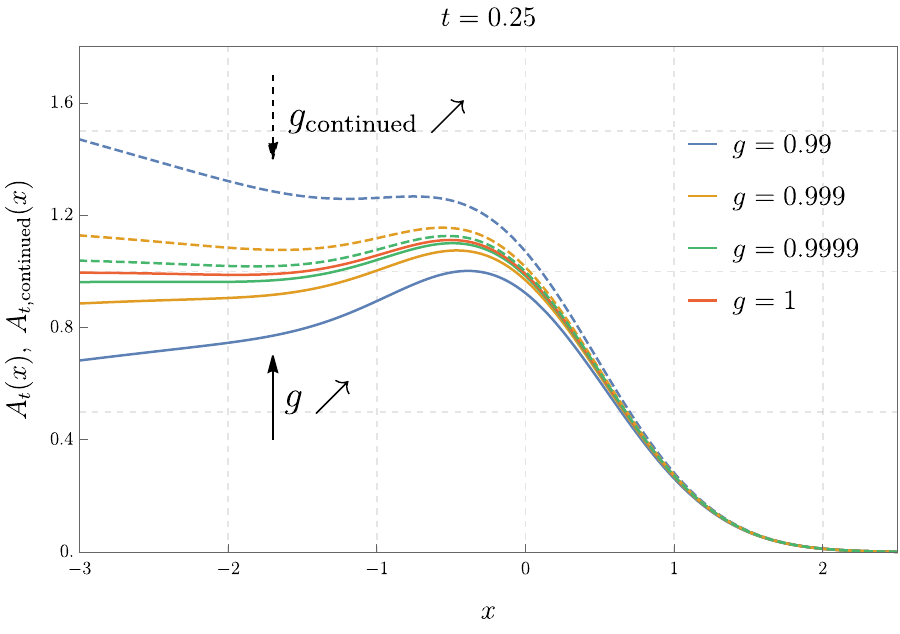}
    \caption{\textbf{Left:} the function $A_t(x)$ for various $t$ and $g=0.95$ ($H=0.03564$ for a critical height $H_c=-0.3052$).
    \textbf{Right:} Both branches of solutions for $A_t(x)$ around the turning point $g=1$ (for clarity the second branch is indicated as
    $A^{\rm continued}_t(x)$) and the related curves are dashed).}
    \label{fig:Acontinued}
\end{figure}

{\bf Conserved quantities.} Let us now evaluate the values of the conserved quantities associated with this new branch of solutions, which
can be written $C_n(g) + \Delta C_n(g)$, where $C_n(g)$ are the values along the first branch. As explained in the text and
in Section \ref{sec:conserved} they are obtained from Laurent expansion of $\log a(k)$ at large $k$. For the droplet initial condition, one has $b(k)=1$, $r(k)=1/a(k)$,
$\log a(k) = -  \I \varphi(k) + \frac{1}{2} \log(1 - g e^{- k^2})$. From the above, the continued version is
obtained by replacing $\varphi(k) \to \varphi(k) + \Delta \varphi(k)$ with 
\begin{equation} \label{Deltaphi} 
    \Delta \varphi(k)= 2 \,  \arctan(\frac{\kappa_0}{k}) 
\end{equation}
where we recall $g=e^{-\kappa_0^2}$. The values of the additional contribution $\Delta C_n(g)$ to the conserved quantities $C_n$ given in
\eqref{eq:conscharges} are obtained 
from the Laurent expansion \eqref{Deltaphi} of $\Delta \varphi(k)$ at large $k$, since $- \I \Delta \varphi(k) = \sum_{n \geq 1} \frac{\Delta C_n(g)}{(\I k)^n}$.
It reads, for general odd integer $n$, and for $n=1,3$
\be 
\Delta C_n(g)=\frac{2}{n}\kappa_0^n  \quad , \quad \Delta C_1(g)=2\kappa_0 \quad , \quad  \Delta C_3(g)=\frac{2}{3}\kappa_0^3 \quad , \quad \kappa_0 = ( \log \frac{1}{g} )^{1/2}
\ee 
Note that $\Delta C_n(g)=0$ for even integer $n$. Since from \eqref{C1C3new} one has $C_3(g)= \frac{1}{2} \Psi(-g)$, this result is consistent with the known
result for the jump in the second branch of $\Psi(z)$, see Eq.~\eqref{2branches} 
\be
 \Delta(z) = 2 \Delta C_3(g) =  \frac{4}{3}\kappa_0^3 = \frac{4}{3}(\log (\frac{-1}{z}))^{3/2} 
\ee
Note the relation $- 2 g \partial_g \Delta C_3(g) = \frac{1}{\kappa_0} \partial_{\kappa_0} \Delta C_3 = 2 \kappa_0 = \Delta C_1$.
\\

\begin{figure}
    \centering
    \includegraphics[scale=0.55]{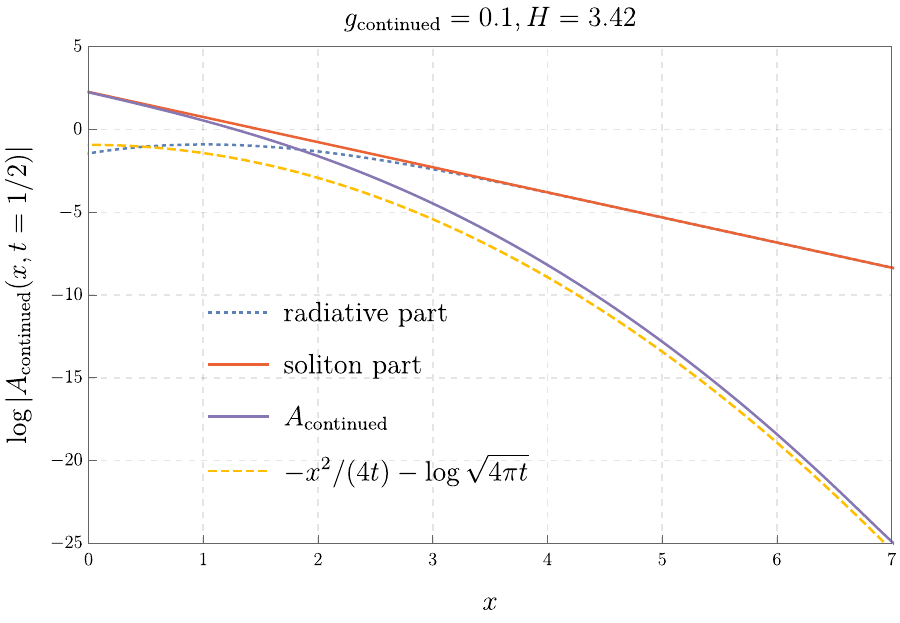}
    \caption{Plot of the logarithm of the absolute values of the radiative part, of the soliton part, and of the sum of the two terms $A_t(x)$ in \eqref{eq:Acont} for $t=1/2$ and 
    $g=0.1$. We see that although the soliton
    exponential decay dominates at small $x$, at large $x$ the two parts cancel and the remainder (the sum) is a fast decaying Gaussian, well fitted by $-x^2/(4 t)$ in log scale.}
    \label{fig:decayA}
\end{figure}

{\bf Limit $H\to +\infty$}

We now study the limit $H\to +\infty$ of our exact solution for the WNT for the droplet initial condition, and show that one recovers the approximate solution
obtained in Ref.~\cite{MeersonParabola} in that limit.  This limit correspond to the second branch of solutions with $g$ vanishing as $g \simeq 2 H^{1/2} e^{-H}$. Indeed, for $H \in [H_c,+\infty)$
the relation between $g$ and $H$ is determined by $\Psi'(z)=e^H$ with $\Psi(z)= - \frac{1}{\sqrt{4 \pi}} {\rm Li}_{5/2}(z) + \frac{4}{3} (\log(\frac{-1}{z}))^{3/2}$,
from Ref.~\cite{le2016exact}, as recalled in \eqref{2branches}.
Setting $z=-g$, this leads to $e^H \simeq \frac{2}{g} (\log(\frac{1}{g}))^{1/2}$ as $g \to 0^+$. \\

In that limit the solitonic part in \eqref{eq:Acont} plays a dominant role. The rapidity parameter $\kappa_0$ is given by $\kappa_0^2 = \log(\frac{1}{g} )
\simeq H - \frac{1}{2} \log(4 H)$, hence $\kappa_0 \to +\infty$. To approach the problem, let us first write the solution of the $\{P,Q\}$ system if one neglects the radiative part in \eqref{eq:Acont}
(for any $\kappa_0$). 
This soliton only solution corresponds to the rank-one soliton described in Section~\ref{app:finiterank}, more precisely \eqref{rank1soliton},
with $\kappa=\mu=\kappa_0$, $\tilde q=2 \kappa_0 e^{ \I \phi(\I \kappa_0)} $ and $\tilde p=2 \kappa_0 e^{\kappa_0^2} e^{ - \I \phi(- \I \kappa_0)}$.
Using \eqref{Q1} we obtain this soliton-only solution, which we denote $P_s(x,t),Q_s(x,t)$, as
\be  \label{solitononly} 
 Q_s(x,t) =  \frac{2 \kappa_0 e^{ \I \phi(\I \kappa_0)} e^{- \kappa_0 x + \kappa_0^2 t} }{1 + g e^{\kappa_0^2} e^{ 2 \I \phi(\I \kappa_0)} 
e^{- 2 \kappa_0 x  } } 
= \frac{\kappa_0 e^{\kappa_0^2 t}}{\cosh(\kappa_0 (x - y_0) ) } \quad , \quad 
g P_s(x,t) Q_s(x,t) =  \frac{\kappa_0^2}{\cosh^2( \kappa_0 (x-y_0))} 
\ee 
where we have used that $g= e^{-\kappa_0^2}$, and where $\kappa_0 y_0=\frac{1}{2} \log ( g e^{\kappa_0^2} e^{ 2 \I \phi(\I \kappa_0)} ) = \I \phi(\I \kappa_0) $,
a quantity which is real and given in \eqref{resid} (recalling that $\varphi(k)$ is an odd function). 
Until now this is valid for any $\kappa_0$, i.e. $P_s(x,t),Q_s(x,t)$ are exact solutions of the $\{P,Q\}$ system, which however
do not obey the boundary data conditions for WNT \eqref{init}. These are ensured only if the radiative part is added. In the
limit $\kappa_0 \simeq H^{1/2} \to +\infty$, and in a certain range of $x$ and $t$, the radiative part is comparatively smaller
and $P_s(x,t),Q_s(x,t)$ are very good
approximations of the exact solution. \\

This can be compared with the solution in Ref.~\cite{MeersonParabola} in the "boundary layer", for $\rho(x,t) = \rho_{\rm bl}(x)$
given in Eq.~(50) there, and ${\sf h}(x,t)= {\sf h}_{\rm bl}(x,t)$ given there in Eq.~(53). Using our equation \eqref{PQ2} and 
the dictionary \eqref{dictionary}, we see that $\rho_{\rm bl}(x)$ is identical with $- 4 g P_s(x,t) Q_s(x,t)$
if one identifies $c=2 \kappa_0^2$.  This identification is correct since $c = - {\sf H}$ (below (53) there)
and $-{\sf H}=2 H \simeq 2 \kappa_0^2$ from the dictionary \eqref{dictionary}. Note that we have neglected the 
shift $y_0$ in \eqref{solitononly} since $\kappa_0 y_0 = \I \phi(\I \kappa_0) \to 0$ in the limit $\kappa_0 \to +\infty$
from \eqref{resid}. We also find that ${\sf h}_{\rm bl}(x,t) = 2 \log \cosh(\sqrt{c/2} x) - c t$ identifies (up to a constant
term $-2 \log \kappa_0$) with 
$-2 h_s(x,t)$ where
$h_s(x,t)= \log Q_s(x,t)$ and $Q_s(x,t)$ given in \eqref{solitononly}.\\

We have checked numerically that the exact solution for large positive $H$ is indeed very close to the above soliton-only
solution for $x = \mathcal{O}(1/\kappa_0)$, i.e. in the boundary layer of Ref.~\cite{MeersonParabola}. In that region it is
time independent except very near $t=0,1$. This can be seen in Fig.~\ref{fig:solitonlargeH} already for a moderately
large value of $H$. In addition, there is a second region, for $x$ outside of this boundary layer
where the decay of $Q(x,t)$ is Gaussian $h(x,t) \sim - x^2/(4 t)$, in agreement with the arguments in Ref.~\cite{MeersonParabola},
see Eq.~(57) there (with $L=0$ there). The way it arises here is also non trivial and as follows. We recall that $Q(x,t) \simeq A_t(x)$ 
as $x \to +\infty$, from \eqref{soluQP} since in $A_x(t)$ and $B_x(t)$ vanish in that limit. We note that the decay at
large $x$ of the soliton part in \eqref{eq:Acont} is exponential, i.e. much slower than Gaussian. However, what happens is that at large positive $x$ the first term in \eqref{eq:Acont}
(the radiative term) cancels the soliton, and what remains is indeed decaying as a Gaussian in $x$. This is illustrated in 
Fig.~\ref{fig:decayA}. 


\subsubsection{General initial condition}

Let us now consider a general initial condition $Q_0(x)$, which we take even in $x$ for simplicity. As we have seen in the text, 
the only difference with the droplet case is that the amplitude $b(k)$ is now a non-trivial even function of $k$. Some of its properties are obtained in Sections \ref{sec:scatt1} and \ref{sec:scattgeneral}. One has now $r(k)= b(k) e^{k^2} \tilde r(-k)$, and now 
$A_t(x)=(\hat b * B_{1-t})(x)$ where $\hat b(x)$ denote the Fourier transform of $b(k)$ and $*$ the convolution. 
\\

{\bf Principal branch}. As shown in
the text, one obtains 
\be \label{ABGeneral} 
 A_t(x)=\int_\R \frac{\rmd k}{2\pi}e^{\I k x-k^2 t+\I \varphi(k)} \frac{b(k)}{\sqrt{1-g b(k) e^{-k^2}}}   \quad , \quad  
 B_t(x)=\int_\R \frac{\rmd k}{2\pi}e^{-\I k x-k^2(1- t)-\I \varphi(k)} \frac{1}{\sqrt{1-g b(k) e^{-k^2}}}
\ee
in agreement with the general relation $A_t(x) = (\hat b * B_{1-t})(x)$, where now the phase reads 
\be 
\varphi(k)= \dashint_\R \frac{\rmd q}{2\pi} \, \frac{k }{q^2-k^2}\log(1-g b(q) e^{-q^2} )
\ee 
We stress that both $b(k)$ and $\varphi(k)$ depend on $g$, and the solution for the droplet initial condition
is recovered setting $b(k)=1$. The associated conserved charges are obtained as before from the
Laurent series
$-\I \varphi(k)=\sum_{n \geq 1} \frac{C_n(g)}{(\I k)^n}$ which leads to 
\be 
C_{2m +1}(g) = (-1)^{m-1} \int_\R\frac{\rmd q}{2 \pi} q^{2m} \log(1-g b(q) e^{-q^2} )
\ee 
which, we can check, recovers \eqref{conservedall} for $b(q)=1$. This leads to 
$C_1(g) = \int_\R \frac{\rmd q}{2 \pi} {\rm Li}_1(g b(q) e^{-q^2})$ since ${\rm Li}_1(y)=- \log(1-y)$. 
From the relation $C_1(g) = g \int_\R\rmd x P Q|_{t=1}=g Q(0,1)=g e^H= - z \Psi'(z)$, with $g=-z$, one obtains
\be
- z \Psi'(z) = \int_\R \frac{\rmd q}{2 \pi} {\rm Li}_1(- z b(q) e^{-q^2}) 
\ee
If $b(q)$ is $z$ independent one can integrate this relation using $z \partial_z {\rm Li}_n = {\rm Li}_{n-1}$,
which leads to forms similar to the one displayed in \cite[Table 7.1]{krajenbrink2019beyond}.
However, this is not the case in general.

Until now this is the principal branch. Since ${\rm Li}_1(y)$ has a branch cut on the real axis for $y>1$, the 
discussion of the possible other branches depends on the behavior of the function $g b(q) e^{-q^2}$ as a function of $q$
(recalling that $b(q)$ is an even function). Let us denote $M(g)= \max_{q \in \mathbb{R}} g b(q) e^{-q^2}$. 
The main branch terminates at $g=g_c$ such that $M(g_c)=1$. 
\\

{\bf Second branch}. We perform the same analysis as for the droplet case. The Laplace transform $\hat B_1(s) = e^{-f(s)}$ with now
$f(s) := \int_\R \frac{\rmd q}{2 \pi} \frac{s}{s^2+ q^2} \log(1 - g b(q) e^{-q^2})$. We will assume that $b(q) e^{-q^2}$ is a decreasing function of $q^2$.
In that case the branching point is at $g_c=1/b(0)$. Taking the derivative $g \partial_g$ it is then possible to perform the
change of variable $y=e^{q^2}/b(q)$, with $\rmd y/y=(2q  - \frac{b'(q)}{b(q)}) \rmd q$, leading to
\begin{equation}
    g\p_g  f(s)=- 2 \int_{1}^\infty \frac{\rmd y}{2 \pi} \frac{s}{s^2+ q(y)^2} \frac{g}{y}\frac{1}{(2q(y)  - \frac{b'(q(y))}{b(q(y))})} \frac{1}{y-g} (1 + \frac{g \partial_g b(q(y))}{b(q(y))}) 
\end{equation}
Since the integral has the form \eqref{delta0} we can apply \eqref{delta1} to obtain $g\p_g  \Delta f(s)$ and, upon 
integration using that $g\p_g =(1 + g \frac{\partial_g b(q(g))}{b(q(g))})/(2 q - \frac{b'(q(g))}{b(q(g))}) \p_q$ we obtain the jump 
\begin{equation}
    \Delta f(s)=2\I \arctan(\frac{q(g)}{s})
\end{equation}
where $q(g)$ is the solution of $g b(q) e^{-q^2}=1$. Similar manipulations as in the droplet case lead to 
\begin{equation}
B_t(x)=\int_\R \frac{\rmd k}{2\pi}e^{-\I k x-k^2(1- t)-\I \varphi(k)}\frac{k-\I \kappa_0}{k+\I \kappa_0}\frac{1}{\sqrt{1-g b(k) e^{-k^2}}}+2 \kappa_0 e^{-\kappa_0 x+\kappa_0^2 (1-t)-\I\varphi(-\I\kappa_0)}
\end{equation}
and
\begin{equation}
A_t(x)=\int_\R \frac{\rmd k}{2\pi}e^{\I k x-k^2 t+\I \varphi(k)}\frac{k+\I \kappa_0}{k-\I \kappa_0}\frac{b(k)}{\sqrt{1-g b(k) e^{-k^2}}}+2 \kappa_0 b(\I \kappa_0) e^{-\kappa_0 x+\kappa_0^2 t+\I\varphi(\I\kappa_0)}
\end{equation}
where $\kappa_0$ is now the solution of the equation
\begin{equation}
    g=\frac{e^{-\kappa_0^2}}{b(\I \kappa_0)}
\end{equation}
This is again in agreement with the general relation $A_t(x) = (\hat b * B_{1-t})(x)$. Since one still has  $\Delta \varphi(k)= 2 \,  \arctan(\frac{\kappa_0}{k})$ 
the shift in the values of the conserved quantities are again $\Delta C_n(g)=\frac{2}{n}\kappa_0^n$, i.e.
$\Delta C_1(g)=2\kappa_0$ and $\Delta C_3(g)=\frac{2}{3}\kappa_0^3$.\\

Note that if $b(q) e^{-q^2}$ is not a decreasing function of $q^2$ the analysis of the various branches of solutions may be more complicated.

\subsection{Numerical calculation of $P,Q,h$ and Fredholm inversions}
\label{sec:numerical} 

In this Section we evaluate numerically the solutions $P(x,t)$ and $Q(x,t)$ of the $\{P,Q\}$ system from the operator inversion formula \eqref{soluQP} in the text. 
In the first part we explain the numerical algorithm to handle formula \eqref{soluQP} for given functions $A_t(x)$, $B_t(x)$. In the second part we
apply it to the WNT boundary data \eqref{init} in the case of the droplet initial condition. In that case 
the operators ${\cal A}_{xt}$ and ${\cal B}_{xt}$ and their associated functions $A_t(x)$, $B_t(x)$ are known 
from the Section \ref{app:continuation}. We plot the optimal height field $h(x,t)=-{\sf h}(x,t)/2$
and the optimal noise field $2 \tilde h(x,t)= 2 g P(x,t) Q(x,t) = - \rho(x,t)/2$ for various values of $g$ and $H$.

\subsubsection{Algorithm for the operator inversion and the numerical evaluation of the solutions of the $\{ P,Q \}$ system}

All operators considered in this work act on $\mathbb{L}^2(\R_+)$. From a numerical analysis point of view, we will approximate these operators as acting on a discrete weighted space defined by a prescribed quadrature similarly to Bornemann's method to evaluate Fredholm determinants of operators acting on continuous space \cite{bornemann2010numerical}. Concretely, since all functions considered have an exponential decay towards $+\infty$, any non-trivial integral over $\R_+$ will be approximated as
\begin{equation}
    \int_{\R_+}\rmd x \, f(x)\simeq \sum_{j=1}^m w_j f(y_j)
\end{equation}
where the set $\{ w_j, y_j \}_{j=1}^m$ defines a quadrature rule over the interval $[0,M]$ with $M\gg 1$, see \cite{bornemann2010numerical} for more details. Recall that the solution $Q(x,t)$ reads
\begin{equation}
    Q(x,t)=\bra{ \delta } {\cal A}_{xt} \frac{I}{I + g {\cal B}_{xt} {\cal A}_{xt}} \ket{\delta}
\end{equation}
we see that two types of integrals are considered. Either we multiply two operators together, e.g. ${\cal B}_{xt}{\cal A}_{xt}$, or we multiply an operator with a bra or a ket, e.g. $\bra{\delta} {\cal A}_{xt}$.\\

\begin{itemize}
    \item The products of type $\bra{\delta} {\cal A}_{xt}$ are considered as trivial since 
\begin{equation}
    (\bra{\delta} {\cal A}_{xt})(v)=\int_{\R_+}\rmd u \delta(u)A_t(x+u+v)=A_t(x+v)
\end{equation}
we only have to enforce the left variable of $\mathcal{A}_{xt}$ to be equal to zero.
\item Operator products are considered as non-trivial and the quadratures will be inserted there as
\begin{equation}
\begin{split}
    ({\cal B}_{xt}{\cal A}_{xt})(u,v)&=\int_{\R_+} \rmd r \, B_t(x+u+r)A_t(x+r+v)\\
    &\simeq \sum_{j=1}^m w_j B_t(x+u+y_j)A_t(x+y_j+v)\\
    &\simeq \sum_{j=1}^m ( B_t(x+u+y_j)\sqrt{w_j})(\sqrt{w_j}A_t(x+y_j+v))
    \end{split}
    \end{equation}
    where we split the weights to keep the operators symmetric.
    
\item In a nutshell, if an operator is not involved in a product with a bra or a ket, its approximation will read e.g. 
    \begin{equation}
    \mathcal{A}_{xt}=\left(\sqrt{w_i}A_t(x+y_i+y_j)\sqrt{w_j}\right)_{i,j=1}^m := \sqrt{w}A_{xt}\sqrt{w}
\end{equation}
and similarly for $\mathcal{B}_{xt}$. We can interpret $w$ as being an $m\times m$ diagonal matrix with entries  $w_i$ and $A_{xt}$ as being a symmetric $m\times m$ matrix with elements $A_t(x+y_i+y_j)$. If an operator is involved in a product with a bra or a ket, its approximation will read e.g.
\begin{equation}
    \bra{\delta} \mathcal{A}_{xt}=\left(A_t(x+y_j)\sqrt{w_j}\right)_{j=1}^m
\end{equation}
\end{itemize}
 Lifting these results from the operators $\{ \mathcal{A}_{xt},\mathcal{B}_{xt}\}$ to the solutions of the $\{ P,Q\}$ system is now simple. Rewriting $Q(x,t)$ as
\begin{equation}
\begin{split}
Q(x,t)&=\bra{ \delta } {\cal A}_{xt}  \ket{\delta}-g \bra{ \delta } {\cal A}_{xt} \frac{I}{I + g {\cal B}_{xt} {\cal A}_{xt}} {\cal B}_{xt} {\cal A}_{xt}\ket{\delta}
\end{split}
\label{eq:Q_reexpression}
\end{equation}
we approximate it as
\begin{equation}
Q(x,t)\simeq  \dbtilde{Q}(x,t)=A_t(x)-g\bra{ \delta } {A}_{xt}\sqrt{w} \frac{I}{I_m + g \sqrt{w}{ B}_{xt} w{ A}_{xt}\sqrt{w}} \sqrt{w}{ B}_{xt} w{ A}_{xt}\ket{\delta}
\end{equation}
With this approximation, the inverse $(I + g {\cal B}_{xt} {\cal A}_{xt})^{-1}$ has been transformed into the inverse of an $m\times m$ dimensional matrix. In Eq.~\eqref{eq:Q_reexpression}, we completed the numerator to avoid the evaluation of the inverse against the ket $\ket{\delta}$. For the purpose of the numerical evaluations of this work, the typical values $m=20$ and $M=10$ are sufficient to obtain solutions of the $\{ P,Q\}$ system up to $\sim 10^{-5}$ precision. We present below the Mathematica implementation of this method used to produce the numerical results of this work. We first present the quadrature implementation for the kernel $K(v,v')=\int_{\R_+}\rmd r B_t(x+v+r)A_t(x+v'+r)$ and then the one for the solution $Q(x,t)$.

\newpage

\begin{lstlisting}[extendedchars=true,language=Mathematica]

Needs["NumericalDifferentialEquationAnalysis`"]
M = 10;
m = 20;

(* we assume the functions A and B to be already defined, 
x is the space, t is the time and g is the coupling constant *)

A[x_,t_,g_]:=A[x,t,g]; 
B[x_,t_,g_]:=B[x,t,g];

K[x_, y_, t_, g_] := Module[{r, w},
  {r, w} = Transpose[GaussianQuadratureWeights[m, 0, M]];
  Flatten[Outer[B, x + r, {t}, {g}] w ].Flatten[Outer[A, y + r, {t}, {g}]]
  ];

Q[x_, t_, g_] := Module[{u, w},
    {u, w} = Transpose[GaussianQuadratureWeights[m, 0, M]];
    w = Sqrt[w];
    A[x, t, g] - g Flatten[ Outer[A, u + x, {t}, {g}] w].
        Inverse[IdentityMatrix[m] + g  Outer[Times, w, w] ArrayFlatten[Outer[K, u + x, u + x, {t}, {g}]]].
        Flatten[w Outer[K, u + x, {x}, {t}, {g}]]
    ];
\end{lstlisting}

\subsubsection{Results for the droplet initial condition} 

We have computed the functions $P(x,t)$ and $Q(x,t)$ using the above numerical scheme for the functions 
$A_t(x)$, $B_t(x)$ defined in Eq.~\eqref{ABdroplet} for the main branch $H<H_c$, and in 
\eqref{eq:Bcont2} and \eqref{eq:Acont} for the second branch $H>H_c$.\\

A first numerical check of our precision is to verify that the functions $P,Q$ are even in $x$, which is quite a non-trivial check given that the formula
a very asymmetric in $x \to -x$. We have verified this fact with the same numerical precision as mentioned in the
previous section.\\

In Fig.~\ref{fig:optimalheight} we have plotted the optimal height $h(x,t)=\log Q(x,t)$ for various values of $g,H$ and time $t$. The black dot on the figure
is the value of $h(0,1)=H$ reached at the observation time $t=1$. Far from the origin the behavior is the parabola $-x^2/(4 t)$, but an important deformation
arises at smaller $x$. \\

\newpage
\begin{figure}[ht!]
    \centering
    \includegraphics[scale=0.5]{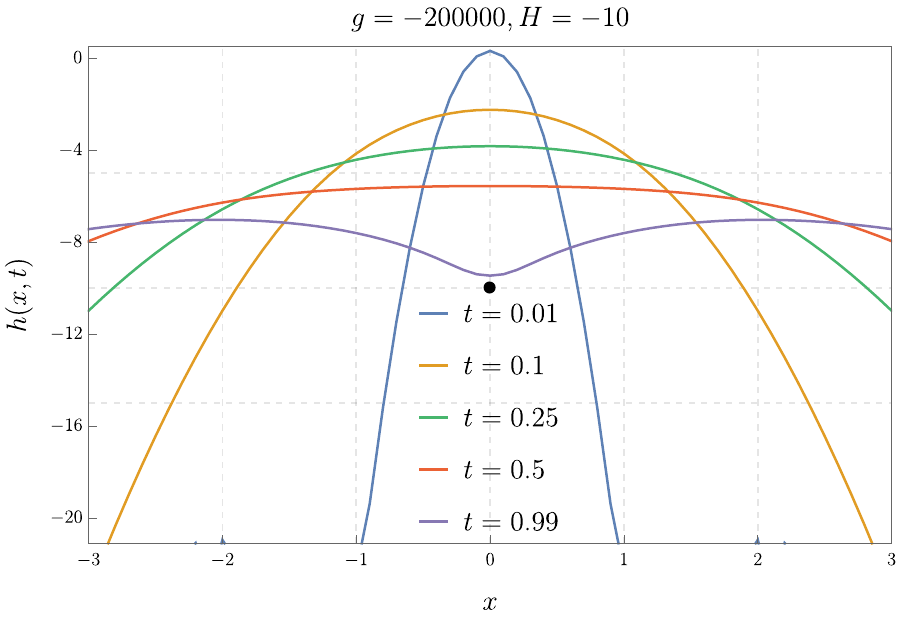}
    \includegraphics[scale=0.5]{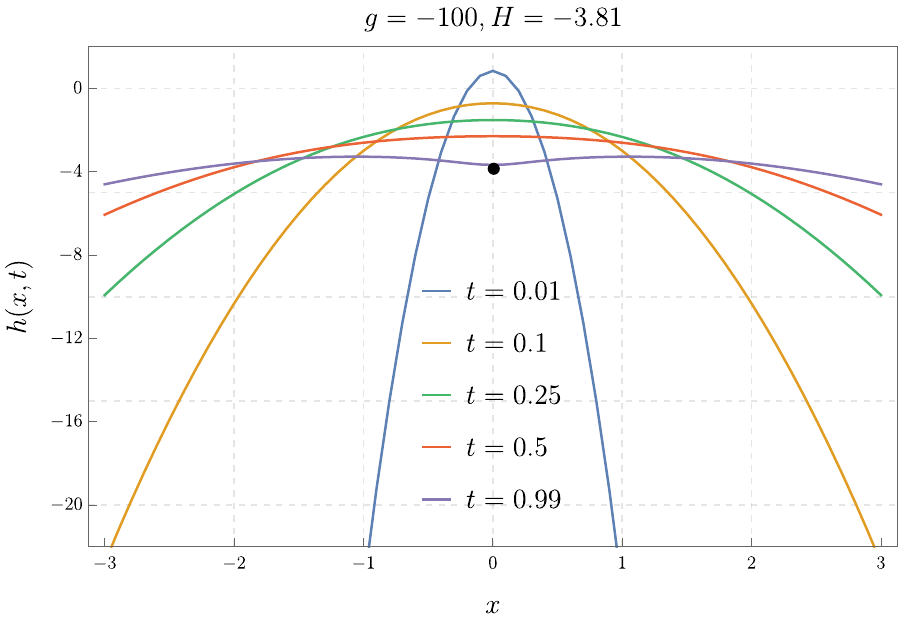}
    \includegraphics[scale=0.5]{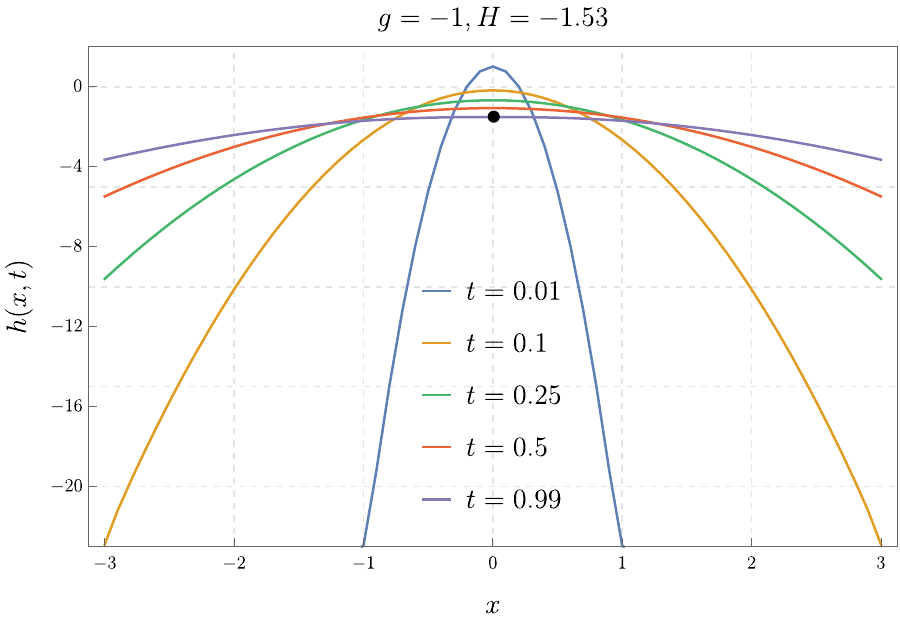}
    \includegraphics[scale=0.5]{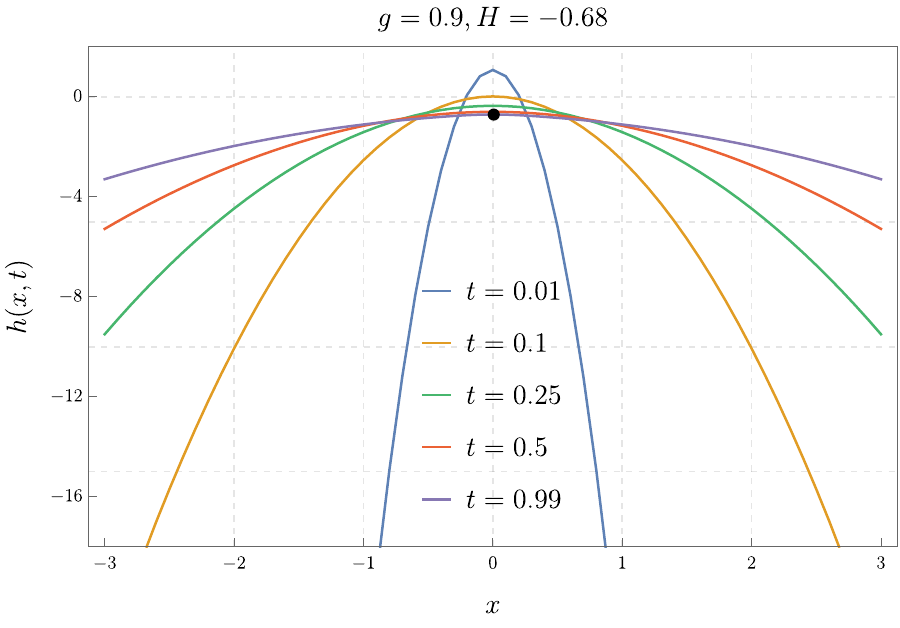}
    \includegraphics[scale=0.5]{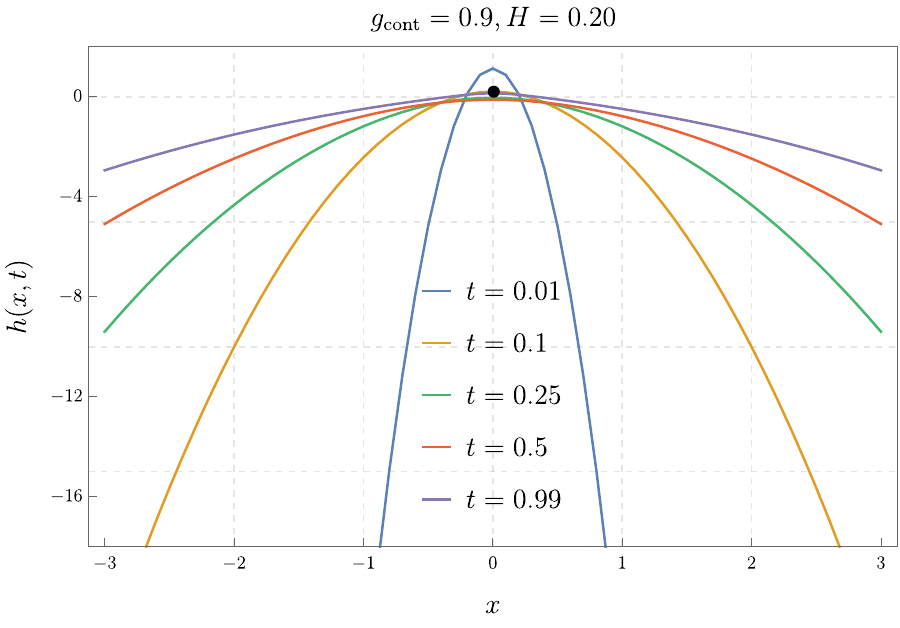}
    \includegraphics[scale=0.5]{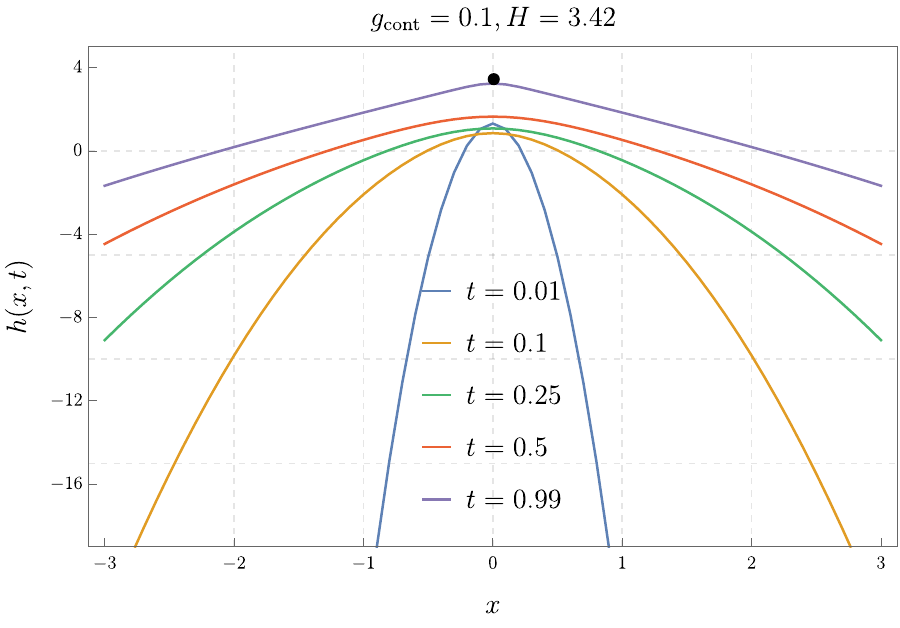}
    \caption{Optimal height $h(x,t)$ evaluated at various times and various final height at the origin $H$. We observe that at short time the height evolves from a parabolic profile to a non-trivial profile at final time $t=1$ which behavior depends on whether $H\gg1$ or $H\ll 1$.}
    \label{fig:optimalheight}
\end{figure}
\newpage



\begin{figure}[h!]
    \centering
    \includegraphics[scale=0.55]{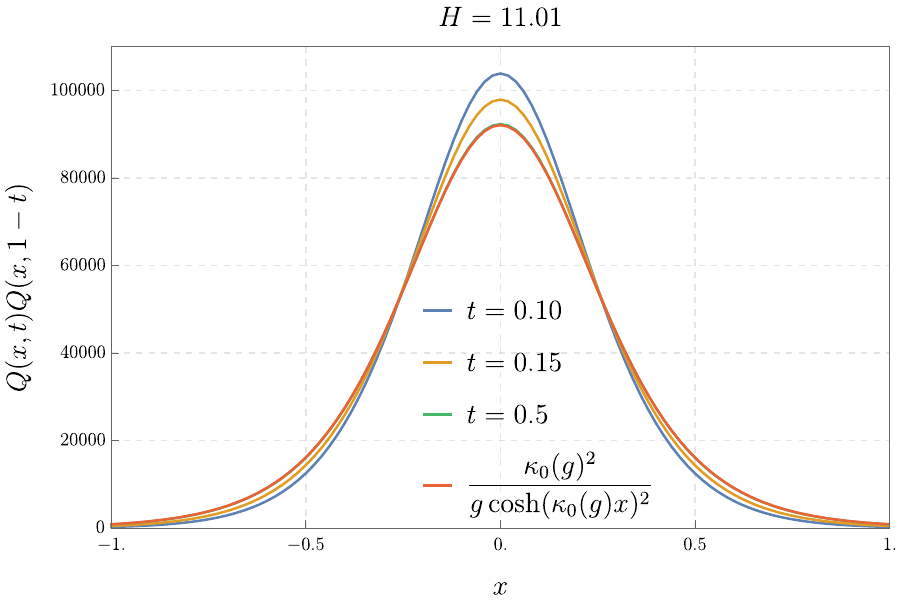}
    \caption{Optimal noise profile $\tilde h(x,t)/g=Q(x,t) Q(x,1-t)$ as a function of $x$ for $H=11.01$ and various times $t$. A weak time dependence can be seen.
    For $t=1/2$ (in green) it already overlaps almost perfectly with the "soliton only" (time independent) solution, valid in the boundary layer $x \sim 1/\kappa_0$
    for $H \to +\infty$ (in red)}
    \label{fig:solitonlargeH}
\end{figure}

In Fig.~\ref{fig:solitonlargeH} we have plotted the optimal noise. The optimal noise can be identified by comparing the right hand side of 
the saddle point equation \eqref{wnth1} with the KPZ equation itself \eqref{KPZ2}, leading to $\sqrt{2} T^{1/4} \tilde \eta_{\rm opt}(x,t) = 2 \tilde h(x,t)$.
For convenience we denote $2 \tilde h(x,t)$ at the saddle point as the optimal noise, i.e. the field $- \rho(x,t)/2$ in
\cite{MeersonParabola}. We would like to point out that for the droplet initial
condition it is directly related to the optimal height by the relation
\be 
2 \tilde h(x,t) = 2 g P(x,t) Q(x,t) = 2 g Q(x,t) Q(x,1-t) = 2 g e^{h(x,t) + h(x,1-t)} 
\ee 
In the Fig.~\ref{fig:solitonlargeH} we have plotted $Q(x,t) Q(x,1-t)$ for various $t$ and for $g=0.0001$ which corresponds to $H=11.01$.
Since it is a moderately large positive value for $H$, we compare it with the soliton only solution given in \eqref{solitononly}, i.e. 
$Q_s(x,t) Q_s(x,1-t) = \frac{\kappa_0^2}{g \cosh(\kappa_0 x)^2}$, where $g=e^{-\kappa_0^2}$. This soliton-only solution is time independent
and is expected to hold for $H \to +\infty$ inside the boundary layer $x \sim 1/\kappa_0$ except very near $t=0$ or $t=1$ \cite{MeersonParabola}.
For the moderately large value of $H$ considered here, we still some small time dependence, while for $t=1/2$ we already see excellent agreement
with the soliton-only solution.

\begin{figure}[h!]
    \centering
    \includegraphics[scale=0.55]{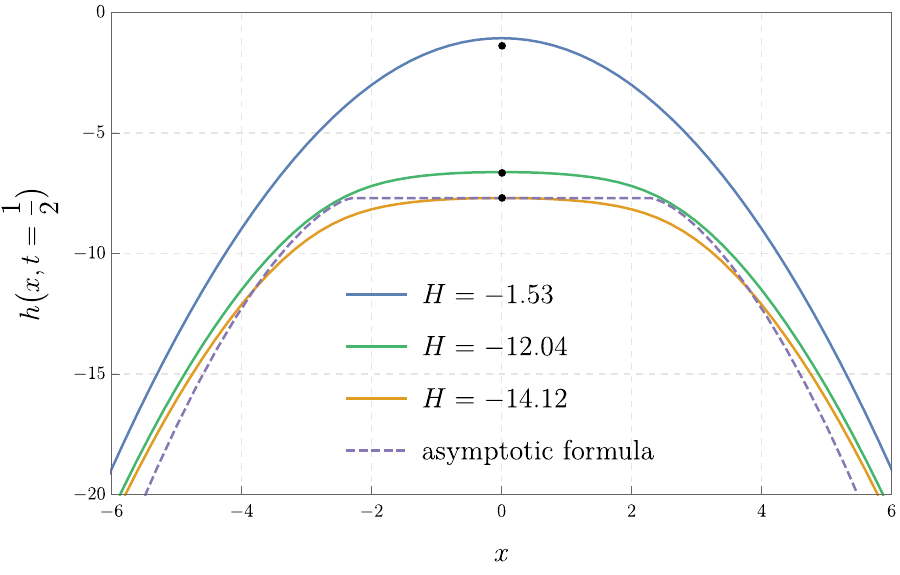}
    \caption{Optimal height $h(x,t=1/2)$ as a function of $x$ for three negative values of the field (indicated by the black dots), i.e. in the left tail regime. 
    As $H$ decreases we can see a plateau forming near the origin. The dashed line is the parametric function $x(u)=\frac{2}{\pi}\sqrt{\vert H-\hat{H}_0 \vert }(1+u\; \arctan u)$, $h(u)=\frac{H+\hat{H}_0}{2}[1+\frac{2}{\pi} (u+(u^2-1)\arctan u]$ for $u\geq 0$, 
  based on the scaling form predicted to hold in the limit $H \to - \infty$ in Ref. \cite{hartmann2019optimal}}
    \label{fig:plateaulargeH}
\end{figure}

In the Fig.~\ref{fig:plateaulargeH} we have plotted $h(x,t=1/2)$ for three different negative values for $H$ (left tail region, $g<0$).
The back dot indicates the value of $h(0,1)=H$ reached at the observation time $t=1$. The prediction from \cite{hartmann2019optimal}
is that for $H \to -\infty$ the optimal height profile develops at $t=1/2$ a plateau of height $h(x,t=1/2)= \frac{H + \hat H_0}{2}$
in a band $|x| < \frac{2}{\pi} \sqrt{\vert H-\hat{H}_0 \vert }$. Furthermore, away from this band the height reaches the parabolic form $-x^2/(4 t)$.
As we can see for the most negative value of $H$, $H=-14.12$, one clearly sees on Fig.~\ref{fig:plateaulargeH} the plateau developing. 
We have also plotted for comparison the prediction of Ref. \cite{hartmann2019optimal} for the crossover function 
between the plateau regime and the parabola regime. For this value of $H$ the asymptotic regime is not yet reached,
however its main features are visible. \\

\begin{figure}[h!]
    \centering
    \includegraphics[scale=0.55]{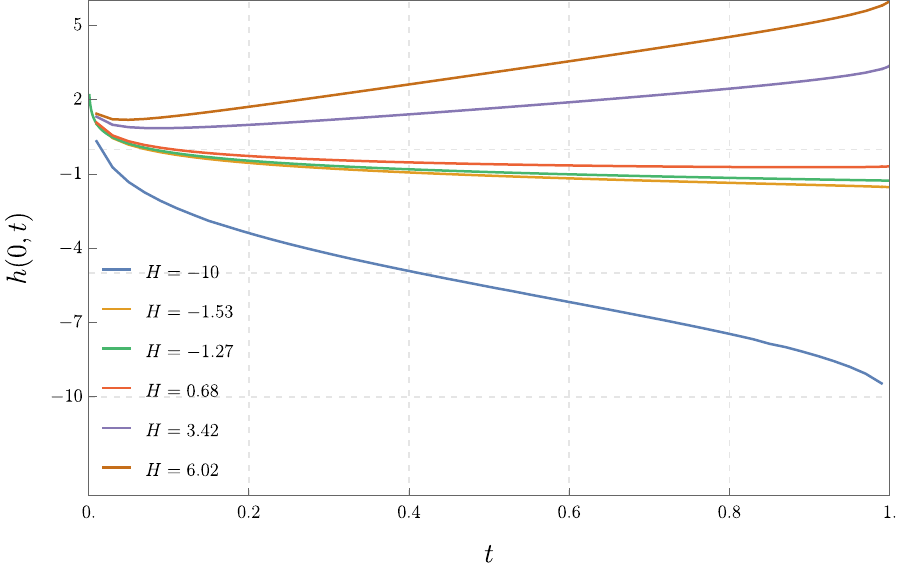}
    \caption{Optimal height at the origin, $h(0,t)$, as a function of $t$ for various values of $H$. The green curve corresponds to the typical height, i.e.
    $H=\hat H_0=-1.27$, see text.}
    \label{fig:h(0,t)}
\end{figure}

Finally in Fig.~\ref{fig:h(0,t)} we have plotted the optimal height at the origin, $h(0,t)$, as a function of $t$ for various values of $H$. 
This shows how the value $h(0,1)=H$ is reached for various $H$ as the time increases. The green curve is the typical height 
$h(0,t)= - \log(\sqrt{4 \pi t})$ and corresponds to $g=0$ and $H=\hat H_0=- \log(\sqrt{4 \pi })=-1.27$. All the curves are asymptotic to that one in the very small
time regime, $t \to 0$
(not shown in details).

\end{widetext}


\begin{thebibliography}{100}
\bibitem{footnote100} This is equivalent to use everywhere the following units of space, time and heights
$ y^*=(2 \nu)^3/(D \lambda_0^2)$,  $\tau^*=2(2 \nu)^5/(D^2 \lambda_0^4)$ and $h^*=\frac{2 \nu}{\lambda_0}$.
\bibitem{footnoteNLS} If $\Psi(x,\tau)$ solves NLS then $Q=\Psi(x,\tau)|_{\tau \to -\I t}$ and
$P=\Psi^*(x,\tau)|_{\tau \to -\I t}$ solves the $\{ P,Q\}$ system.
\bibitem{footnoteP} The equation for $P$ should thus be solved backward in time.
\bibitem{footnoteWronskien} The Wronskian $W=\phi_1 \bar{\phi_2}-\phi_2 \bar{\phi_1}$ is
space and time independent since
$\p_x W=\Tr(U_1) W=0$ and $\p_t W=\Tr(U_2) W=0$. It is
$W=-1$ at $x=-\infty$ and $W=-a(k) \tilde{a}(k)-b(k)\tilde{b}(k)$
from \eqref{eq:plusinfinity}
at $x=+\infty$.
\bibitem{footnotereal} For the present application to WNT, these functions are real.
\bibitem{footnoteProlhac} Note that in Ref.~\cite{ProlhacKrajenbrink} the Sparre-Andersen theorem also appears in the short time 
large deviation expansion, using a different route (based on the exact droplet solution at all times).
\bibitem{footnoteeven} Note that all even conserved quantities $C_{2p}=0$ by parity of the initial condition $Q_0(x)=\delta(x)$.
\bibitem{footnotePsi} From \eqref{Legendre} one has $\Psi'(z)=e^H$ and in particular
$\Psi'(0)=e^{\hat H_0}$ 
where $\hat H_0$ is the most probable value of $H$, i.e. 
such that $\Phi'(\hat H_0)=0$. Since one has, to $\mathcal{O}(z)$, 
$\Psi(z) \simeq z \overline{e^H}$, it implies from 
\cite{footnoteH0} that $\hat H_0= \log \overline{e^H}  = 
\log  \int_\R \frac{\rmd x'}{\sqrt{4 \pi}}  e^{-(x')^2/(4)} Q_0(x')$
for a general initial condition, and $\hat H_0=- \frac{1}{2} \log (4 \pi)$
for the droplet IC.
\bibitem{footnoteb} where we recall that $b(q)$ depends on $z$.
\bibitem{footnoteprove} Deriving it from the WNT requires to extend our method to a non-decaying
$Q_0(x)$, a work in progress.
\bibitem{footnoteHz} Note that with our notations $H_{z=0}$ equals $\hat H_0$ and not $H_0$.
\end{thebibliography}
\end{document}